\documentclass[aps,prb,twocolumn, longbibliography]{revtex4-2} 
\usepackage{natbib}
\usepackage{epsfig}
\usepackage{graphicx}
\usepackage{xcolor}
\usepackage{amsmath}
\usepackage{amsfonts}
\usepackage{amssymb}
\usepackage{wrapfig}
\usepackage{textcomp}
\usepackage[colorlinks=true, linkcolor=blue]{hyperref}
\usepackage[utf8]{inputenc} 
\usepackage{textgreek}

\newcommand{\be}{\begin{equation}}
\newcommand{\ee}{\end{equation}}
\newcommand{\bml}{\begin{multline}}
\newcommand{\eml}{\end{multline}}

\usepackage{mathtools}
\usepackage{esint}
\usepackage{comment}

\begin{document}

\title{
Digital-Analog Simulations of Schr\" odinger Cat States in the Dicke-Ising Model
}
\author{Dmitriy S. Shapiro$^{1}$}
\email[]{d.shapiro@fz-juelich.de}
\author{Yannik Weber$^{2}$}
\author{Tim Bode$^{1}$}
\author{Frank K. Wilhelm$^{1,2}$}
\author{Dmitry Bagrets$^{1,3}$}

   \affiliation{$^1$Peter Gr\"unberg Institute, Quantum Computing Analytics (PGI-12), Forschungszentrum J\"ulich, 52425 J\"ulich, Germany}
\affiliation{$^2$Theoretical Physics, Universit\"at des Saarlandes, 66123 Saarbr\"ucken, Germany}
 \affiliation{$^3$Institute for Theoretical Physics, University of Cologne, 50937 K\"oln, Germany}

\begin{abstract}

The Dicke-Ising model, one of the few paradigmatic models of matter-light interaction, exhibits a superradiant quantum phase transition above a critical coupling strength. However, in natural optical systems, its experimental validation is hindered by a ``no-go theorem''. Here, we propose a digital-analog quantum simulator for this model based on an ensemble of interacting qubits coupled to a single-mode photonic resonator. We analyze the system's free energy landscape using field-theoretical methods and develop a digital-analog quantum algorithm that disentangles qubit and photon degrees of freedom through a parity-measurement protocol. This disentangling enables the emulation of a photonic Schrödinger cat state, which is a hallmark of the superradiant ground state in finite-size systems and can be unambiguously probed through the Wigner tomography of the resonator's field.
\end{abstract}

\maketitle

\section{Introduction}

The Dicke-Ising model has garnered significant attention in recent years due to the richness of its quantum phases~\cite{PhysRevLett.93.083001,Gammelmark_2011,Zhang:2014aa, 10.21468/SciPostPhys.1.1.004,PhysRevResearch.2.023131, 10.21468/SciPostPhysCore.7.3.038, langheld2024quantumphasediagramsdickeising, PhysRevLett.133.106901}. A central feature of this model is the competition between spin-spin interactions, which tend to drive an Ising transition, and the collective Dicke coupling, which leads to superradiant photon condensation. This interplay results in a more complex superradiant quantum phase transition (QPT) compared to that in the conventional Dicke model~\cite{popov1988functional,emary2003chaos, eastham2001bose,dalla2013keldysh, PhysRevA.94.061802, kirton2018introduction,PhysRevA.102.023703}.

In natural optical systems, the superradiant QPT is generally considered forbidden by a no-go theorem, which asserts that the diamagnetic term—proportional to the square of the vector potential $\hat{\mathbf{A}}^2$—prevents photon condensation. However, this constraint can be circumvented~\cite{PhysRevA.75.013804,PhysRevLett.104.023601,PhysRevLett.107.113602} in quantum simulators such as cold atom lattices~\cite{Baumann:2010aa,PhysRevLett.110.090402,PhysRevLett.113.020408,doi:10.1073/pnas.1417132112, PhysRevLett.121.040503, PhysRevLett.115.230403,Ferioli2023, PhysRevX.14.011020} or circuit QED setups~\cite{PhysRevLett.103.083601,Feng2015,Yoshihara2017} (see also review articles~\cite{RevModPhys.91.025005,FriskKockum2019,RevModPhys.93.025005,QIN20241}), where the theorem is overcome on the physical level.

Another possibility to simulate the class of Dicke-like models could potentially be offered by quantum hardware, which relies on the Trotterized approximation of the evolution operator via quantum circuits. This approach, known as digital quantum simulation~\cite{Feynman1982, Lloyd:1996,  Weimer2010, BassmanOftelie_2021,Bravyi2024,PRXQuantum.5.040320}, has recently become a popular line of research and is considered one of the few promising future applications of quantum computers. Its validity has been successfully verified in numerous experiments, see recent reviews such as Ref.~\cite{Fauseweh2024}. However, simulating many-body correlated systems with spin-boson or fermion-boson interactions presents a distinct challenge: encoding bosonic fields using qubits. Specifically, encoding a single bosonic mode with a finite occupation-number cutoff $N_{\rm max}$ requires $\log_2 N_{\rm max}$ qubits~\cite{Macridin:2018, Macridin:2018a}. To address this problem, one may work within the alternative quantum digital-analog framework~\cite{Mezzacapo:2014aa, Langford:2017aa}. The latter employs bosonic degrees of freedom as a computational resource and enables boson-qubit entanglement at the hardware level. Within this scheme, the dynamics of the quantum Rabi model in the strongly coupled limit was simulated using a single transmon qubit coupled to a resonator~\cite{Langford:2017aa}. {\color{black} More recently, this concept has been suggested for simulating the   lattice gauge theories~\cite{crane2024} and Hubbard–Holstein model~\cite{Kumar2025}.} Our study leverages this methodology to simulate the dynamics governed by the more elaborate Dicke-Ising Hamiltonian, with a focus on novel qubit-boson architectures that are experimentally feasible within the context of circuit QED~\cite{Langford:2017aa, huber2024parametricmultielementcouplingarchitecture}. {\color{black} Superconducting platforms are particularly well-suited for this purpose due to the long coherence times of their resonators (on the order up to milliseconds~\cite{PhysRevB.94.014506,Ganjam2024}), which significantly exceed the timescales of analog evolution (hundred of nanoseconds~\cite{Langford:2017aa}). In contrast, implementing alternative systems such as trapped ions coupled to vibrational modes  is more challenging in this regard. The  decay time  (on the order of tens of milliseconds) are only about two  orders of magnitude greater than the relevant Rabi periods~\cite{Um2016, PhysRevLett.125.150505}. }

A particularly compelling aspect of our digital-analog approach is twofold: (i) the potential to simulate the transition into the superradiant phase via a quench protocol, and (ii) the ability to disentangle the photon condensate and the qubit degrees of freedom in the many-body density matrix. This gives the proposed simulation strategy a striking advantage over fully analog simulators~\cite{PhysRevLett.121.040503}, where only macroscopic parameters of the condensate have been available for direct measurement. Remarkably, this disentanglement of condensed photons can enable the emulation of Schrödinger cat states, which are a hallmark of the superradiant ground state in finite-size systems.

The paper is organized as follows. In Section~\ref{protocol}, we present the model and the main idea of creating a cat-state density matrix with the help of qubit-parity measurements.
In Section~\ref{Sec_QPT}, we provide a field-theory description of the superradiant QPT for different limits of the model. We introduce  the method of deriving the free energy using  path integrals in~\ref{Sec_free_en}. The mean-field results for  the conventional Dicke model and the Dicke-Ising model with spin-1/2 are discussed in~\ref{Sec_free_en_D} and \ref{Sec_free_en_DI}, respectively. 
The role of quantum fluctuations near the instanton trajectory and the relation to the Kibble-Zurek mechanism are addressed in~\ref{Sec_KZ}. The generalization of the Dicke-Ising model to spins larger than 1/2, via angular bosonization, is provided in~\ref{Sec_angular}.  A quasi-classical approach for angular fluctuations is presented in~\ref{angular_quasicl}. In Section~\ref{Sec_algorithm}, we present the quantum simulation algorithm. We discuss the idea of the superradiant ground state approximation via the quench in~\ref{quench_sec}. In~\ref{gates}, we present  digital-analog quantum circuits for Jaynes-Cummings, Rabi and Dicke gates; in~\ref{overview} we give an overview of the algorithm. In Section~\ref{discussion},  we discuss our results and present data for the exact and Trotterized dynamics; in the ending section~\ref{conclusions} we conclude.

{\color{black}The numerical code and research data supporting this study are publicly available  online~\cite{shapiro_2025_16581022}.}

\section{Main idea}
\label{protocol}

The  Dicke-Ising model ($\hbar=k_{\rm B}=1$ hereafter),
\be
\hat H_{\rm DI}=\hat H_{\rm D}-J \sum_{j=1}^{N-1}\hat\sigma_j^z\hat\sigma_{j+1}^z  , \label{H_DI}
\ee
is  a combination of the standard Ising model and the Dicke  Hamiltonian 
\be
\hat H_{\rm D}=\omega_0 \hat a^\dagger \hat a -  \omega_z \sum_{j=1}^N\hat \sigma^z_j+   \frac{g}{\sqrt N} (\hat a^\dagger+ \hat a) \sum_{j=1}^N\hat \sigma^x_j , \label{H_D}
\ee
which describes an ensemble of $N$ spin-$s$ degrees of freedom coupled to a common photon mode. In the Ising part of Eq.~\eqref{H_DI}, a positive coupling $J>0$ corresponds to a ferromagnetic spin-spin interaction. Through the Dicke part of Eq.~\eqref{H_DI}, the spins obtain excitation frequencies $\omega_z>0$, while  the photon mode has frequency $\omega_0$ and is described by the bosonic {\color{black} annihilation and creation} operators{\color{black}, $\hat a$ and $\hat a^\dagger$, commuting as} $[\hat a$, $\hat a^\dagger]=1$ {\color{black} and acting in the space of  photonic Fock states $|n\rangle$ as $\hat a|n\rangle = \sqrt{n}| n{-}1 \rangle$ and $\hat a^\dagger|n\rangle = \sqrt{n{+}1}| n{+}1 \rangle$ with $n$ being a photon number}; the  qubit-resonator coupling strength is denoted by $g$. For qubits, which correspond to spin $s=1/2$, we associate the logical $|0\rangle_j$ of qubit $j=1, ..., N$ to the eigenstate $(1, 0)^T$ of $\hat \sigma^z_j$ with the eigenvalue $1$.

At zero temperature,  $g$  plays the role of a control parameter of the superradiant QPT.  If $g$ is less than the critical value, $g_{\rm c}$, the system is in its normal phase, with a \textit{ferromagnetic} ground state
\be
|{\rm FM}\rangle = |0\rangle\otimes\prod_{j=1}^N |0_j \rangle .
\ee
In the \textit{superradiant} phase, where $g>g_{\rm c}$, there exist two quasi-degenerate superradiant many-body states $|\Psi_{ R}\rangle$ and $|\Psi_{L}\rangle$, and the highly entangled ground state becomes the superposition
\be
|{\rm SR}\rangle= \frac{1}{\sqrt2}\Big( |\Psi_{ L}\rangle  +  |\Psi_{ R}\rangle \Big) 
\ee
In a large spin ensemble, these wave functions are given by the direct products $|\Psi_{  R}\rangle= |{-}\alpha\rangle {\otimes}  |R\rangle$ and $|\Psi_{  L}\rangle=|\alpha\rangle{\otimes} |L\rangle $. Here, $|\!\pm\!\alpha\rangle \!=\!e^{-\frac{1}{2}|\alpha|^2} \!\sum\limits_{n\geq 0}\!\frac{(\pm\alpha)^n}{\sqrt{n!}}|n\rangle  $ are  photon coherent states  with opposite phases. The number of photons stored in these states, $|\alpha|^2$, can be macroscopically large. In the mean-field picture, the value of $\alpha$ is given by a free energy minimum. The qubit states $|R(L)\rangle=\prod_{j=1}^N (|0\rangle_j \pm |1\rangle_j)/\sqrt2$ are anti-parallel to each other (on their respective single-particle Bloch spheres).

Let us remember that in circuit QED, Schr\"odinger's cat state is the nonclassical state
\be
|{\rm cat}\rangle= \frac{1}{\sqrt2}\Big(   |\alpha\rangle + |\!-\!\alpha\rangle \Big), \label{cat_state}
\ee
which is a promising candidate for qubit encoding due to its non-locality in phase space~\cite{PhysRevA.87.042315, doi:10.1126/science.1243289, Grimm2020,Lescanne2020}, rendering it stable against  local perturbations provided the photon number is large. 

The central idea of our work is to disentangle $|\pm\alpha\rangle$ and $|R(L)\rangle$ from the joint many-body density matrix, $ \hat \rho_{\rm SR}=|{\rm SR}\rangle\hspace{-0.8mm}\langle{\rm SR}|$, 
thus emulating the cat state density matrix in the  photon basis, i.e.,
\be\label{eq:rho_cat}
  \hat \rho_{\rm cat}=  \frac{1}{2}\Big(|{-}\alpha\rangle\hspace{-0.8mm}\langle-\alpha| + |\alpha\rangle\hspace{-0.8mm}\langle\alpha| +|\alpha\rangle\hspace{-0.8mm}\langle{-}\alpha|+|{-}\alpha\rangle\hspace{-0.8mm}\langle\alpha|\Big).
\ee
Note that simply taking the trace over the qubit degrees of freedom in $\hat\rho_{\rm SR}$ results in a mixed-state density matrix 
\be
 \hat \rho_{\rm mix} = {\rm tr}_\sigma[\hat\rho_{\rm SR}]
 =   \frac{1}{2}\Big(|{-}\alpha\rangle\hspace{-0.8mm}\langle-\alpha| + |\alpha\rangle\hspace{-0.8mm}\langle\alpha| \Big), \label{rho_mix}
\ee
which lacks the coherent cross terms $ |{\pm}\alpha\rangle\hspace{-0.8mm}\langle{\mp}\alpha|$ that appear in Eq.~\eqref{eq:rho_cat}.  To obtain these cross terms, we select one half of the qubit states  corresponding to a given value of the total qubit parity. This selective parity measurement can be defined as
\be  \hat \rho_{+} = {\rm tr}_\sigma[ \hat \rho_{\rm SR} \hat P_+] \label{rho_P},
\ee
 where the {\color{black}positive parity  operator} is
 \be
 \hat P_+=\frac{1}{2}\left(\hat 1+\prod_{j=1}^N \hat \sigma^z_j\right),
 \ee
with $\hat P_+|R\rangle=\hat P_+|L\rangle= (|R\rangle+|L\rangle)/\sqrt{2}$. The trace with the $\hat P_+$ in Eq.~\eqref{rho_P} provides the desired result, i.e.\ $\hat  \rho_{+}=\hat  \rho_{\rm cat}$. Performing a  series of projective measurements following Eq.~\eqref{rho_P} enables us to observe non-classical cat state signatures in a subsequent Wigner tomography~\cite{PhysRevA.87.042315, doi:10.1126/science.1243289} of the photon mode.

Consider the Wigner function $W(x,p)$  corresponding to a reduced density matrix ${\hat \rho}={\rm tr}_\sigma[\hat{ \mathcal{O}}\hat \rho_{\rm SR}]$,
\be
W (x,p) = \sum\limits_{n,m=0}^\infty  \langle n| \hat \rho|m\rangle V_{n,m}(x,p)  , \label{W_xp}
\ee 
where
\be
    V_{n,m}(x,p) =  \int dy \frac{ H_n (x{-}y)H_m (x{+}y) e^{2ipy - (x^2 + y^2)}}{\sqrt{\pi^3  2^{n+m} n!m!}}  
\ee
are the harmonic oscillator eigenfunctions overlap integrals, and $H_n(x)\!=\!(-1)^n e^{x^2} d^n e^{-x^2}/dx^n$ are Hermite polynomials. 
Following to the definitions~(\ref{rho_mix}) and~(\ref{rho_P}), we have $\hat {\mathcal{O}}=\hat 1$  for the mixed state and $\hat {\mathcal{O}}=\hat P_+$ for the cat state.

\begin{figure}[b!]
\center\includegraphics[width=\linewidth]{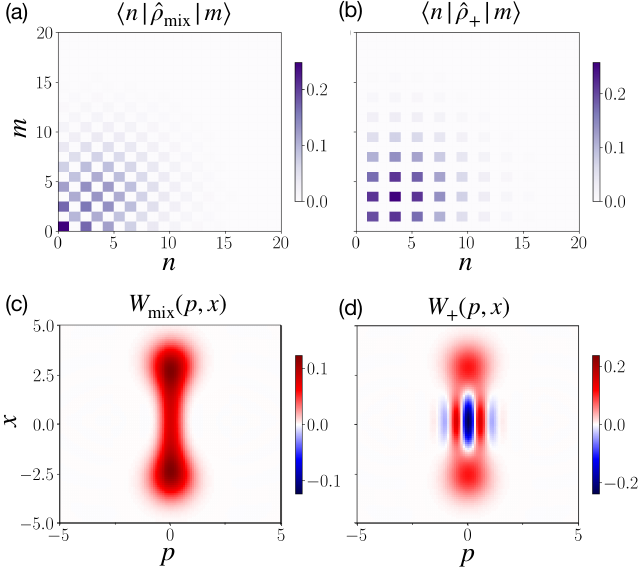} 
\caption{
Reduced density matrices (a) for the mixed state  $ \hat \rho_{\rm mix} = {\rm tr}_\sigma [\hat \rho_{\rm SR}]$ and (b) when projected to the positive-parity subspace $\hat\rho_+={\rm tr}_\sigma[\hat \rho_{\rm SR}\hat P_+]$. (c) Wigner function of the mixed state, and (d) of the projected state showing non-classical features indicative of a cat state.
The photon Hilbert space has a cutoff of 20 photons. The   coupling $g=0.9\sqrt{\omega_0J}$ is near the critical value $\tilde g_{\rm c}$, the chain has open ends and comprises $N=7$ qubits. The other parameters  are $J=\omega_0$ and $\omega_z=0.05\omega_0$.
}
\label{wigner_f} 
\end{figure}

The momentum-integrated Wigner function yields the photon probability distribution $w(x)=\int dp W(x,p) $, which reduces to 
\be
w(x,t)=  e^{-x^2} \sum_{n, m\geq 0} \frac{H_n(x)H_m(x)}{\sqrt{\pi 2^{n+m}n!m!}}  \langle n| \hat \rho(t)|m\rangle .
\ee 
In Fig.~\ref{wigner_f}(a) and (b) we present, respectively,  illustrations of  the matrix elements of   $\hat \rho_{\rm mix}$ and $\hat \rho_{+}$ for a finite system in the superradinant phase. The many-body density matrix $\hat\rho_{\rm SR}$ is found numerically via exact diagonalization of the Dicke-Ising Hamiltonian. One can observe that $\langle n|\hat\rho| m \rangle \neq 0$ if both of Fock state numbers $n$, $m$ are odd. In the thermodynamic limit, $W_{\rm mix}(x,p)$ would have two singular points at $x=\pm\sqrt{2}\alpha$; in a finite system near the critical point,   $W_{\rm mix}(x,p)$ has two linked blobs as shown in~\ref{wigner_f}~(c).
 
The projected Wigner function  $W_+(x,p)$ calculated  from $\hat\rho_+$ is shown  in Fig.~\ref{wigner_f}(d). The signatures of  Schr\"odinger's cat state are visible as fringes of  negative quasi-probability, $W_+(x,p){<}0$, which is a benchmark for the presence of cat states in the output of our algorithm given below.

{\color{black}We note that the Wigner function for negative parity, $ W_{-} = W_{\rm mix} - W_{+}$, where the corresponding projector is $\hat{P}_{-} = \hat{1} - \hat{P}_{+} $, also exhibits cat-state signatures, albeit less prominently. The choice of positive parity in our protocol follows from the fact that in the $g = 0$ limit, the system has positive parity (i.e., $\langle \hat{P}_{+} \rangle = 1$ and $\langle \hat{P}_{-} \rangle = 0$), which is a feature of the normal \textit{ferromagnetic} state supported by $ J > 0$ and $\omega_z > 0$. Near the critical coupling $ g_{\rm c}$, or under quench dynamics, the system becomes partially superradiant \textit{paramagnetic}, and the expectation value satisfies $1/2 < \langle \hat{P}_{+} \rangle < 1$. Deep in the superradiant phase, $\langle \hat{P}_{+} \rangle$ and $\langle \hat{P}_{-} \rangle$ approach $1/2$, indicating that the probabilities to observe positive and negative parity become balanced.
}

The remainder of the paper has two main threads: (i) Applying methods of statistical physics to derive a profile of the free energy. This sheds light on the order of the QPT as well as on the quantum fluctuations around the mean-field solutions and the associated macroscopic quantum tunneling   between the superradiant states $|\Psi_{\rm L}\rangle$ and $|\Psi_{\rm R}\rangle$.  (ii) Based on this understanding of the free energy profile, we then perform quantum-circuit simulations of the real-time dynamics of the photon distribution in the corresponding effective potential, culminating in the sought-for cat states.

\section{Superradiant quantum phase transitions}
\label{Sec_QPT}
In this section, we start with a recapitulation of the well-known result about the second-order QPT in the conventional Dicke model.  After that, we turn to the Dicke-Ising Hamiltonian Eq.~\eqref{H_DI} with $J\neq0$ and $\omega_z=0$; the limit $\omega_z\to 0$ enables an exact  calculation of the trace over the spins. It can be performed for qubits ($s=1/2$) via the usual Jordan-Wigner transformation, resulting in a mean-field solution for the free energy that predicts a first-order QPT.  Besides that, the free energy determines an instanton trajectory in Matsubara imaginary time and, therefore, the rate of macroscopic quantum tunneling.

For $\omega_z\ll J$, in the superradiant phase, the order-parameter fluctuations are critical because the magnon excitations become gapless (Eq.~\eqref{magnon_spec}). 
If the fluctuations are unstable, the Gaussian approximation is not sufficient. If $s$ is large (qudit case), and the photon mode has a low frequency, then fluctuations are suppressed; the mean-field solution becomes asymptotically exact in this case. 
At finite $\omega_z\sim J$, integrating out the magnons exactly via the Jordan-Wigner transformation is more challenging. As an alternative approach, we therefore suggest the angular representation of spins, which is valid for arbitrary $s$.

\subsection{Methods}
\label{Sec_free_en}
To calculate the free energy $F$ as a function  of the superradiant order parameter, we recall the relation between $F$ and the partition function $Z={\rm tr} e^{-\hat H/T}$ at finite temperature $T$,
\be
Z=e^{-F/T},
\ee
where the trace is taken over all degrees of freedom. Field-theory methods enable one to represent the trace as a path integral over complex bosonic fields, $a$ and $\bar a$, while the fields $ \vec\sigma$ parameterize the spin sector, i.e. the partition function may be written as
\be
Z=\int  d [a, \bar a , \vec\sigma] e^{-S[a, \bar a , \vec\sigma]}. 
\ee
Upon transformation from a Hamiltonian to the path integral, the  real part of the photon operators becomes a trajectory $u(\tau)$ on the Matsubara time interval $\tau\in[0,1/T]$,
\be
 \frac{1}{\sqrt{N}}(\hat a +\hat a^\dagger) \to u (\tau) .
\ee
The trajectories $u(\tau)$ are slow if $N$ is large (thermodynamic limit) and the photon frequency $\omega_0$ is small. This corresponds to the mean-field limit, where $u$ can be associated with the  superradiant order parameter. The photon position operator, $\hat x=\frac{1}{\sqrt 2}(\hat a +\hat a^\dagger)$, is related to  $u$ by $\hat x\to\sqrt{N/2} u$.

The idea now is to calculate the  path integral  over all \textit{fast} fields and represent the partition function as a single path integral over the \textit{slow} quantum field $u$, i.e.\ $Z=\int D[u] \exp{(-S_{\rm eff}[u])}$; the new functional in the exponential is the effective action for $u$. The free energy follows from $S_{\rm eff}$ if we neglect the slow time dependence of $u(\tau)$ to obtain 
\be
F(u)=TS_{\rm eff}[u={\rm const.}].
\ee
The low-temperature action $S_{\rm eff}$ is proportional to $1/T$; hence, $T$ drops from all the formulas for the free energy.

Note that  the  momentum operator $\hat p=\frac{i}{\sqrt2}(\hat a -\hat a^\dagger)$ does not appear in the interacting part of the Hamiltonian~\eqref{H_DI}. Therefore, in the path integral, the real field $v$ corresponding to $i\hat p$ appears only in the free photon Matsubara action, namely
\be
    S_0 =\int\limits_0^{1/T} \! d\tau\, \bar a( \partial_\tau +\omega_0) a,
\ee
where $a(\tau)$ and $\bar a(\tau)$ are complex bosonic fields. If we make a rotation to the real fields,  $u=(a+\bar a)/\sqrt{N}$ and $v=i(a-\bar a)/\sqrt{N}$, and integrate out the field $v$, we arrive at the  free action for the order parameter,
\be
S_0[u]=N\int\limits_0^{1/T}\! d\tau\, \mathcal{L}[u(\tau)] ,
\ee
with the Lagrangian 
\be
 \mathcal{L}=\frac{\left(\partial_\tau u\right)^2}{4\omega_0}+\frac{\omega_0}{4} u^2.
\ee
This is the sum of the kinetic term $\sim(\partial_\tau u)^2$ and the potential energy $\mathcal{F}_0= \omega_0u^2/4$, which determines the parabolic free energy profile. In the next Section, we show how the interaction with the spins contributes additional terms to $\mathcal{F}$.

\subsection{Free energy in the Dicke model} 

\label{Sec_free_en_D}

Consider the  conventional Dicke Hamiltonian (\ref{H_D}). In what follows, we work with the normalized   free energy, $\mathcal{F}=F/N$, which, in the thermodynamic limit, reads
\be
 \mathcal{F}_{\rm D}(u)=\frac{\omega_0}{4} u^2- \omega_z\left(\sqrt{1+\frac{g^2}{\omega_z^2}u^2}-1\right) . \label{F_D}
\ee
This result has been derived by integrating out the qubit states, which can be done via different spin representations such as Holstein-Primakoff bosonization~\cite{emary2003chaos} or bilinear combinations of fermion fields~\cite{popov1988functional,eastham2001bose,PhysRevA.102.023703}. The Dicke free energy $\mathcal{F}_{\rm D}(u)$ exhibits a second-order QPT, as shown in Fig.~\ref{free_en}~(a-c). In the normal phase $g<g_{\rm c} = \sqrt{\omega_0\omega_z/2}$ below the critical coupling, there is only one minimum at $u=0$. At  $g=g_{\rm c}$, the QPT occurs. Finally, there is a superradiant phase at $g>g_{\rm c}$, which means that $\mathcal{F}_{\rm D}(u)$ acquires two minima at $u=\pm u_0$, $u_0>0$, and the system  spontaneously relaxes to one of them.

\subsection{Free energy in the  Dicke-Ising model at $\omega_z=0$ and $s=1/2$}

\label{Sec_free_en_DI}
\begin{figure}[htb!]
\center\includegraphics[width=\linewidth]{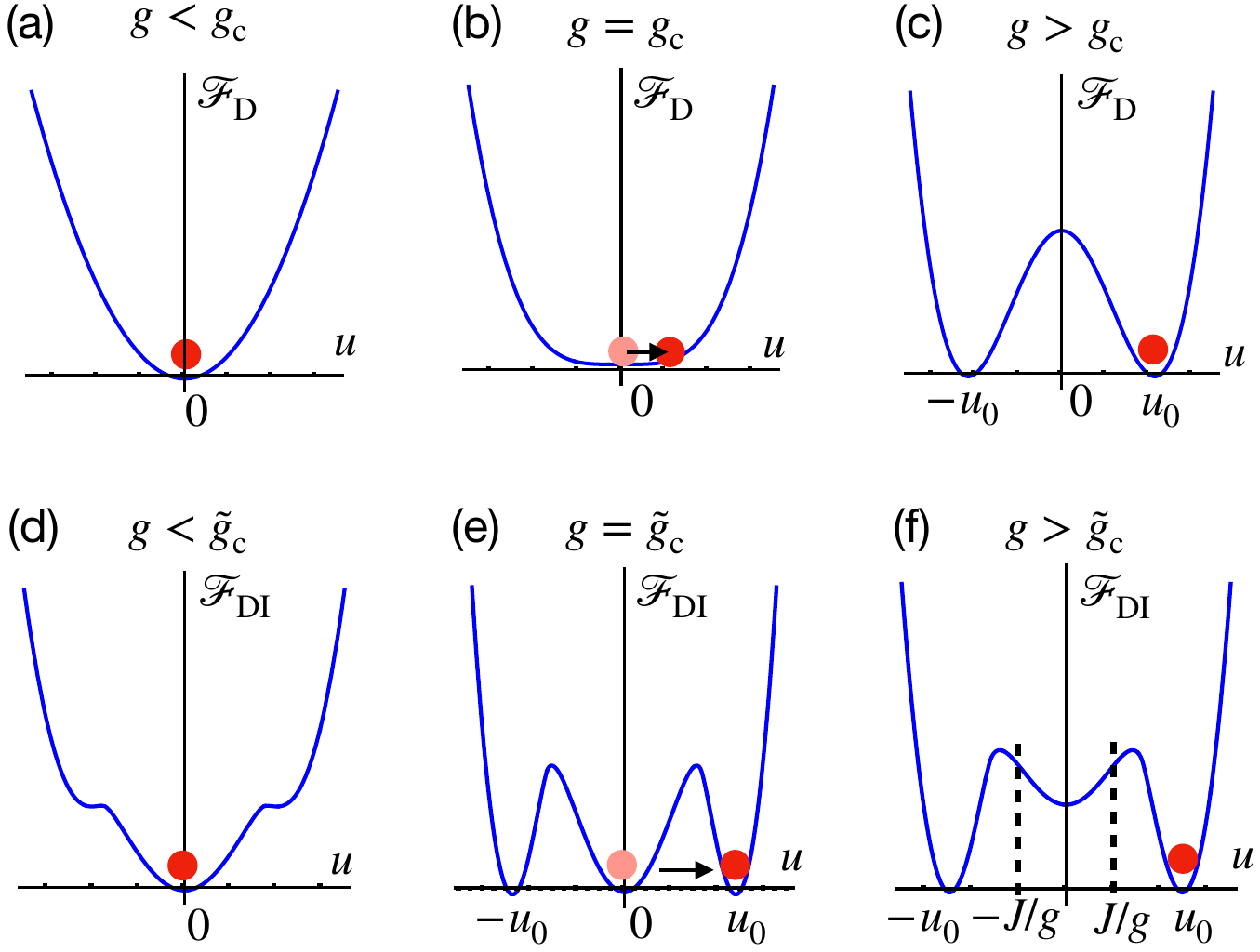} 
\caption{
{Sketch of the free energies as functions of the superradiant order parameter for (a-c) the  Dicke   and (d-f) the  Dicke-Ising  model.  (a, d) Normal phases. Critical points   (b) of the second-order  and (e) first-order QPTs. (d, f) Superradiant phases. The values $u=\pm J/g$ in (f) correspond to the critical Ising chain.}
}
\label{free_en} 
\end{figure}

Coming back to the Dicke-Ising Hamiltonian with $J>0$, we consider the limit of $\omega_z \to 0$. In Appendix~\ref{App_JW}, we apply the Jordan-Wigner transformation for spin operators and perform the subsequent integral over the fermion fields. As a result, we obtain the mean-field free energy  
\be
\mathcal{F}_{\rm DI}(u)=\frac{\omega_0}{4} u^2- \frac{2}{\pi }  (g|u|+J) {\rm E}\left(\frac{4 g|u| J}{(g |u| +J)^2}\right), \label{F_DI}
\ee
where the elliptic function ${\rm E}(x)$ results from an integral over the quasi-momentum in the Brillouin zone ${\bf k}\in (-\pi, \pi)$. One obtains two bands of Ising-chain magnons in the Brillouin zone, and their spectrum is
\begin{equation}
\epsilon({\bf k})=\pm2  \sqrt{  g^2 u^2(\tau )+ J^2-2 J g u \cos {\bf k}} . \label{magnon_spec}
\end{equation}    
In contrast to $\mathcal{F}_{\rm D}$, the function $\mathcal{F}_{\rm DI}$ has three minima in a certain range of $g$ around the critical $\tilde g_{\rm c}= c_0\sqrt{\omega_0J}$, $c_0\approx 0.9$. The first-order QPT occurs when the two side minima become lower than the central minimum at $u=0$ (see Fig.~\ref{free_en}(d-f)). Note that our approach is complementary to previous studies of phase transitions in this model~\cite{PhysRevLett.93.083001,Gammelmark_2011,Zhang:2014aa, 10.21468/SciPostPhys.1.1.004,PhysRevResearch.2.023131, 10.21468/SciPostPhysCore.7.3.038, langheld2024quantumphasediagramsdickeising, PhysRevLett.133.106901}.

\subsection{Instanton approach. Relation to Ising transition and Kibble-Zurek mechanism}
\label{Sec_KZ}

The free energy  $\mathcal{F}_{\rm DI}$  given by   (\ref{F_DI}) is  part of a mean-field Matsubara action where $u(\tau)$ is a trajectory in imaginary time,
\begin{equation}
    S_{\rm mf}= N\int\limits_0^{1/T}d\tau \left(\frac{\left(\partial_\tau u\right)^2}{4\omega_0} +\mathcal{F}_{\rm DI}(u) \right)  . \label{S_mf_app}
\end{equation} 
Variation of this action yields an instanton equation that describes  macroscopic quantum tunneling   between the minima of the free energy at  $u= \pm u_0$, see Appendix~\ref{App_instanton}. The solution of the instanton trajectory, $u_{\rm inst}(\tau)$, can be defined implicitly via 
\begin{equation}
    \tau = \int\limits_{-u_0}^{u_{\rm inst}(\tau)} \frac{ d  u}{ \sqrt{ 2 \omega_0 (\mathcal{F}_{\rm DI}( u) -\mathcal{F}_{\rm DI}(-u_0))}}  . \label{instanton}
\end{equation}
We find that in the superradiant phase, the instanton trajectory always  crosses two special   points $u=\pm J/g$, see Fig.~\ref{free_en}(f).  According to Eq.~\eqref{magnon_spec} for the magnon spectrum, the Ising chain becomes critical  due to  the  gap closing at these  points. This crossing  occurs  because $u_{\rm inst}$ has support in the interval $[-u_0; u_0]$, which includes these special points  since  $|J/g| < u_0$, as implied by Eq.~\eqref{F_DI}.   We conclude that the fluctuations above the QPT are non-vanishing; in other words, the system remains critical in the superradiant phase. This  behavior contrasts with the conventional Dicke model, which is critical only at the transition point. We can also draw an imaginary-time analogy of the Kibble-Zurek mechanism around the second-order Ising QPT. In our case, the Ising transition is virtual and hidden in the superradiant phase.

\subsection{Angular representation at $\omega_z\neq 0$ and $s \geq 1/2$}
\label{Sec_angular}

As mentioned above, the presence of special points on the instanton trajectory indicates critical fluctuations  at $\omega_z\!=\!0$ in this model, attributed to the gapless spectrum.
However, the gap re-emerges at finite $\omega_z$.
The analytic derivation of $\mathcal{F}_{\rm DI}$ for this general case of $\omega_z\neq 0$ is more involved: Neither the Jordan-Wigner nor the Majorana representation of the Pauli operators  yields a quadratic action over fermions. Therefore, the exact Gaussian integration over spin states used in  the derivations of Eq.~(\ref{F_D}) and (\ref{F_DI}), does not apply.

\begin{figure}[htb!]
\center\includegraphics[width=0.99\linewidth]{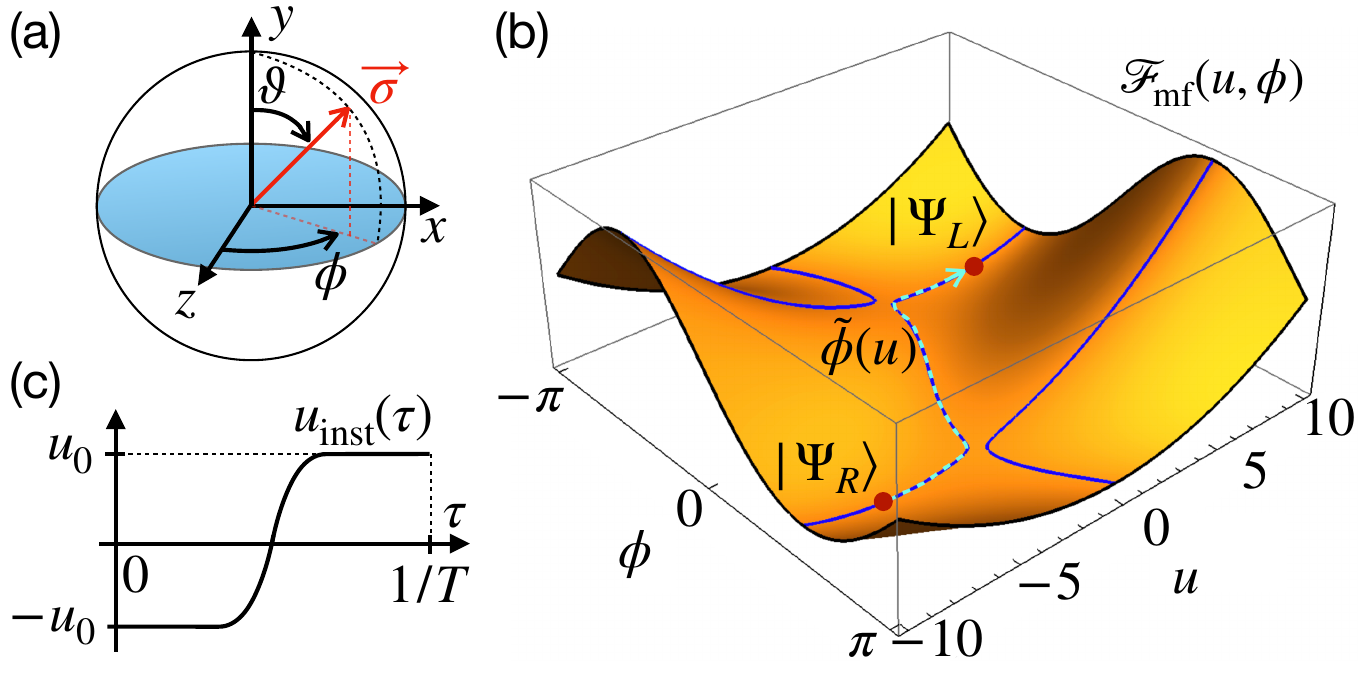}
\caption{
{(a) Angular representation of qubit states.  The $xz$-plane contributing most to the mean-field solution is shown in blue. (b) The effective potential in the mean-field approximation. The red dots are the two minima representing the superradiant states $|\Psi_{R,L}\rangle$. The dashed curve is an instanton trajectory. (c) Schematic representation of the  instanton trajectory $u_{\rm inst}(\tau)$.}
}
\label{angular} 
\end{figure}

An alternative representation of spins with arbitrary $s$ is provided by (sketch shown in Fig.~\ref{angular}(a))
\begin{align}
    \begin{split}
         \hat\sigma^x_j &\to 2s\sin\vartheta_j\sin\phi_j , \\
         \hat\sigma^y_j &\to 2s\cos\vartheta_j, \\
         \hat\sigma^z_j &\to 2s\sin\vartheta_j \cos\phi_j.\label{angular_repr}
    \end{split}
\end{align}
As the number of excited states is  $2s$,  qubits correspond to $s=1/2$, \textit{qutrits} to $s=1$, and so on. In the path integral for the partition function, $\vartheta_j$ and $\phi_j$ are real bosonic fields,
\be
Z=\int  d [\{\phi, \vartheta\}_j^N, u]\exp(-S[\{\phi, \vartheta\}_j^N,u]),
\ee
and the full Matsubara action reads
\be
S=S_{\rm WZNW} + \int\limits_0^{1/T}    d \tau\, (\bar a\partial_\tau a+H_{\rm DI}). \label{WZNW}
\ee
This is the sum of the Wess-Zumino-Novikov-Witten action 
\be
S_{\rm WZNW}=- is\sum_j \int\limits_0^{1/T} \! d \tau \dot \phi_j (1-\cos\vartheta_j) ,
\ee 
i.e.\ the integral over the spin Berry phase, the kinetic term for the  photon field $\sim \bar a \partial_\tau a$, and finally the Dicke-Ising Hamiltonian parameterized by Eq.~(\ref{angular_repr}).

In the limit of small photon frequency, $\omega_0{\ll} J,\, g$ and finite $\omega_z$, a mean-field solution can be found as this regime of energy scales allows one to consider $u(\tau)$ as a \textit{slow} variable. 
The trajectories contributing to $Z$ then  effectively reside near the $xz$-plane, with small, rapid out-of-plane fluctuations, i.e.
\be
    \vartheta_j(\tau)= \frac{\pi}{2}+\Delta\vartheta_j(\tau).
\ee
The corresponding geometric interpretation is shown in Fig.~\ref{angular}(a). The logic of separation into slow and fast components is also applicable to the in-plane components 
\be
\phi_j(\tau)=\phi(\tau)+\Delta\phi_j(\tau) ,
\ee
with a slow collective angle $\phi(\tau)$ and rapid fluctuations $\Delta\phi_j(\tau)$.

\subsection{Quasi-classical approach for fluctuations}
\label{angular_quasicl}

Further progress can be achieved in the quasi-classical limit where one assumes that the fluctuations near the mean-field trajectory are small. We start from the mean-field solution for free energy, neglecting all angle fluctuations, i.e.\ $\Delta\vartheta_j(\tau)=\Delta\phi_j(\tau)=0$. We also assume that $\omega_0$ is smaller than other energy scales, which guarantees that $u(\tau)$ and $\phi(\tau)$ are slow. The resulting mean-field action reads
\be
S_{\rm mf}=N\int\limits_0^\infty  d \tau \left(\frac{(\partial_\tau u)^2}{4\omega_0}  + \mathcal{F}_{\rm mf}(u,\phi)\right ),
\ee
where the mean-field free energy is 
\be
\mathcal{F}_{\rm mf}(u,\phi)=\frac{1}{4}\omega_0u^2 +h(u,\phi). \label{h}
\ee
Here, $h(u,\phi)$ corresponds to the spin part of the Hamiltonian with homogeneous configurations of $\phi_j=\phi$ and $\vartheta_j=\pi/2$,
\be
h(u,\phi)=2s\left(-\omega_z\cos\phi+gu\sin\phi -2sJ\cos^2\phi\right) . \label{F_mf}
\ee
The  profile of (\ref{F_mf}) in the superradiant phase is shown in Fig.~\ref{angular}(b), where the two minima correspond to the distinct macroscopic superradiant states $|\Psi_{\rm R}\rangle$ and $|\Psi_{\rm L}\rangle$. 
As long as there is no time-derivative term for $\phi$ in $S_{\rm mf}$,  non-trivial contributions to $Z$ are given only by the single quantum trajectory $\tilde \phi(u_{\rm inst})$ that connects these two minima, i.e.
\be
Z\sim e^{-S_{\rm mf}[u_{\rm inst}(\tau)]}.
\ee
The value of $Z$ then determines the amplitude of macroscopic quantum tunneling. 

The angular dependence $\tilde \phi(u)$ is determined by the condition $\partial_\phi \mathcal{F}_{\rm mf}(u,\phi)=0$. The motion of $u(\tau)$ along the quantum trajectory is described by a modified instanton equation
\be
 \frac{\ddot u}{2\omega_0} +\partial_u \mathcal{F}_{\rm mf}\left(u,\tilde\phi(u)\right)=0  \label{instanton_eq}
\ee
with boundary conditions $u(0)\!=\!-u(1/T)\!=\!-u_0$. Again, a sketch of the instanton solution is shown in Fig.~\ref{angular}(c). 

{
The next step of our quasi-classical approach is to calculate the quantum corrections $\mathcal{F}_{\rm fl}$ to the mean-field potential $\mathcal{F}_{\rm mf}$ caused by the Gaussian fluctuations of $\Delta\vartheta_j(\tau)$ and $\Delta\phi_j(\tau)$ neglected previously. To this end, we expand $S$  given by (\ref{WZNW}) to second order in $\Delta\vartheta_j(\tau)$ and $\Delta\phi_j(\tau)$,
\be
S=S_{\rm mf}[u,\phi]+S_{\rm G}[u,\phi,\Delta\vartheta ,\Delta\phi]. \label{S_2}
\ee
 This  is a sum of the mean-field action (\ref{F_mf}) and the Gaussian part $S_{\rm G}$ given by
\be
\!\!\!S_{\rm G} \!=\!\frac{1}{2}  \!\!\int\limits_{0}^{1/T}\!\!\! d \tau\!\!\!\int\limits_{-\pi}^\pi\!\!\frac{ d {\bf k}}{2\pi} \!\begin{bmatrix}
\Delta \vartheta &\! \!\Delta \phi
\end{bmatrix}_{{-}{\bf k}} \!\!
\begin{bmatrix}
\mathcal{A}(u,\phi) & \!\!- is \partial_\tau \\ 
 is \partial_\tau & \!\!\mathcal{B}_{\bf k}(u,\phi)
\end{bmatrix}\!\!
\begin{bmatrix}
\!\Delta \vartheta \\  \!\Delta \phi
\end{bmatrix}_{\! \bf k}, \!\!\label{S_G}
\ee
where the matrix elements depend on the slow trajectories. Note that the action does not have a linear contribution because  we assume that  $\sum_j\Delta\vartheta_j=\sum_j\Delta\phi_j=0$. For the term ${\sim}\Delta\vartheta^2$, the   element $\mathcal{A}$ partially  coincides with $h$ from (\ref{h}),
\be
\mathcal{A}(u,\phi)=4Js^2\cos^2\phi-h(u,\phi).
\ee
The amplitude of the term ${\sim}\Delta\phi^2$  involves the momentum dependence
\be
\mathcal{B}_{\bf k}(u,\phi) = \mathcal{A}(u,\phi) -8Js^2\sin^2\phi\cos{\bf k}.
\ee
The stability of the action (\ref{S_G}) along the trajectory  $\tilde \phi(u)$ is provided by the joint condition 
$\mathcal{A}(u,\phi){>}0$ and $\mathcal{B}_{\bf k}(u,\phi){>}0$   for all $\bf k$. It is     equivalent to 
\be
4sJ\cos2\phi+\omega_z\cos\phi-gu\sin\phi>0.\label{cond}
\ee
Assuming that (\ref{cond}) is satisfied and $u$ and $\phi$ are slow (adiabatic limit), the  Gaussian integration over the fields $\Delta \vartheta_{\bf k}$ and $\Delta \phi_{\bf k}$ yields the fluctuation correction $S_{\rm fl}[u,\phi]$ to the mean-field action. This correction reads
\be
S_{\rm fl}= N\int\limits_{-\pi}^\pi\!\!\frac{ d {\bf k}}{2\pi}\ln \prod_{n\geq 1}\left(1+\frac{\mathcal{A}(u,\phi)\mathcal{B}_{\bf k}(u,\phi)}{(2\pi sT n)^2}\right).
\ee
We calculate the infinite product over the Matsubara index $n\geq 1$ using the identity~(\ref{product}) from the Appendix.  After that, we take the limit $T\to 0$ in the definition for the free energy correction $\mathcal{F}_{\rm fl}=N^{-1}TS_{\rm fl}$ and find 
\be
\mathcal{F}_{\rm fl}(u,\phi)=\frac{1}{2s}\int\limits_{-\pi}^\pi\frac{ d {\bf k}}{2\pi}\sqrt{\mathcal{A}(u,\phi)\mathcal{B}_{\bf k}(u,\phi)}.
\ee
The integral over $\bf k$ yields
\begin{multline}
\mathcal{F}_{\rm fl}(u,\phi)=\frac{1}{\pi s} \sqrt{\!\mathcal{A}(u,\phi)\left(\mathcal{A}(u,\phi)\!+\!8s^2J\sin^2\!\phi\right)} \ \times \\
\times{\rm E}\Big(\frac{16s^2J\sin^2\phi}{\mathcal{A}(u,\phi)+8s^2J\sin^2\phi}\Big).
\end{multline}
In the Dicke-model limit, $J\!=\!0$, we use $\mathcal{A}\!=\!-h$ and  obtain 
\be
\mathcal{F}_{\rm fl}{=} -\frac{1}{2s}h(u,\phi).
\ee
The full free energy then reads
\be
\mathcal{F}_{\rm mf}+\mathcal{F}_{\rm fl}= \frac{1}{4}\omega_0u^2+\left(1-\frac{1}{2s}\right)h(u,\phi). \label{f_mf_fl_D}
\ee
One can see from (\ref{f_mf_fl_D}) that
the fluctuation  correction is small at  $s\!\gg\!1 $. 
The  Dicke model for large spins has been studied recently in Ref.~\cite{PhysRevA.104.043708} where the authors predicted  multicritical behavior at QPT.
For nonzero $J$, one also finds  $\mathcal{F}_{\rm fl}/\mathcal{F}_{\rm mf}\sim s^{-1}$. Therefore, in the large-spin limit, the fluctuations are small and the quasi-classical approach  is legitimate. 

In the following, based on our understanding of the free energy profile $\mathcal{F}$, we formulate a quench protocol for the simulation of the condensate dynamics.}

\section{Hybrid quantum circuit}
\label{Sec_algorithm}
\subsection{Approximation of the superradiant ground state via quench}
\label{quench_sec}

Our remaining objective is to obtain an approximate superradiant state  from a finite quantum circuit, which takes the form of unitary evolution on the time interval $t\in [0;t_{\rm f}]$ starting from the trivial ferromagnetic state for spin $s=1/2$, {\color{black}$|{\rm FM}\rangle$}. Note that $|{\rm FM}\rangle$ is an eigenstate of $\hat H_{\rm DI}$ at $g=0$. The final state  $|\Psi{\color{black}(t)}\rangle=e^{-i \hat H_{\rm DI}t}{\color{black}|{\rm FM}\rangle}$ is supposed to be close to the exact eigenstate $|{\rm SR}\rangle$. The evolution with $\hat H_{\rm DI}$ can be understood as a quench after the coupling $g$ is switched on at $t=0$.

\begin{figure}[h!]
\center\includegraphics[width=0.9\linewidth]{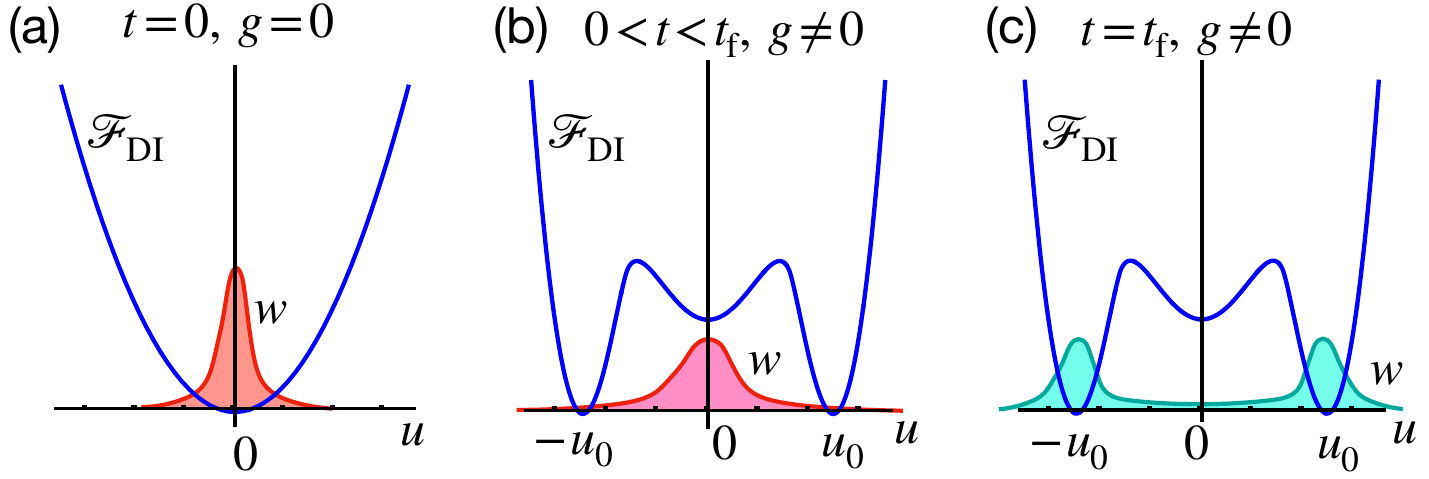} 
\caption{
{Sketch illustrating the quench dynamics of the photon probability distribution  $w(x,t)$ in the potential formed by the free energy $\mathcal{F}_{\rm DI}(u)$. (a) Gaussian     $w(x,t)$ at $t=0$ when $g=0$. (b) Evolution of $w(x,t)$ after the quench of $g$ from 0  to $g\approx g_c$. (c) End of the  evolution at $t=t_{\rm f}$ shows two maxima of  $w(x,t_{\rm f})$  at $u=\pm u_0$    corresponding to superradiant  condensates.}
}
\label{quench} 
\end{figure}

In coordinate representation, the photon distribution at $t=0$ is a Gaussian wave packet, i.e.
\be
    w(u, t\!=\!0)=\frac{1}{\sqrt\pi}e^{-u^2 N},
\ee
which can be interpreted as an eigenstate of a particle in the parabolic  free energy profile $\mathcal{F}_{\rm DI}(u)$ at $g=0$, see Fig.~\ref{quench}(a). We note that, when sending $N\to \infty$, the wave packet $w(u,t)$ tends to a $\delta$-singularity, which is the classical limit. 

As sketched in Fig.~\ref{quench}(b), the quench induces an instantaneous change to $\mathcal{F}_{\rm DI}(u)$, and the Gaussian packet tunnels into the side minima; the evolution should stop at the moment $t=t_{\rm f}$ when $w(u, t_{\rm f})$ is concentrated in either of these minima (Fig.~\ref{quench}~(c)). The many-body wave function $|\Psi(t)\rangle$ then contains a substantial amount of condensed photons, which is used in the further protocol detailed below.

\subsection{Rabi, Jaynes-Cummings, and Dicke gates}\label{gates}

\begin{figure}[t!]
\center\includegraphics[width=0.99\linewidth]{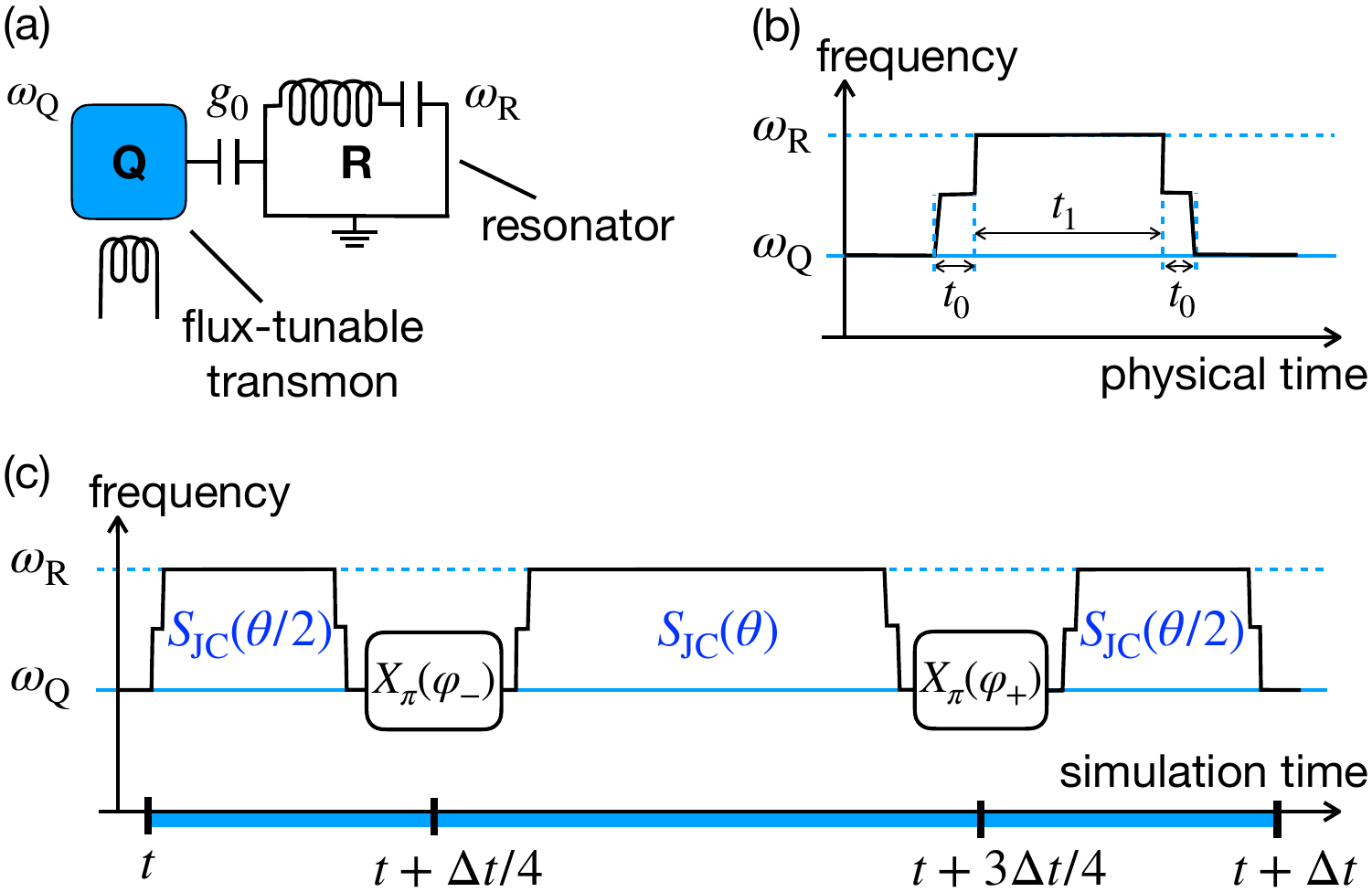} 
\caption{
{(a) Qubit-resonator architecture for the Rabi model. (b) Resonant pulse for the Jaynes-Cummings gate. (c) Pulse sequence representing the Rabi gate.}
}
\label{rabi_pulse} 
\end{figure}

We take inspiration from the { digital-analog} approach of Ref.~\cite{Mezzacapo:2014aa, Langford:2017aa}, where the authors suggested to simulate the quantum Rabi model
\be
\hat H_{\rm R}=\omega_0\hat a^\dagger \hat a   + g(\hat a +\hat a ^\dagger)\hat \sigma^x 
\label{HR}
\ee
through a combination of single-qubit rotations and a hybrid Jaynes-Cummings (JC) gate 
\be
\hat S_{\rm JC}(\theta)= \exp(-i\theta(\hat a^\dagger\hat \sigma^-+\hat a\hat \sigma^+)) , \label{JC}
\ee
where {$\hat \sigma^\pm=\frac{1}{2}(\hat \sigma^x\mp i \hat \sigma^y)$.} The JC gate enables  efficient rotations in the joint qubit-resonator Hilbert space. 
We have in mind an architecture as shown in Fig.~\ref{rabi_pulse}(a), where a tunable transmon qubit (Q) with  physical frequency $\omega_{\rm Q}$  is coupled to a superconducting resonator (R), modeled as an $LC$-circuit with fundamental frequency $\omega_{\rm R}$. The JC gate can be implemented as a flux pulse applied to Q, similar to two-qubit $XY$-gates~\cite{PhysRevX.5.021027}. A sketch of the pulse envelope is shown in Fig.~\ref{rabi_pulse}(b). The pulse tunes $\omega_{\rm Q}$ into a resonance with  $\omega_{\rm R}$ during the \textit{physical} time $t_1$, which enables the system to  acquire the desired relative phase $\theta=g\Delta t=g_0 t_1$ where $\Delta t$ is the Trotterization time \textit{in the simulation}, and $g_0$ is the \textit{physical} qubit-resonator coupling. The additional buffer steps $t_0$ may be used to gauge out dynamic phases in $\hat S_{\rm JC}$.

\begin{figure}[htb!]
\center\includegraphics[width=0.99\linewidth]{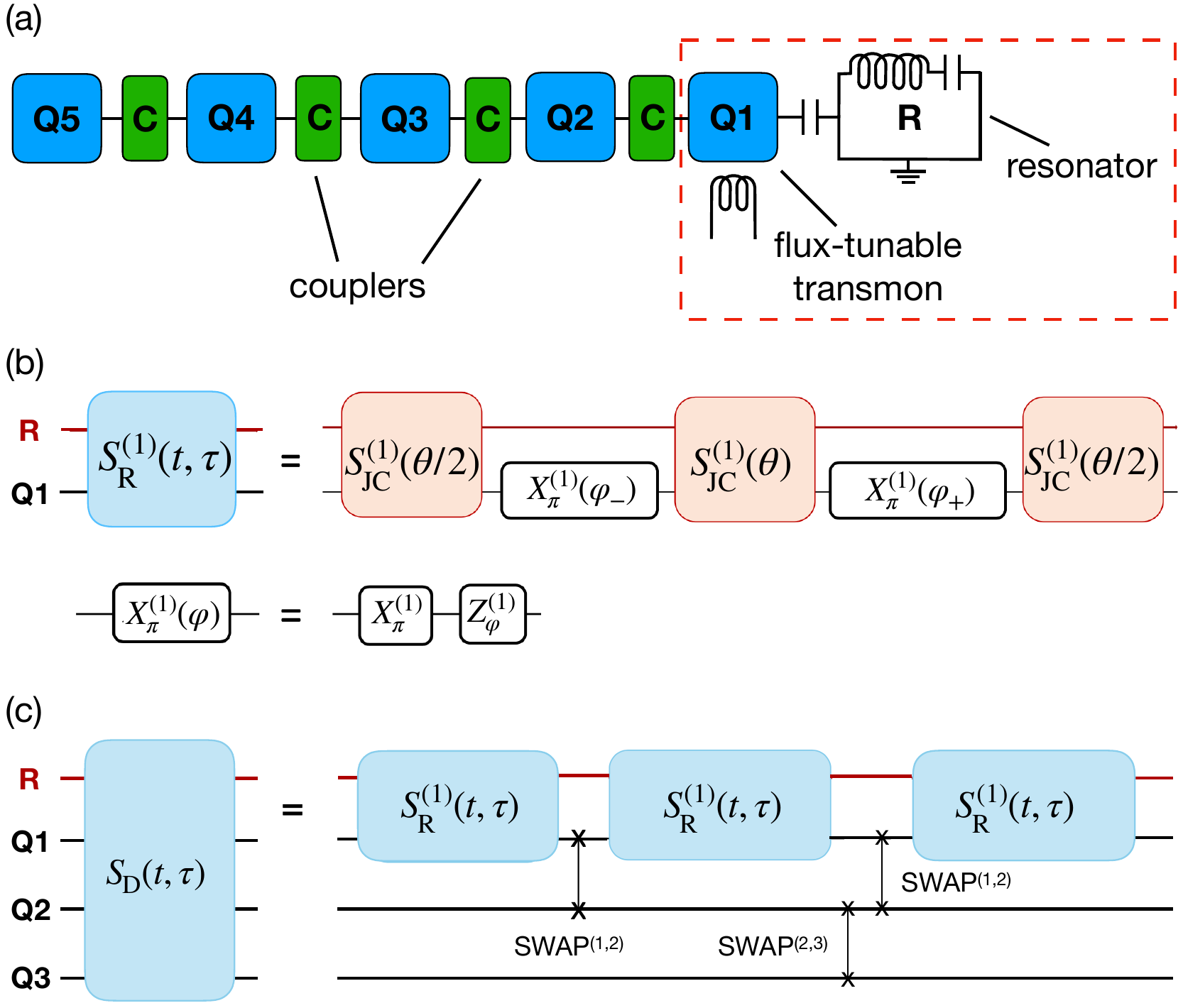} 
\caption{
{(a) Qubit-boson architecture for the Dicke-Ising model. Dashed red box: Rabi gate block for the first qubit.  Also shown are quantum circuits representing  (b) the Rabi and (c) Dicke  gates. The  gate $X_\pi^{(1)}(\varphi)$ is acting to the first qubit.}
}
\label{dicke} 
\end{figure}

Reproducing the logic of Ref.~\cite{Langford:2017aa}, we now derive the Rabi gate from the JC gate and then generalize the former to a Dicke gate. The original idea is to decompose  Eq.~(\ref{HR}) into
\be
\hat H_{\rm R}=\frac{1}{2}\big(  \hat  H_{\rm JC} +  \hat  H_{\rm AJC} \big)
\ee
where 
\be
   \hat  H_{\rm JC}=\hat H_0+ 2g(\hat a^\dagger\hat \sigma^- + \hat a\hat \sigma^+)
 \ee
 is the Jaynes-Cummings Hamiltonian and 
\be 
\hat  H_{\rm AJC}=\hat \sigma^x   \hat  H_{\rm JC} \hat \sigma^x 
\ee
is the corresponding counter-rotating interaction term. The free part can be chosen as
\be
\label{eq:H_0_def}
    \hat H_0=\omega_0(\hat a^\dagger \hat a-\hat \sigma^z/2 ). 
\ee
The exact Trotter step $e^{-i \hat H_{\rm R}\Delta t}$ on the simulation time interval $[t; t+\Delta t]$ is approximated to second order by 
\be
    \hat U_{\rm R}(t + \Delta t, t) = e^{-i \hat H_{\rm JC}\Delta t/4}e^{-i \hat H_{\rm AJC}\Delta t/2}e^{-i \hat H_{\rm JC}\Delta t/4}, \label{rabi_gate_evol}
  \ee
which has a discretization error of $\mathcal{O}(\Delta t^3)$. Moving to the  frame rotating with $\hat H_0$,  one obtains
\be
\label{eq:Int_Rep}
 \hat U_{\rm R}(t + \Delta t, t) = e^{-i\hat H_0 (t+\Delta t)} \hat S_{\rm R}(t + \Delta t, t) e^{i\hat H_0 t} 
\ee
where $\hat S_{\rm R}$ is  the hybrid qubit-resonator gate sequence
\be
  \hat S_{\rm R}= \hat S_{\rm JC}(\theta/2)\hat X_\pi(\varphi_+) \hat S_{\rm JC}(\theta)\hat X_\pi(\varphi_-) \hat S_{\rm JC}(\theta/2)  \label{SRabi}
\ee
where the phases are $\varphi_-\!=\!\omega_0(t\!+\!\Delta t/4)$ and $\varphi_+\!=\!\omega_0(t\!+\!3\Delta t/4)$. {\color{black} For explicit derivation of the latter equation, see Appendix~\ref{s_di_gate_app}.} This gate has been realized experimentally as a pulse sequence~\cite{Langford:2017aa}, which  is  shown schematically in Fig.~\ref{rabi_pulse}(c). It consists of three analog JC gates separated by single-qubit gates, which we have combined in the definition 
\be
 \hat X_{\pi}(\varphi_\pm)= \exp(-i\varphi_\pm\hat\sigma^z)\hat\sigma^x. \label{x_gate}
\ee
These single-qubit gates encode the counter-rotating evolution due to $\hat H_{\rm AJC}$  as well as the necessary rotating-frame transformations.

\begin{figure}[t!]
\center\includegraphics[width=\linewidth]{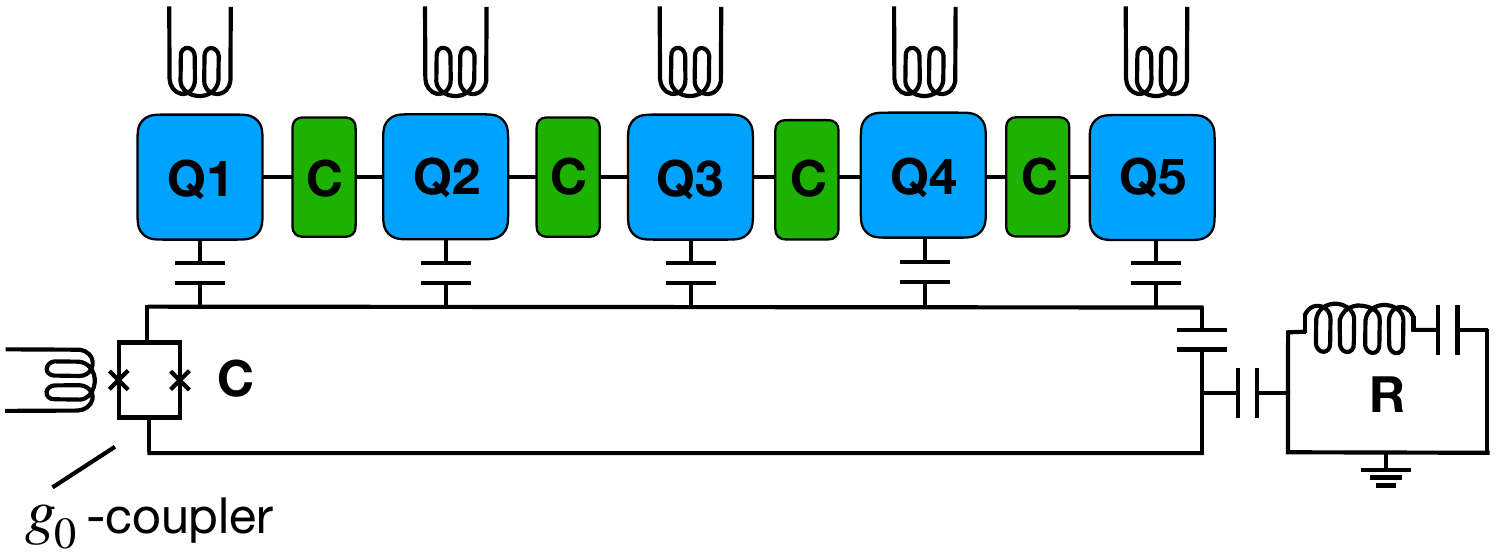} 
\caption{Alternative qubit-boson architecture with all qubits tunable and coupled to the resonator. The auxiliary  qubit acts as a coupler (C) to the resonator with tunable $g_0$. }
\label{dicke_2} 
\end{figure}

With the multi-qubit architecture of Fig.~\ref{dicke}(a) in mind, where only one qubit is \textit{physically} coupled to the resonator, we propose a Dicke gate $\hat S_{\rm D}$
implemented by applying the Rabi gate Fig.~\ref{dicke}(b)  to Q1 only, while digital SWAP gates mediate the interaction to the other qubits Q2 and Q3 (Fig.~\ref{dicke}(c)). 
An alternative architecture analogous to the experimental setting of Ref.~\cite{huber2024parametricmultielementcouplingarchitecture} is suggested in Fig.~\ref{dicke_2}. Compared to Fig.~\ref{dicke}(a), all qubits are tunable and directly coupled to the resonator via the additional $g_0$-coupler. Rabi gates can then be applied to each of the qubits without the need for additional SWAP gates.

\subsection{Algorithm overview}\label{overview}

\begin{figure}[t!]
\center\includegraphics[width=\linewidth]{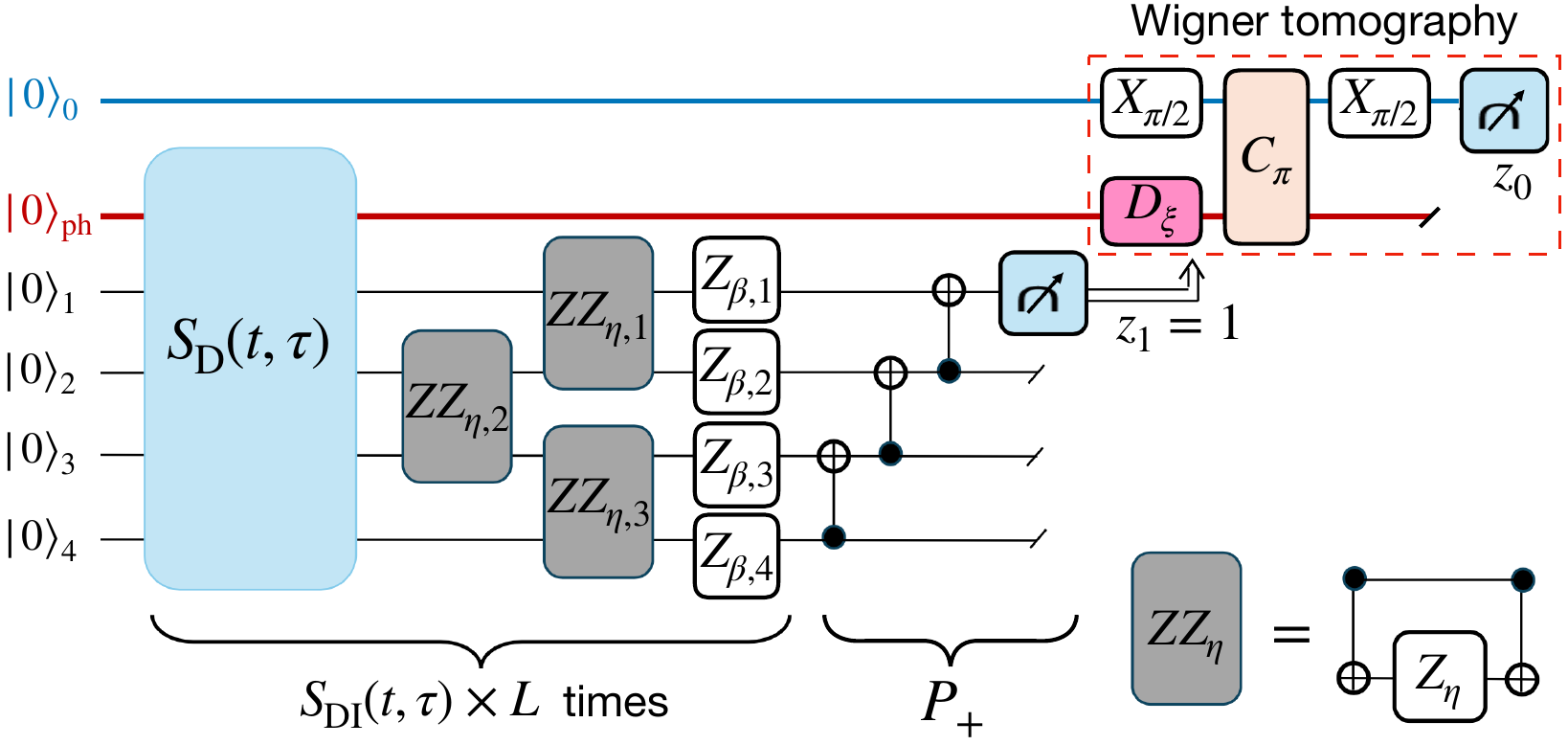} 
\caption{Full quantum circuit implementing the cat state and Wigner tomography. The circuit with $L$ Trotter steps and $N$ qubits requires $3L N$ JC gates and $(2L{+}1)(N{-}1)$ two-qubit CNOT-gates. Assuming the architecture  shown in Fig.~\ref{dicke}(a), the algorithm involves  $\frac{1}{2}LN(N-1)$ SWAPs. The alternative architecture  from Fig.~\ref{dicke_2} does not require SWAPs. }
\label{ising_dicke} 
\end{figure}

The full algorithm, which starts from {\color{black}$|\rm FM\rangle$}, is shown in Fig.~\ref{ising_dicke}. In each Trotter step, the Dicke gate $\hat S_{\rm D} $ is followed by a set of single-qubit $Z$-gates and two-qubit $ZZ$-gates, which simulate the on-site frequencies $\omega_z$ and the Ising interactions in Eq.~\eqref{H_DI} via
\be
ZZ_{\eta,j}=e^{i \eta \hat \sigma_j^z\hat\sigma_{j+1}^z} , \ Z_{\beta,j}=e^{i \beta \hat \sigma_j^z} , \label{ZZ_Z_gates}
\ee
where the phases are given by $\eta=J \Delta t$ and $\beta=\omega_z \Delta t$. Together, these gates constitute our Dicke-Ising gate $\hat S_{\rm DI}$, 
{\color{black} 
which governs the Trotterized evolution under the Hamiltonian $\hat H_{\rm DI}$ in the photon-coupled interaction picture:
\be
 \hat U^\dagger_{0}(t)e^{-i \hat H_{\rm DI} t  }
\approx  \prod\limits_{k=1}^L \hat S_{\rm DI}(t_{k}, \Delta t)
. \label{trotterization}
\ee
Here $\hat U_{0}(t) = e^{-i H_0 t }$ is the free unitary evolution with the Hamiltonian 
\be
\label{eq:H_free}
\hat H_0= \omega_0 \hat a^\dagger \hat a- \frac 12 \omega_0  \sum_{j=1}^N \hat \sigma^z_j, 
\ee
which form is chosen similar to Eq.~\eqref{eq:H_0_def} used previously.
The Trotter step index $k$ runs from 1 to $L$, with the simulation times given by $t_k = (k{-}1)\Delta t$ and the Trotter step size $\Delta t = t/L$.
The Dicke-Ising gate appearing in (\ref{trotterization}) is constructed as
\be
\!\hat S_{\rm DI}(t_{k}, \Delta t){=}\!\left(\prod\limits_{j=1}^{N-1} ZZ_{\eta, j}\!\right)
\left(\prod\limits_{j=1}^{N} Z_{\beta, j}\!\right) \hat S_{\rm D}(t_{k}, \Delta t). \label{S_DI_gate0}
\ee
Depending on the architecture shown either in Fig.~\ref{dicke}~(a) or in Fig.~\ref{dicke_2}, the Dicke gate $\hat S_{\rm D}$ in (\ref{S_DI_gate0}) takes the form
\be
\color{black}
\!\!\hat S_{\rm D}
=\begin{cases}
\prod\limits_{j=1}^{N-1} \prod\limits_{l=1}^{j}\!\left(  {\rm SWAP}^{(N-l,N-l+1)} \hat S_{{\rm R}}^{(1)}
\right)\!, & \!\! \! \!\! \text{[Fig.\ref{dicke}(a)]};\\ \\
\prod\limits_{j=1}^{N} \hat S_{{\rm R}}^{(j)},
& \!\! \text{[Fig.\ref{dicke_2}]}.
\end{cases}\!\! \label{S_D_gate0}
\ee
The Rabi gates $\hat S_{\rm R}^{(j)}$ are defined in Eq.~(\ref{SRabi}). The associated phases are given by $\varphi_{+}(t_k) = \omega_0 (t_k + 3\Delta t/4)$, 
$\varphi_{-}(t_k) =  \omega_0 ( t_k + \Delta t/4)$, and the coupling angle is $\theta = \frac{g\Delta t}{\sqrt{N}}$.
A full derivation of the decomposition in Eqs.~(\ref{trotterization})--(\ref{S_D_gate0}) is provided in Appendix~\ref{s_di_gate_app}.
}

At the end of the full Trotter evolution, we arrive at the many-body state $|\Psi{\color{black}(t_{\rm f})}\rangle \approx |{\rm SR}\rangle$ that approximates the exact superradiant state. The Trotter evolution is followed by a CNOT sequence (emulating $\hat P_+$) and measurement of the first qubit. As proposed in~\cite{PhysRevA.92.062318}, this CNOT sequence yields the parity by measuring only a single qubit instead of all of them. If the measurement result is $z_1=1$, then one performs Wigner tomography of the resonator. If $z_1=-1$, the tomography is not performed. This completes our algorithm.

\begin{figure*}[t!]
\center\includegraphics[width=\linewidth]{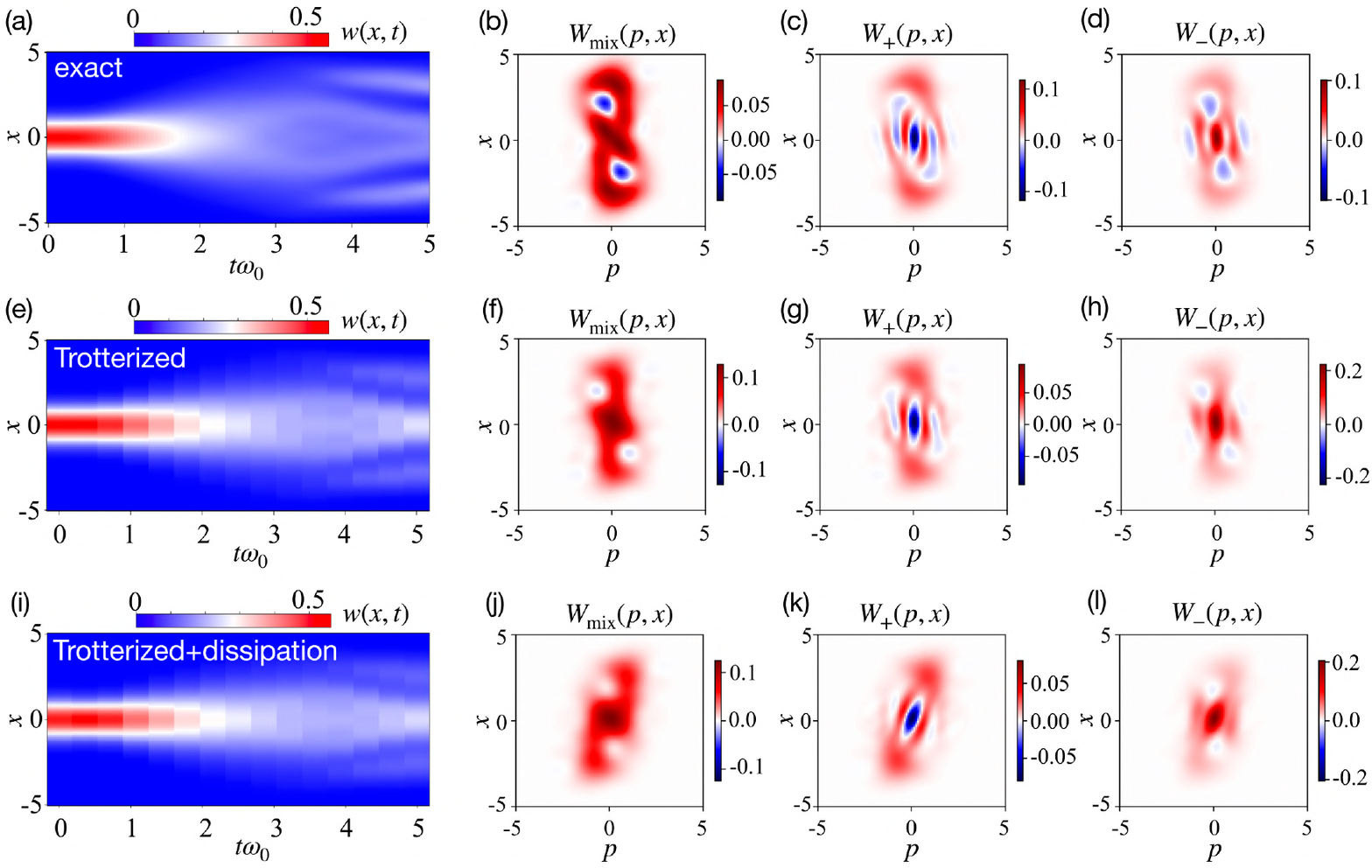} 
\caption{Numerical simulations of exact and Trotter evolutions{\color{black}, without and with dissipation}. (a) Exact dynamics of the photon probability distribution $w(x,t)$ after the quench during time $t\in [0, \ t_{\rm f}]$.  Wigner functions at the end of evolution $t=t_{\rm f}$: (b) $W_{\rm mix}$  without parity selection,  (c) $W_+$ for the positive parity, which shows cat state signatures, {\color{black} and (d) $W_-$ for the negative parity sector. (e-h): The data for $w(x,t)$, $W_{\rm mix}$, and $W_\pm$}  emulated by means of our quantum algorithm  as depicted in Fig.~\ref{ising_dicke} with  $L=15$ Trotter steps. The  parameters are $N=5$ qubits, a Fock-space cutoff at 20 photons, $\omega_z/\omega_0=0.05$, $J/\omega_0=1$ and $g/\omega_0=0.9$, which places the system slightly above the QPT. {\color{black}(i-l): Data for emulation   with a Lindbladian dissipation (\ref{lindblad}) added after each of the Dicke-Ising gates. The {\it physical} duration time  of each constituent Rabi gate is  assumed to be $\tau_{\rm Rabi}=100$ns; the decay rates are  $\kappa=2\pi\times 1$kHz for the resonator, and $\Gamma_\phi=\Gamma_1=2\pi\times 5$kHz for the qubits.}}\label{numerics} \end{figure*}

The Wigner tomography circuit is  shown in Fig.~\ref{ising_dicke} inside the dashed contour. We  follow the ideas of the measurement protocols suggested in Refs.~\cite{PhysRevLett.78.2547, doi:10.1126/science.1243289,Langford:2017aa}.  There is a representation of the Wigner function equivalent to Eq.~(\ref{W_xp}) that reads
\be
W_\xi=\frac{2}{\pi}{\rm tr}\big(\hat \Pi \hat D^\dagger_\xi \hat \rho_{\rm ph} \hat D_\xi \big),
\ee
where the photon-state density matrix $\rho_{\rm ph}$  is given by the projection of the full many-body state on a certain spin configuration.
Here, $\hat \Pi = e^{i\pi \hat a^\dagger\hat a } $ is  the photon parity operator,  $\hat D_\xi=e^{\xi \hat a^\dagger-\xi^*\hat a}$ is the standard displacement operator with the complex phase $\xi=x+ip$  parameterized by $x$ and $p$. The displacement operator can be implemented as a drive pulse applied to the resonator. It is depicted as the gate $D_\xi$ in Fig.~\ref{ising_dicke}. The density matrix $\hat\rho_{\rm ph}$ becomes $\hat D^\dagger_\xi \hat\rho_{\rm ph} \hat D_\xi $ after this pulse. The parity operator $\hat \Pi $ is implemented (i) via the gate $C_\pi$ entangling the resonator with an off-resonant  ancilla qubit and (ii) via two $X_{\pi/2}$ gates to perform Ramsey interferometry. Measuring the ancilla in the computational basis yields the photon parity $\Pi=\pm 1$ via the measurement value  $z_0$. The measurement is performed repeatedly to obtain the respective probabilities $\mathcal{P}(z_0\!=\!\pm 1)$, the difference of which yields the Wigner function value $W_\xi$.    To implement this protocol on a physical level, we have in mind the standard dispersive Hamiltonian for the resonator and the  ancilla qubit with frequency $\omega^{(0)}_{\rm Q}$,
\be
\hat H_{\rm disp}=\omega_{\rm R} \hat a ^\dagger \hat a +\omega^{(0)}_{\rm Q} |1\rangle\!\langle 1|-\chi \hat a^\dagger \hat a |1\rangle\!\langle 1| . \label{H_disp}
\ee
The evolution with $\hat H_{\rm disp}$ during time interval $\delta t$ yields the operator
$\hat U_\Phi= |0\rangle\!\langle 0| + e^{i\Phi \hat a^\dagger\hat a} |1\rangle\!\langle 1|$ with phase $\Phi=\chi \delta  t$. Before the tomography, the ancilla qubit is  in the state $|\psi_0\rangle=|0\rangle$; the $X_{\pi/2}$ gate brings it into the superposition $|\psi_1\rangle=\frac{1}{\sqrt 2}\big(|0\rangle+i |1\rangle\big)$. After that, by tuning the duration $\delta t$ of the off-resonant evolution such that $\Phi=\pi$, the entangling gate $C_\pi$ can be realized. Depending on the photon parity, $C_\pi$ rotates the qubit state either over the angle $\pi$ or $2\pi$ along the Bloch-sphere equator. The ancilla wave function then becomes $|\psi_2\rangle=\frac{1}{\sqrt 2}\big(|0\rangle+i\Pi|1\rangle\big)$. The second $X_{\pi/2}$ gate and subsequent $z_0$ measurement finalize the Ramsey interferometry. As a result, one measures the state  $|\psi_3\rangle =|1\rangle$ if the parity is even ($\Pi=1$) or $|\psi_3\rangle =|0\rangle$ if the parity is odd ($\Pi=-1$).

\section{Discussion}\label{discussion}

Illustrative results of our numerical simulations of the Dicke-Ising Hamiltonian with $N=5$ qubits are shown in  Fig.~\ref{numerics}. Our data include temporal evolution of the photon probability distribution  $w(x,t)$ as well as the Wigner functions $W_{\rm mix}$, $W_{+}$ at the very end of the time evolution. In Fig.~\ref{numerics}(a) we show the exact dynamics of $w(x,t)$ when the evolution starts with a Gaussian distribution  $w(x, 0)=e^{-x^2}/\sqrt{\pi}$, which corresponds the many-body wave function $|\Psi{\color{black}(t{=}0)}\rangle=|{\rm FM}\rangle$. At the end of the evolution ($ \omega_0 t_{\rm f}=5$), the wave function $|\Psi_{t_{\rm f}}\rangle$ is supposed to be  similar to $|{\rm SR}\rangle$. The distribution $w(x, t_{\rm f})$  indeed has well-defined side peaks around $x\approx\pm 4$, which correspond to a finite amount of condensed photons.

While the Wigner functions $W_{\rm mix}$ and $W_{+}$ at $t=t_{\rm f}$ (Fig.~\ref{numerics}(b, c)) look  distorted, they  are qualitatively  similar to the ideal distributions shown in Fig.~\ref{wigner_f}(c, d). It is important to note that $W_+$ in Fig.~\ref{numerics}(c)  retains clear cat state signatures, visible as   blue stripes of negative quasi-probabilities. {\color{black}}
In Fig.~\ref{numerics}(e-h), we present the equivalent  simulation of the Trotterized dynamics with 15 steps according to our { digital-analog} algorithm given in Fig.~\ref{ising_dicke}.  
The Wigner functions found through the Trotter evolution (Fig.~\ref{numerics}~(f,g)) are in good agreement with the exact simulation results in Fig.~\ref{numerics}(b, c), respectively. {\color{black}The Wigner functions $W_-$ (Fig.~\ref{numerics}(d, h)) also exhibit negative values, although these are less pronounced than in $W_+$.  }

{\color{black} In Fig.~\ref{numerics}~(i-l), we show simulation data for    Trotterized evolution in the presence of noise in the circuit, which is modeled by the Lindbladian  dynamics of the density matrix;   details are provided in Appendix \ref{noise_model}. This model  simulates dissipative processes  of individual qubits and  photons during the {\it physical} operation time of Rabi gates. In our simulations, we choose eperimentally relevant  qubit decay rates of  $\Gamma_\phi=\Gamma_1=2\pi \times 5$kHz~\cite{huber2024parametricmultielementcouplingarchitecture} and Rabi gate time of $\tau_{\rm Rabi}=100$ns~\cite{Langford:2017aa}. As the simulations show, the signatures of cat states are well resolved for high-quality resonators with decay rates $\kappa$ of just a few kiloherz~\cite{Ganjam2024}. 
}

\begin{figure}[t!]
\center\includegraphics[width=0.9\linewidth]{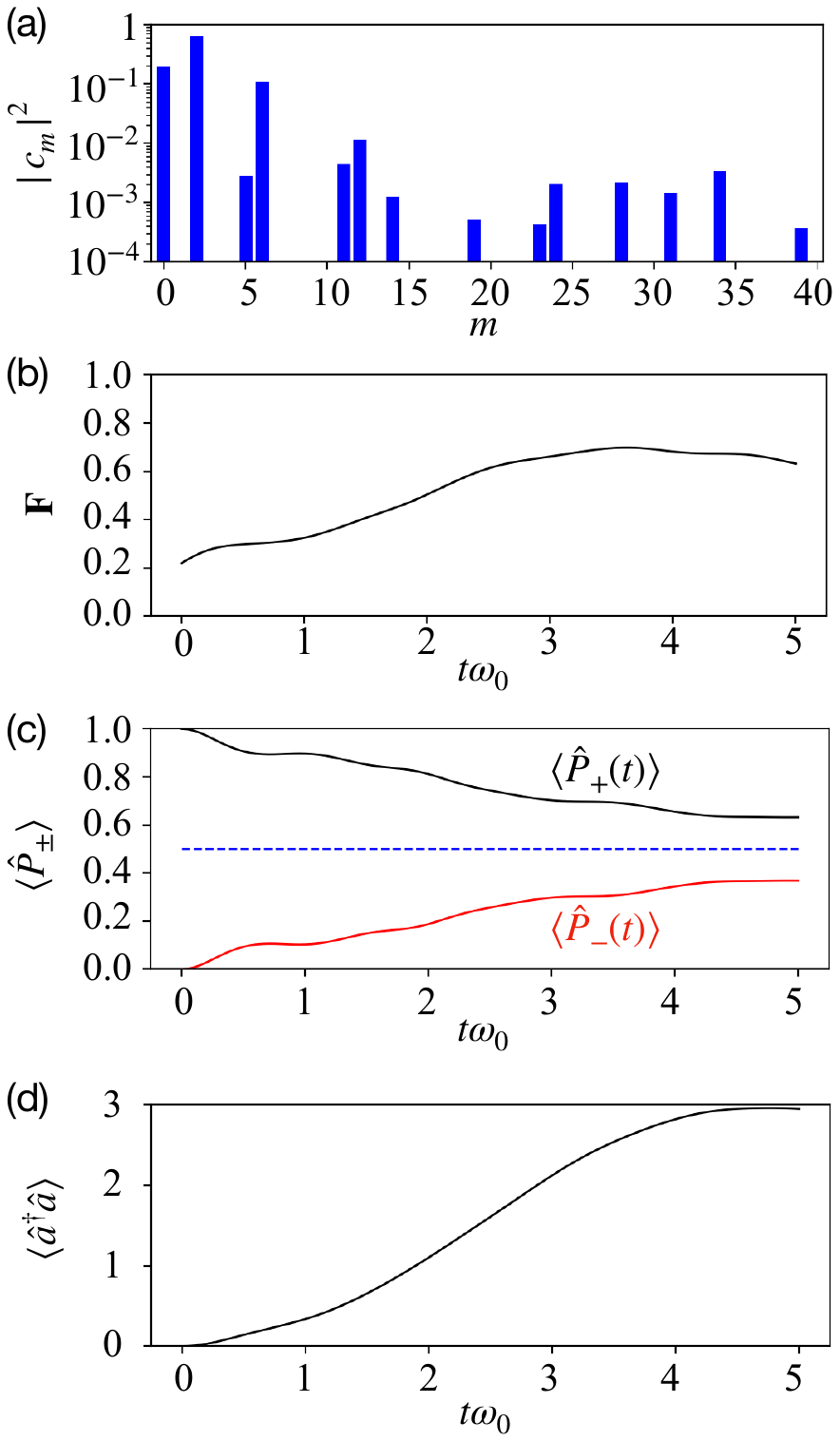} 
\caption{(a) Expansion coefficients $|c_m|^2$ of the {\it ferromagnetic} initial state over the eigenbasis of $\hat H_{\rm DI}$. {\color{black}(b--d) Results for the exact time evolution after the quench, with Hamiltonian parameters identical to those used in Fig.~\ref{numerics}.} (b) Fidelity $\mathbf F(t)$ between the reduced density matrices for photons corresponding to the  simulated wavefunction $\rho(t)$ and the ground state of $\hat H_{\rm DI}$,  $\rho_{\rm SR}$. (c) Time dependence of the probabilities to measure positive and negative qubit parities $\langle\hat P_\pm(t)\rangle$. {\color{black}(d) Time dependence of the photon number $\langle \hat{a}^\dagger \hat{a} \rangle$.  }
}
\label{fidelity} 
\end{figure}

Under exact quench dynamics, the many-body wavefunction $|\Psi(t)\rangle$ evolves as a superposition of  the eigenstates $|\Psi_m\rangle$ of the  Hamiltonian
 $\hat H_{\rm DI}$:
\be
|\Psi(t)\rangle=\sum_m e^{-i\varepsilon_m t} c_m |\Psi_m\rangle . 
\ee
Here, $m$ enumerates all many-body eigenstates, with $m=0$ corresponding to the ground state and $m\geq 1$ to the excited states. The expansion coefficients     $c_m=\langle\Psi_m|  {\rm FM}\rangle$ are the overlaps of the eigenstates with  the initial   {\it ferromagnetic} state $|\rm FM\rangle$. 
In Fig.~\ref{fidelity}~(a), we show the squared overlap coefficients $|c_m|^2$ for $0\leq m\leq 40$. The nonvanishing overlap between the {\it ferromagnetic} initial state and the exact superradiant ground state  $|\rm SR\rangle=|\Psi_0\rangle$ is given by $|c_0|^2\approx 0.2$, which reflects the small underbarrier tunnel probability in the ground state slightly above the critical point.

{\color{black}In Figs.~\ref{fidelity}(b--d) we show results for the exact time evolution after the quench.}
Figure~\ref{fidelity}(b) shows the temporal dependence of the  fidelity ${\mathbf F}(t)=\big({\rm tr}\sqrt{\sqrt{\rho_{\rm SR}}\rho(t)\sqrt{\rho_{\rm SR}}}\big)^2$ between the   reduced density matrices  $\rho(t)={\rm tr}_\sigma\Big[|\Psi(t)\rangle\!\langle \Psi(t)|\Big]$ and $\rho_{\rm SR}={\rm tr}_\sigma\Big[|\rm SR\rangle\!\langle \rm SR|\Big]$. The  nonzero overlap and fidelity indicate that more than  half  of the total photon population resides in the superradiant condensate. Also in {\color{black}Fig.~\ref{fidelity}(c)}, we present the time evolution of the probabilities to measure even or odd qubit parity, $\langle\hat P_\pm(t)\rangle$.  These begin at $\langle\hat P_+\rangle=1$ and $\langle\hat P_-\rangle=0$ at $t=0$,  reflecting fully imbalanced initial probabilities,  and approach  1/2 at the final time $t=t_{\rm f}$. This parity equalization provides an additional signature of superradiance in this protocol.
{\color{black}In Fig.~\ref{fidelity}(d) we show the time dependence of the photon number $\langle \hat{a}^\dagger \hat{a} \rangle$ after the quench. It starts from zero in the normal phase and increases to $\approx 3$ at the final time, confirming that boundary effects from the 20-photon Fock-space cutoff are negligible. 
}

For the total gate count, we obtain the following estimation. 
Each Rabi gate $S_{\rm R}$ involves three JC gates. {\color{black}Assuming the 1D qubit-boson architecture from Fig.~\ref{dicke}(a), where the qubit at one end of the chain is coupled to the resonator,} the Dicke gate $S_{\rm D} $ has $N$ $S_{\rm R}$-gates and $\frac 1 2 {N(N{-}1)}$ SWAPs. {\color{black}For the star-chain hybrid geometry  from Fig.~\ref{dicke_2}, SWAPs are not required}. The Dicke-Ising $S_{\rm DI}$-gate has one $S_{\rm D}$-gate and  $2(N{-}1)$ CNOTs. Parity selection involves $(N{-}1)$ CNOTs. To summarize, the circuit with $L$ Trotter steps requires:
\begin{itemize}
\item 
{\color{black}
$\begin{cases}
\frac{1}{2} L N(N-1) & \!\!\text{[1D chain Fig.~\ref{dicke}(a)]} \\ \\ 
0 & \!\! \text{[star-chain Fig.~\ref{dicke_2}]}
\end{cases} $} \ SWAPs,
\item $(2L+1)(N-1)$ \ CNOTs,
\item $3LN $  \ JC gates.
\end{itemize}
Based on these numbers, we conjecture that  the  algorithm is capable of simulating the QPT after a quench as well as the cat state preparation with a finite depth quantum circuit without fine-tuning parameters.

{\color{black}
The Wigner tomography  represents a substantial experimental overhead. It requires careful calibration of the dispersive coupling strength ($\chi$) and evolution time ($\delta t$).
To achieve an adequate resolution in the phase space, a two-dimensional grid of displacement complex parameter $\xi$ must be sampled, typically to $\sim
10^2\times10^2$ points~\cite{doi:10.1126/science.1243289}. At each point, repeated ancilla qubit measurements (on the order of  $10^3$  shots) are necessary to accurately estimate the parity expectation value, resulting in total counts around several million. Ideally,  the dispersive interaction  enables quantum non-demolition measurements of the photon  parity.  In practice, the readout process introduces partial backaction on the resonator's state. The qubit measurement    does not fully collapse the resonator  into a specific Fock state, but rather into a state with a  given parity that preserves some coherence within the corresponding Hilbert subspace.

While implementing the continuous-variable representation with $x$ and $p$, one  concerns about resonator's leakage. In the superconducting platform, the resonator coherence times are sufficiently long compared to the operation times guaranteed by the condition $\chi\gg \langle \hat a ^\dagger \hat a \rangle\kappa, \Gamma_\phi, \Gamma_1$, which makes the leakage effects negligible. The experimental value $\chi\sim2.4$MHz~\cite{doi:10.1126/science.1243289}, and   the average photon number  $\langle \hat a ^\dagger \hat a \rangle\approx 3$ is estimated  from simulations data  in {\color{black}Fig.~\ref{fidelity}~(d)}. We note that in systems with shorter-lived modes, such as trapped-ion implementations, mitigating decoherence would indeed be a critical requirement.}

\section{Conclusion and outlook}\label{conclusions}

In this work, we proposed a digital-analog quantum algorithm for simulating the superradiant QPT in the Dicke-Ising model, where individual qubits interact with each other and with a common photon mode. The algorithm features a sequence of analog Jaynes-Cummings gates combined with standard digital single-qubit and two-qubit rotations. This quantum circuit is capable of simulating quench dynamics and the QPT between the normal and superradiant phases. 
We proposed a protocol based on qubit-parity measurements that allows one to obtain a Schrödinger cat state as the output of resonator Wigner tomography. Additionally, we applied a path-integral description to the model via a bosonic angular representation of the spin operators and formulated the quasi-classical description of fluctuations in the large-spin limit. This approach can be useful for further studies of macroscopic quantum tunneling.
Finally, we found that the qubit-qubit interaction leads to an emergent Ising transition driven by the Kibble-Zurek mechanism in imaginary time. The qubit subsystem becomes critical for certain quantum trajectories of the photon field, making the fluctuations in the superradiant phase non-trivial, in contrast to the conventional Dicke model.
As an outlook for future investigations in this direction, an interesting question to address is the behavior of the entanglement entropy and the value of the central charge at the QPT and in the superradiant phase.

\vspace{1.5cm}

\begin{acknowledgments}
This work is supported by DPG under Germany’s Excellence Strategy – Cluster of Excellence Matter and Light for Quantum Computing (ML4Q) EXC 2004/1 – 390534769.
We further acknowledge support from the German Federal Ministry of Education and Research (BMBF) in the funding program 
"Quantum technologies – from basic research to market”, contract numbers 13N15584 (project DAQC) and 13N16149 (project QSolid).  
We are also grateful to the funding from the Deutsche Forschungsgemeinschaft (DFG, German Research Foundation) - Project-ID 429529648 - TRR 306 QuCoLiMa (”Quantum Cooperativity of Light and Matter”). 
The authors thank Alessandro Ciani and Alexander Shnirman for their fruitful discussions.
\end{acknowledgments}


\begin{thebibliography}{60}%
\makeatletter
\providecommand \@ifxundefined [1]{%
 \@ifx{#1\undefined}
}%
\providecommand \@ifnum [1]{%
 \ifnum #1\expandafter \@firstoftwo
 \else \expandafter \@secondoftwo
 \fi
}%
\providecommand \@ifx [1]{%
 \ifx #1\expandafter \@firstoftwo
 \else \expandafter \@secondoftwo
 \fi
}%
\providecommand \natexlab [1]{#1}%
\providecommand \enquote  [1]{``#1''}%
\providecommand \bibnamefont  [1]{#1}%
\providecommand \bibfnamefont [1]{#1}%
\providecommand \citenamefont [1]{#1}%
\providecommand \href@noop [0]{\@secondoftwo}%
\providecommand \href [0]{\begingroup \@sanitize@url \@href}%
\providecommand \@href[1]{\@@startlink{#1}\@@href}%
\providecommand \@@href[1]{\endgroup#1\@@endlink}%
\providecommand \@sanitize@url [0]{\catcode `\\12\catcode `\$12\catcode
  `\&12\catcode `\#12\catcode `\^12\catcode `\_12\catcode `\%12\relax}%
\providecommand \@@startlink[1]{}%
\providecommand \@@endlink[0]{}%
\providecommand \url  [0]{\begingroup\@sanitize@url \@url }%
\providecommand \@url [1]{\endgroup\@href {#1}{\urlprefix }}%
\providecommand \urlprefix  [0]{URL }%
\providecommand \Eprint [0]{\href }%
\providecommand \doibase [0]{https://doi.org/}%
\providecommand \selectlanguage [0]{\@gobble}%
\providecommand \bibinfo  [0]{\@secondoftwo}%
\providecommand \bibfield  [0]{\@secondoftwo}%
\providecommand \translation [1]{[#1]}%
\providecommand \BibitemOpen [0]{}%
\providecommand \bibitemStop [0]{}%
\providecommand \bibitemNoStop [0]{.\EOS\space}%
\providecommand \EOS [0]{\spacefactor3000\relax}%
\providecommand \BibitemShut  [1]{\csname bibitem#1\endcsname}%
\let\auto@bib@innerbib\@empty
\bibitem [{\citenamefont {Lee}\ and\ \citenamefont
  {Johnson}(2004)}]{PhysRevLett.93.083001}%
  \BibitemOpen
  \bibfield  {author} {\bibinfo {author} {\bibfnamefont {C.~F.}\ \bibnamefont
  {Lee}}\ and\ \bibinfo {author} {\bibfnamefont {N.~F.}\ \bibnamefont
  {Johnson}},\ }\bibfield  {title} {\bibinfo {title} {{First-Order Superradiant
  Phase Transitions in a Multiqubit Cavity System}},\ }\href
  {https://doi.org/10.1103/PhysRevLett.93.083001} {\bibfield  {journal}
  {\bibinfo  {journal} {Phys. Rev. Lett.}\ }\textbf {\bibinfo {volume} {93}},\
  \bibinfo {pages} {083001} (\bibinfo {year} {2004})}\BibitemShut {NoStop}%
\bibitem [{\citenamefont {Gammelmark}\ and\ \citenamefont
  {Mølmer}(2011)}]{Gammelmark_2011}%
  \BibitemOpen
  \bibfield  {author} {\bibinfo {author} {\bibfnamefont {S.}~\bibnamefont
  {Gammelmark}}\ and\ \bibinfo {author} {\bibfnamefont {K.}~\bibnamefont
  {Mølmer}},\ }\bibfield  {title} {\bibinfo {title} {{Phase transitions and
  Heisenberg limited metrology in an Ising chain interacting with a single-mode
  cavity field}},\ }\href {https://doi.org/10.1088/1367-2630/13/5/053035}
  {\bibfield  {journal} {\bibinfo  {journal} {New Journal of Physics}\ }\textbf
  {\bibinfo {volume} {13}},\ \bibinfo {pages} {053035} (\bibinfo {year}
  {2011})}\BibitemShut {NoStop}%
\bibitem [{\citenamefont {Zhang}\ \emph {et~al.}(2014)\citenamefont {Zhang},
  \citenamefont {Yu}, \citenamefont {Liang}, \citenamefont {Chen},
  \citenamefont {Jia},\ and\ \citenamefont {Nori}}]{Zhang:2014aa}%
  \BibitemOpen
  \bibfield  {author} {\bibinfo {author} {\bibfnamefont {Y.}~\bibnamefont
  {Zhang}}, \bibinfo {author} {\bibfnamefont {L.}~\bibnamefont {Yu}}, \bibinfo
  {author} {\bibfnamefont {J.~Q.}\ \bibnamefont {Liang}}, \bibinfo {author}
  {\bibfnamefont {G.}~\bibnamefont {Chen}}, \bibinfo {author} {\bibfnamefont
  {S.}~\bibnamefont {Jia}},\ and\ \bibinfo {author} {\bibfnamefont
  {F.}~\bibnamefont {Nori}},\ }\bibfield  {title} {\bibinfo {title} {{Quantum
  phases in circuit QED with a superconducting qubit array}},\ }\href
  {https://doi.org/10.1038/srep04083} {\bibfield  {journal} {\bibinfo
  {journal} {Scientific Reports}\ }\textbf {\bibinfo {volume} {4}},\ \bibinfo
  {pages} {4083} (\bibinfo {year} {2014})}\BibitemShut {NoStop}%
\bibitem [{\citenamefont {Gelhausen}\ \emph {et~al.}(2016)\citenamefont
  {Gelhausen}, \citenamefont {Buchhold}, \citenamefont {Rosch},\ and\
  \citenamefont {Strack}}]{10.21468/SciPostPhys.1.1.004}%
  \BibitemOpen
  \bibfield  {author} {\bibinfo {author} {\bibfnamefont {J.}~\bibnamefont
  {Gelhausen}}, \bibinfo {author} {\bibfnamefont {M.}~\bibnamefont {Buchhold}},
  \bibinfo {author} {\bibfnamefont {A.}~\bibnamefont {Rosch}},\ and\ \bibinfo
  {author} {\bibfnamefont {P.}~\bibnamefont {Strack}},\ }\bibfield  {title}
  {\bibinfo {title} {{Quantum-optical magnets with competing short- and
  long-range interactions: Rydberg-dressed spin lattice in an optical
  cavity}},\ }\href {https://doi.org/10.21468/SciPostPhys.1.1.004} {\bibfield
  {journal} {\bibinfo  {journal} {SciPost Phys.}\ }\textbf {\bibinfo {volume}
  {1}},\ \bibinfo {pages} {004} (\bibinfo {year} {2016})}\BibitemShut {NoStop}%
\bibitem [{\citenamefont {Rohn}\ \emph {et~al.}(2020)\citenamefont {Rohn},
  \citenamefont {H\"ormann}, \citenamefont {Genes},\ and\ \citenamefont
  {Schmidt}}]{PhysRevResearch.2.023131}%
  \BibitemOpen
  \bibfield  {author} {\bibinfo {author} {\bibfnamefont {J.}~\bibnamefont
  {Rohn}}, \bibinfo {author} {\bibfnamefont {M.}~\bibnamefont {H\"ormann}},
  \bibinfo {author} {\bibfnamefont {C.}~\bibnamefont {Genes}},\ and\ \bibinfo
  {author} {\bibfnamefont {K.~P.}\ \bibnamefont {Schmidt}},\ }\bibfield
  {title} {\bibinfo {title} {{Ising model in a light-induced quantized
  transverse field}},\ }\href
  {https://doi.org/10.1103/PhysRevResearch.2.023131} {\bibfield  {journal}
  {\bibinfo  {journal} {Phys. Rev. Res.}\ }\textbf {\bibinfo {volume} {2}},\
  \bibinfo {pages} {023131} (\bibinfo {year} {2020})}\BibitemShut {NoStop}%
\bibitem [{\citenamefont {Schellenberger}\ and\ \citenamefont
  {Schmidt}(2024)}]{10.21468/SciPostPhysCore.7.3.038}%
  \BibitemOpen
  \bibfield  {author} {\bibinfo {author} {\bibfnamefont {A.}~\bibnamefont
  {Schellenberger}}\ and\ \bibinfo {author} {\bibfnamefont {K.~P.}\
  \bibnamefont {Schmidt}},\ }\bibfield  {title} {\bibinfo {title} {{(Almost)
  everything is a Dicke model - Mapping non-superradiant correlated
  light-matter systems to the exactly solvable Dicke model}},\ }\href
  {https://doi.org/10.21468/SciPostPhysCore.7.3.038} {\bibfield  {journal}
  {\bibinfo  {journal} {SciPost Phys. Core}\ }\textbf {\bibinfo {volume} {7}},\
  \bibinfo {pages} {038} (\bibinfo {year} {2024})}\BibitemShut {NoStop}%
\bibitem [{\citenamefont {Langheld}\ \emph {et~al.}(2024)\citenamefont
  {Langheld}, \citenamefont {H\"ormann},\ and\ \citenamefont
  {Schmidt}}]{langheld2024quantumphasediagramsdickeising}%
  \BibitemOpen
  \bibfield  {author} {\bibinfo {author} {\bibfnamefont {A.}~\bibnamefont
  {Langheld}}, \bibinfo {author} {\bibfnamefont {M.}~\bibnamefont
  {H\"ormann}},\ and\ \bibinfo {author} {\bibfnamefont {K.~P.}\ \bibnamefont
  {Schmidt}},\ }\href {https://arxiv.org/abs/2409.15082} {\bibinfo {title}
  {{Quantum phase diagrams of Dicke-Ising models by a wormhole algorithm}}}
  (\bibinfo {year} {2024}),\ \Eprint {https://arxiv.org/abs/2409.15082}
  {arXiv:2409.15082 [cond-mat.str-el]} \BibitemShut {NoStop}%
\bibitem [{\citenamefont {Puel}\ and\ \citenamefont
  {Macr\`{\i}}(2024)}]{PhysRevLett.133.106901}%
  \BibitemOpen
  \bibfield  {author} {\bibinfo {author} {\bibfnamefont {T.~O.}\ \bibnamefont
  {Puel}}\ and\ \bibinfo {author} {\bibfnamefont {T.}~\bibnamefont
  {Macr\`{\i}}},\ }\bibfield  {title} {\bibinfo {title} {{Confined Meson
  Excitations in Rydberg-Atom Arrays Coupled to a Cavity Field}},\ }\href
  {https://doi.org/10.1103/PhysRevLett.133.106901} {\bibfield  {journal}
  {\bibinfo  {journal} {Phys. Rev. Lett.}\ }\textbf {\bibinfo {volume} {133}},\
  \bibinfo {pages} {106901} (\bibinfo {year} {2024})}\BibitemShut {NoStop}%
\bibitem [{\citenamefont {Popov}\ and\ \citenamefont
  {Fedotov}(1988)}]{popov1988functional}%
  \BibitemOpen
  \bibfield  {author} {\bibinfo {author} {\bibfnamefont {V.~N.}\ \bibnamefont
  {Popov}}\ and\ \bibinfo {author} {\bibfnamefont {S.}~\bibnamefont
  {Fedotov}},\ }\bibfield  {title} {\bibinfo {title} {{The functional
  integration method and diagram technique for spin systems}},\ }\href
  {http://jetp.ras.ru/cgi-bin/dn/e_067_03_0535.pdf} {\bibfield  {journal}
  {\bibinfo  {journal} {Sov. Phys. JETP}\ }\textbf {\bibinfo {volume} {67}},\
  \bibinfo {pages} {535} (\bibinfo {year} {1988})}\BibitemShut {NoStop}%
\bibitem [{\citenamefont {Emary}\ and\ \citenamefont
  {Brandes}(2003)}]{emary2003chaos}%
  \BibitemOpen
  \bibfield  {author} {\bibinfo {author} {\bibfnamefont {C.}~\bibnamefont
  {Emary}}\ and\ \bibinfo {author} {\bibfnamefont {T.}~\bibnamefont
  {Brandes}},\ }\bibfield  {title} {\bibinfo {title} {{Chaos and the quantum
  phase transition in the Dicke model}},\ }\href
  {https://doi.org/10.1103/PhysRevE.67.066203} {\bibfield  {journal} {\bibinfo
  {journal} {Physical Review E}\ }\textbf {\bibinfo {volume} {67}},\ \bibinfo
  {pages} {066203} (\bibinfo {year} {2003})}\BibitemShut {NoStop}%
\bibitem [{\citenamefont {Eastham}\ and\ \citenamefont
  {Littlewood}(2001)}]{eastham2001bose}%
  \BibitemOpen
  \bibfield  {author} {\bibinfo {author} {\bibfnamefont {P.}~\bibnamefont
  {Eastham}}\ and\ \bibinfo {author} {\bibfnamefont {P.}~\bibnamefont
  {Littlewood}},\ }\bibfield  {title} {\bibinfo {title} {{Bose condensation of
  cavity polaritons beyond the linear regime: The thermal equilibrium of a
  model microcavity}},\ }\href {https://doi.org/10.1103/PhysRevB.64.235101}
  {\bibfield  {journal} {\bibinfo  {journal} {Physical Review B}\ }\textbf
  {\bibinfo {volume} {64}},\ \bibinfo {pages} {235101} (\bibinfo {year}
  {2001})}\BibitemShut {NoStop}%
\bibitem [{\citenamefont {Dalla~Torre}\ \emph {et~al.}(2013)\citenamefont
  {Dalla~Torre}, \citenamefont {Diehl}, \citenamefont {Lukin}, \citenamefont
  {Sachdev},\ and\ \citenamefont {Strack}}]{dalla2013keldysh}%
  \BibitemOpen
  \bibfield  {author} {\bibinfo {author} {\bibfnamefont {E.~G.}\ \bibnamefont
  {Dalla~Torre}}, \bibinfo {author} {\bibfnamefont {S.}~\bibnamefont {Diehl}},
  \bibinfo {author} {\bibfnamefont {M.~D.}\ \bibnamefont {Lukin}}, \bibinfo
  {author} {\bibfnamefont {S.}~\bibnamefont {Sachdev}},\ and\ \bibinfo {author}
  {\bibfnamefont {P.}~\bibnamefont {Strack}},\ }\bibfield  {title} {\bibinfo
  {title} {{Keldysh approach for nonequilibrium phase transitions in quantum
  optics: Beyond the Dicke model in optical cavities}},\ }\href
  {https://doi.org/10.1103/PhysRevA.87.023831} {\bibfield  {journal} {\bibinfo
  {journal} {Physical Review A}\ }\textbf {\bibinfo {volume} {87}},\ \bibinfo
  {pages} {023831} (\bibinfo {year} {2013})}\BibitemShut {NoStop}%
\bibitem [{\citenamefont {Dalla~Torre}\ \emph {et~al.}(2016)\citenamefont
  {Dalla~Torre}, \citenamefont {Shchadilova}, \citenamefont {Wilner},
  \citenamefont {Lukin},\ and\ \citenamefont {Demler}}]{PhysRevA.94.061802}%
  \BibitemOpen
  \bibfield  {author} {\bibinfo {author} {\bibfnamefont {E.~G.}\ \bibnamefont
  {Dalla~Torre}}, \bibinfo {author} {\bibfnamefont {Y.}~\bibnamefont
  {Shchadilova}}, \bibinfo {author} {\bibfnamefont {E.~Y.}\ \bibnamefont
  {Wilner}}, \bibinfo {author} {\bibfnamefont {M.~D.}\ \bibnamefont {Lukin}},\
  and\ \bibinfo {author} {\bibfnamefont {E.}~\bibnamefont {Demler}},\
  }\bibfield  {title} {\bibinfo {title} {{Dicke phase transition without total
  spin conservation}},\ }\href {https://doi.org/10.1103/PhysRevA.94.061802}
  {\bibfield  {journal} {\bibinfo  {journal} {Phys. Rev. A}\ }\textbf {\bibinfo
  {volume} {94}},\ \bibinfo {pages} {061802} (\bibinfo {year}
  {2016})}\BibitemShut {NoStop}%
\bibitem [{\citenamefont {Kirton}\ \emph {et~al.}(2019)\citenamefont {Kirton},
  \citenamefont {Roses}, \citenamefont {Keeling},\ and\ \citenamefont {{Dalla
  Torre}}}]{kirton2018introduction}%
  \BibitemOpen
  \bibfield  {author} {\bibinfo {author} {\bibfnamefont {P.}~\bibnamefont
  {Kirton}}, \bibinfo {author} {\bibfnamefont {M.~M.}\ \bibnamefont {Roses}},
  \bibinfo {author} {\bibfnamefont {J.}~\bibnamefont {Keeling}},\ and\ \bibinfo
  {author} {\bibfnamefont {E.~G.}\ \bibnamefont {{Dalla Torre}}},\ }\bibfield
  {title} {\bibinfo {title} {{Introduction to the Dicke Model: From Equilibrium
  to Nonequilibrium, and Vice Versa}},\ }\href
  {https://doi.org/10.1002/qute.201800043} {\bibfield  {journal} {\bibinfo
  {journal} {Adv. Quantum Technol.}\ }\textbf {\bibinfo {volume} {2}},\
  \bibinfo {pages} {1800043} (\bibinfo {year} {2019})}\BibitemShut {NoStop}%
\bibitem [{\citenamefont {Shapiro}\ \emph {et~al.}(2020)\citenamefont
  {Shapiro}, \citenamefont {Pogosov},\ and\ \citenamefont
  {Lozovik}}]{PhysRevA.102.023703}%
  \BibitemOpen
  \bibfield  {author} {\bibinfo {author} {\bibfnamefont {D.~S.}\ \bibnamefont
  {Shapiro}}, \bibinfo {author} {\bibfnamefont {W.~V.}\ \bibnamefont
  {Pogosov}},\ and\ \bibinfo {author} {\bibfnamefont {Y.~E.}\ \bibnamefont
  {Lozovik}},\ }\bibfield  {title} {\bibinfo {title} {{Universal fluctuations
  and squeezing in a generalized Dicke model near the superradiant phase
  transition}},\ }\href {https://doi.org/10.1103/PhysRevA.102.023703}
  {\bibfield  {journal} {\bibinfo  {journal} {Phys. Rev. A}\ }\textbf {\bibinfo
  {volume} {102}},\ \bibinfo {pages} {023703} (\bibinfo {year}
  {2020})}\BibitemShut {NoStop}%
\bibitem [{\citenamefont {Dimer}\ \emph {et~al.}(2007)\citenamefont {Dimer},
  \citenamefont {Estienne}, \citenamefont {Parkins},\ and\ \citenamefont
  {Carmichael}}]{PhysRevA.75.013804}%
  \BibitemOpen
  \bibfield  {author} {\bibinfo {author} {\bibfnamefont {F.}~\bibnamefont
  {Dimer}}, \bibinfo {author} {\bibfnamefont {B.}~\bibnamefont {Estienne}},
  \bibinfo {author} {\bibfnamefont {A.~S.}\ \bibnamefont {Parkins}},\ and\
  \bibinfo {author} {\bibfnamefont {H.~J.}\ \bibnamefont {Carmichael}},\
  }\bibfield  {title} {\bibinfo {title} {{Proposed realization of the
  Dicke-model quantum phase transition in an optical cavity QED system}},\
  }\href {https://doi.org/10.1103/PhysRevA.75.013804} {\bibfield  {journal}
  {\bibinfo  {journal} {Phys. Rev. A}\ }\textbf {\bibinfo {volume} {75}},\
  \bibinfo {pages} {013804} (\bibinfo {year} {2007})}\BibitemShut {NoStop}%
\bibitem [{\citenamefont {Nataf}\ and\ \citenamefont
  {Ciuti}(2010)}]{PhysRevLett.104.023601}%
  \BibitemOpen
  \bibfield  {author} {\bibinfo {author} {\bibfnamefont {P.}~\bibnamefont
  {Nataf}}\ and\ \bibinfo {author} {\bibfnamefont {C.}~\bibnamefont {Ciuti}},\
  }\bibfield  {title} {\bibinfo {title} {{Vacuum Degeneracy of a Circuit QED
  System in the Ultrastrong Coupling Regime}},\ }\href
  {https://doi.org/10.1103/PhysRevLett.104.023601} {\bibfield  {journal}
  {\bibinfo  {journal} {Phys. Rev. Lett.}\ }\textbf {\bibinfo {volume} {104}},\
  \bibinfo {pages} {023601} (\bibinfo {year} {2010})}\BibitemShut {NoStop}%
\bibitem [{\citenamefont {Viehmann}\ \emph {et~al.}(2011)\citenamefont
  {Viehmann}, \citenamefont {von Delft},\ and\ \citenamefont
  {Marquardt}}]{PhysRevLett.107.113602}%
  \BibitemOpen
  \bibfield  {author} {\bibinfo {author} {\bibfnamefont {O.}~\bibnamefont
  {Viehmann}}, \bibinfo {author} {\bibfnamefont {J.}~\bibnamefont {von
  Delft}},\ and\ \bibinfo {author} {\bibfnamefont {F.}~\bibnamefont
  {Marquardt}},\ }\bibfield  {title} {\bibinfo {title} {{Superradiant Phase
  Transitions and the Standard Description of Circuit QED}},\ }\href
  {https://doi.org/10.1103/PhysRevLett.107.113602} {\bibfield  {journal}
  {\bibinfo  {journal} {Phys. Rev. Lett.}\ }\textbf {\bibinfo {volume} {107}},\
  \bibinfo {pages} {113602} (\bibinfo {year} {2011})}\BibitemShut {NoStop}%
\bibitem [{\citenamefont {Baumann}\ \emph {et~al.}(2010)\citenamefont
  {Baumann}, \citenamefont {Guerlin}, \citenamefont {Brennecke},\ and\
  \citenamefont {Esslinger}}]{Baumann:2010aa}%
  \BibitemOpen
  \bibfield  {author} {\bibinfo {author} {\bibfnamefont {K.}~\bibnamefont
  {Baumann}}, \bibinfo {author} {\bibfnamefont {C.}~\bibnamefont {Guerlin}},
  \bibinfo {author} {\bibfnamefont {F.}~\bibnamefont {Brennecke}},\ and\
  \bibinfo {author} {\bibfnamefont {T.}~\bibnamefont {Esslinger}},\ }\bibfield
  {title} {\bibinfo {title} {{Dicke quantum phase transition with a superfluid
  gas in an optical cavity}},\ }\href {https://doi.org/10.1038/nature09009}
  {\bibfield  {journal} {\bibinfo  {journal} {Nature}\ }\textbf {\bibinfo
  {volume} {464}},\ \bibinfo {pages} {1301} (\bibinfo {year}
  {2010})}\BibitemShut {NoStop}%
\bibitem [{\citenamefont {Zhang}\ \emph {et~al.}(2013)\citenamefont {Zhang},
  \citenamefont {Sun}, \citenamefont {Wen}, \citenamefont {Liu}, \citenamefont
  {Eggert},\ and\ \citenamefont {Ji}}]{PhysRevLett.110.090402}%
  \BibitemOpen
  \bibfield  {author} {\bibinfo {author} {\bibfnamefont {X.-F.}\ \bibnamefont
  {Zhang}}, \bibinfo {author} {\bibfnamefont {Q.}~\bibnamefont {Sun}}, \bibinfo
  {author} {\bibfnamefont {Y.-C.}\ \bibnamefont {Wen}}, \bibinfo {author}
  {\bibfnamefont {W.-M.}\ \bibnamefont {Liu}}, \bibinfo {author} {\bibfnamefont
  {S.}~\bibnamefont {Eggert}},\ and\ \bibinfo {author} {\bibfnamefont {A.-C.}\
  \bibnamefont {Ji}},\ }\bibfield  {title} {\bibinfo {title} {{Rydberg
  Polaritons in a Cavity: A Superradiant Solid}},\ }\href
  {https://doi.org/10.1103/PhysRevLett.110.090402} {\bibfield  {journal}
  {\bibinfo  {journal} {Phys. Rev. Lett.}\ }\textbf {\bibinfo {volume} {110}},\
  \bibinfo {pages} {090402} (\bibinfo {year} {2013})}\BibitemShut {NoStop}%
\bibitem [{\citenamefont {Baden}\ \emph {et~al.}(2014)\citenamefont {Baden},
  \citenamefont {Arnold}, \citenamefont {Grimsmo}, \citenamefont {Parkins},\
  and\ \citenamefont {Barrett}}]{PhysRevLett.113.020408}%
  \BibitemOpen
  \bibfield  {author} {\bibinfo {author} {\bibfnamefont {M.~P.}\ \bibnamefont
  {Baden}}, \bibinfo {author} {\bibfnamefont {K.~J.}\ \bibnamefont {Arnold}},
  \bibinfo {author} {\bibfnamefont {A.~L.}\ \bibnamefont {Grimsmo}}, \bibinfo
  {author} {\bibfnamefont {S.}~\bibnamefont {Parkins}},\ and\ \bibinfo {author}
  {\bibfnamefont {M.~D.}\ \bibnamefont {Barrett}},\ }\bibfield  {title}
  {\bibinfo {title} {Realization of the dicke model using cavity-assisted raman
  transitions},\ }\href {https://doi.org/10.1103/PhysRevLett.113.020408}
  {\bibfield  {journal} {\bibinfo  {journal} {Phys. Rev. Lett.}\ }\textbf
  {\bibinfo {volume} {113}},\ \bibinfo {pages} {020408} (\bibinfo {year}
  {2014})}\BibitemShut {NoStop}%
\bibitem [{\citenamefont {Klinder}\ \emph
  {et~al.}(2015{\natexlab{a}})\citenamefont {Klinder}, \citenamefont {Keßler},
  \citenamefont {Wolke}, \citenamefont {Mathey},\ and\ \citenamefont
  {Hemmerich}}]{doi:10.1073/pnas.1417132112}%
  \BibitemOpen
  \bibfield  {author} {\bibinfo {author} {\bibfnamefont {J.}~\bibnamefont
  {Klinder}}, \bibinfo {author} {\bibfnamefont {H.}~\bibnamefont {Keßler}},
  \bibinfo {author} {\bibfnamefont {M.}~\bibnamefont {Wolke}}, \bibinfo
  {author} {\bibfnamefont {L.}~\bibnamefont {Mathey}},\ and\ \bibinfo {author}
  {\bibfnamefont {A.}~\bibnamefont {Hemmerich}},\ }\bibfield  {title} {\bibinfo
  {title} {{Dynamical phase transition in the open Dicke model}},\ }\href
  {https://doi.org/10.1073/pnas.1417132112} {\bibfield  {journal} {\bibinfo
  {journal} {Proceedings of the National Academy of Sciences}\ }\textbf
  {\bibinfo {volume} {112}},\ \bibinfo {pages} {3290} (\bibinfo {year}
  {2015}{\natexlab{a}})},\ \Eprint
  {https://arxiv.org/abs/https://www.pnas.org/doi/pdf/10.1073/pnas.1417132112}
  {https://www.pnas.org/doi/pdf/10.1073/pnas.1417132112} \BibitemShut {NoStop}%
\bibitem [{\citenamefont {Safavi-Naini}\ \emph {et~al.}(2018)\citenamefont
  {Safavi-Naini}, \citenamefont {Lewis-Swan}, \citenamefont {Bohnet},
  \citenamefont {G\"arttner}, \citenamefont {Gilmore}, \citenamefont {Jordan},
  \citenamefont {Cohn}, \citenamefont {Freericks}, \citenamefont {Rey},\ and\
  \citenamefont {Bollinger}}]{PhysRevLett.121.040503}%
  \BibitemOpen
  \bibfield  {author} {\bibinfo {author} {\bibfnamefont {A.}~\bibnamefont
  {Safavi-Naini}}, \bibinfo {author} {\bibfnamefont {R.~J.}\ \bibnamefont
  {Lewis-Swan}}, \bibinfo {author} {\bibfnamefont {J.~G.}\ \bibnamefont
  {Bohnet}}, \bibinfo {author} {\bibfnamefont {M.}~\bibnamefont {G\"arttner}},
  \bibinfo {author} {\bibfnamefont {K.~A.}\ \bibnamefont {Gilmore}}, \bibinfo
  {author} {\bibfnamefont {J.~E.}\ \bibnamefont {Jordan}}, \bibinfo {author}
  {\bibfnamefont {J.}~\bibnamefont {Cohn}}, \bibinfo {author} {\bibfnamefont
  {J.~K.}\ \bibnamefont {Freericks}}, \bibinfo {author} {\bibfnamefont {A.~M.}\
  \bibnamefont {Rey}},\ and\ \bibinfo {author} {\bibfnamefont {J.~J.}\
  \bibnamefont {Bollinger}},\ }\bibfield  {title} {\bibinfo {title}
  {{Verification of a Many-Ion Simulator of the Dicke Model Through Slow
  Quenches across a Phase Transition}},\ }\href
  {https://doi.org/10.1103/PhysRevLett.121.040503} {\bibfield  {journal}
  {\bibinfo  {journal} {Phys. Rev. Lett.}\ }\textbf {\bibinfo {volume} {121}},\
  \bibinfo {pages} {040503} (\bibinfo {year} {2018})}\BibitemShut {NoStop}%
\bibitem [{\citenamefont {Klinder}\ \emph
  {et~al.}(2015{\natexlab{b}})\citenamefont {Klinder}, \citenamefont
  {Ke\ss{}ler}, \citenamefont {Bakhtiari}, \citenamefont {Thorwart},\ and\
  \citenamefont {Hemmerich}}]{PhysRevLett.115.230403}%
  \BibitemOpen
  \bibfield  {author} {\bibinfo {author} {\bibfnamefont {J.}~\bibnamefont
  {Klinder}}, \bibinfo {author} {\bibfnamefont {H.}~\bibnamefont {Ke\ss{}ler}},
  \bibinfo {author} {\bibfnamefont {M.~R.}\ \bibnamefont {Bakhtiari}}, \bibinfo
  {author} {\bibfnamefont {M.}~\bibnamefont {Thorwart}},\ and\ \bibinfo
  {author} {\bibfnamefont {A.}~\bibnamefont {Hemmerich}},\ }\bibfield  {title}
  {\bibinfo {title} {{Observation of a Superradiant Mott Insulator in the
  Dicke-Hubbard Model}},\ }\href
  {https://doi.org/10.1103/PhysRevLett.115.230403} {\bibfield  {journal}
  {\bibinfo  {journal} {Phys. Rev. Lett.}\ }\textbf {\bibinfo {volume} {115}},\
  \bibinfo {pages} {230403} (\bibinfo {year} {2015}{\natexlab{b}})}\BibitemShut
  {NoStop}%
\bibitem [{\citenamefont {Ferioli}\ \emph {et~al.}(2023)\citenamefont
  {Ferioli}, \citenamefont {Glicenstein}, \citenamefont {Ferrier-Barbut},\ and\
  \citenamefont {Browaeys}}]{Ferioli2023}%
  \BibitemOpen
  \bibfield  {author} {\bibinfo {author} {\bibfnamefont {G.}~\bibnamefont
  {Ferioli}}, \bibinfo {author} {\bibfnamefont {A.}~\bibnamefont
  {Glicenstein}}, \bibinfo {author} {\bibfnamefont {I.}~\bibnamefont
  {Ferrier-Barbut}},\ and\ \bibinfo {author} {\bibfnamefont {A.}~\bibnamefont
  {Browaeys}},\ }\bibfield  {title} {\bibinfo {title} {{A non-equilibrium
  superradiant phase transition in free space}},\ }\href
  {https://doi.org/10.1038/s41567-023-02064-w} {\bibfield  {journal} {\bibinfo
  {journal} {Nature Physics}\ }\textbf {\bibinfo {volume} {19}},\ \bibinfo
  {pages} {1345} (\bibinfo {year} {2023})}\BibitemShut {NoStop}%
\bibitem [{\citenamefont {Liedl}\ \emph {et~al.}(2024)\citenamefont {Liedl},
  \citenamefont {Tebbenjohanns}, \citenamefont {Bach}, \citenamefont {Pucher},
  \citenamefont {Rauschenbeutel},\ and\ \citenamefont
  {Schneeweiss}}]{PhysRevX.14.011020}%
  \BibitemOpen
  \bibfield  {author} {\bibinfo {author} {\bibfnamefont {C.}~\bibnamefont
  {Liedl}}, \bibinfo {author} {\bibfnamefont {F.}~\bibnamefont
  {Tebbenjohanns}}, \bibinfo {author} {\bibfnamefont {C.}~\bibnamefont {Bach}},
  \bibinfo {author} {\bibfnamefont {S.}~\bibnamefont {Pucher}}, \bibinfo
  {author} {\bibfnamefont {A.}~\bibnamefont {Rauschenbeutel}},\ and\ \bibinfo
  {author} {\bibfnamefont {P.}~\bibnamefont {Schneeweiss}},\ }\bibfield
  {title} {\bibinfo {title} {{Observation of Superradiant Bursts in a Cascaded
  Quantum System}},\ }\href {https://doi.org/10.1103/PhysRevX.14.011020}
  {\bibfield  {journal} {\bibinfo  {journal} {Phys. Rev. X}\ }\textbf {\bibinfo
  {volume} {14}},\ \bibinfo {pages} {011020} (\bibinfo {year}
  {2024})}\BibitemShut {NoStop}%
\bibitem [{\citenamefont {Fink}\ \emph {et~al.}(2009)\citenamefont {Fink},
  \citenamefont {Bianchetti}, \citenamefont {Baur}, \citenamefont {G\"oppl},
  \citenamefont {Steffen}, \citenamefont {Filipp}, \citenamefont {Leek},
  \citenamefont {Blais},\ and\ \citenamefont
  {Wallraff}}]{PhysRevLett.103.083601}%
  \BibitemOpen
  \bibfield  {author} {\bibinfo {author} {\bibfnamefont {J.~M.}\ \bibnamefont
  {Fink}}, \bibinfo {author} {\bibfnamefont {R.}~\bibnamefont {Bianchetti}},
  \bibinfo {author} {\bibfnamefont {M.}~\bibnamefont {Baur}}, \bibinfo {author}
  {\bibfnamefont {M.}~\bibnamefont {G\"oppl}}, \bibinfo {author} {\bibfnamefont
  {L.}~\bibnamefont {Steffen}}, \bibinfo {author} {\bibfnamefont
  {S.}~\bibnamefont {Filipp}}, \bibinfo {author} {\bibfnamefont {P.~J.}\
  \bibnamefont {Leek}}, \bibinfo {author} {\bibfnamefont {A.}~\bibnamefont
  {Blais}},\ and\ \bibinfo {author} {\bibfnamefont {A.}~\bibnamefont
  {Wallraff}},\ }\bibfield  {title} {\bibinfo {title} {{Dressed Collective
  Qubit States and the Tavis-Cummings Model in Circuit QED}},\ }\href
  {https://doi.org/10.1103/PhysRevLett.103.083601} {\bibfield  {journal}
  {\bibinfo  {journal} {Phys. Rev. Lett.}\ }\textbf {\bibinfo {volume} {103}},\
  \bibinfo {pages} {083601} (\bibinfo {year} {2009})}\BibitemShut {NoStop}%
\bibitem [{\citenamefont {Feng}\ \emph {et~al.}(2015)\citenamefont {Feng},
  \citenamefont {Zhong}, \citenamefont {Liu}, \citenamefont {Yan},
  \citenamefont {Yang}, \citenamefont {Twamley},\ and\ \citenamefont
  {Wang}}]{Feng2015}%
  \BibitemOpen
  \bibfield  {author} {\bibinfo {author} {\bibfnamefont {M.}~\bibnamefont
  {Feng}}, \bibinfo {author} {\bibfnamefont {Y.~P.}\ \bibnamefont {Zhong}},
  \bibinfo {author} {\bibfnamefont {T.}~\bibnamefont {Liu}}, \bibinfo {author}
  {\bibfnamefont {L.~L.}\ \bibnamefont {Yan}}, \bibinfo {author} {\bibfnamefont
  {W.~L.}\ \bibnamefont {Yang}}, \bibinfo {author} {\bibfnamefont
  {J.}~\bibnamefont {Twamley}},\ and\ \bibinfo {author} {\bibfnamefont
  {H.}~\bibnamefont {Wang}},\ }\bibfield  {title} {\bibinfo {title} {{Exploring
  the quantum critical behaviour in a driven Tavis--Cummings circuit}},\ }\href
  {https://doi.org/10.1038/ncomms8111} {\bibfield  {journal} {\bibinfo
  {journal} {Nature Communications}\ }\textbf {\bibinfo {volume} {6}},\
  \bibinfo {pages} {7111} (\bibinfo {year} {2015})}\BibitemShut {NoStop}%
\bibitem [{\citenamefont {Yoshihara}\ \emph {et~al.}(2017)\citenamefont
  {Yoshihara}, \citenamefont {Fuse}, \citenamefont {Ashhab}, \citenamefont
  {Kakuyanagi}, \citenamefont {Saito},\ and\ \citenamefont
  {Semba}}]{Yoshihara2017}%
  \BibitemOpen
  \bibfield  {author} {\bibinfo {author} {\bibfnamefont {F.}~\bibnamefont
  {Yoshihara}}, \bibinfo {author} {\bibfnamefont {T.}~\bibnamefont {Fuse}},
  \bibinfo {author} {\bibfnamefont {S.}~\bibnamefont {Ashhab}}, \bibinfo
  {author} {\bibfnamefont {K.}~\bibnamefont {Kakuyanagi}}, \bibinfo {author}
  {\bibfnamefont {S.}~\bibnamefont {Saito}},\ and\ \bibinfo {author}
  {\bibfnamefont {K.}~\bibnamefont {Semba}},\ }\bibfield  {title} {\bibinfo
  {title} {{Superconducting qubit--oscillator circuit beyond the
  ultrastrong-coupling regime}},\ }\href {https://doi.org/10.1038/nphys3906}
  {\bibfield  {journal} {\bibinfo  {journal} {Nature Physics}\ }\textbf
  {\bibinfo {volume} {13}},\ \bibinfo {pages} {44} (\bibinfo {year}
  {2017})}\BibitemShut {NoStop}%
\bibitem [{\citenamefont {Forn-D\'{\i}az}\ \emph {et~al.}(2019)\citenamefont
  {Forn-D\'{\i}az}, \citenamefont {Lamata}, \citenamefont {Rico}, \citenamefont
  {Kono},\ and\ \citenamefont {Solano}}]{RevModPhys.91.025005}%
  \BibitemOpen
  \bibfield  {author} {\bibinfo {author} {\bibfnamefont {P.}~\bibnamefont
  {Forn-D\'{\i}az}}, \bibinfo {author} {\bibfnamefont {L.}~\bibnamefont
  {Lamata}}, \bibinfo {author} {\bibfnamefont {E.}~\bibnamefont {Rico}},
  \bibinfo {author} {\bibfnamefont {J.}~\bibnamefont {Kono}},\ and\ \bibinfo
  {author} {\bibfnamefont {E.}~\bibnamefont {Solano}},\ }\bibfield  {title}
  {\bibinfo {title} {{Ultrastrong coupling regimes of light-matter
  interaction}},\ }\href {https://doi.org/10.1103/RevModPhys.91.025005}
  {\bibfield  {journal} {\bibinfo  {journal} {Rev. Mod. Phys.}\ }\textbf
  {\bibinfo {volume} {91}},\ \bibinfo {pages} {025005} (\bibinfo {year}
  {2019})}\BibitemShut {NoStop}%
\bibitem [{\citenamefont {Frisk~Kockum}\ \emph {et~al.}(2019)\citenamefont
  {Frisk~Kockum}, \citenamefont {Miranowicz}, \citenamefont {De~Liberato},
  \citenamefont {Savasta},\ and\ \citenamefont {Nori}}]{FriskKockum2019}%
  \BibitemOpen
  \bibfield  {author} {\bibinfo {author} {\bibfnamefont {A.}~\bibnamefont
  {Frisk~Kockum}}, \bibinfo {author} {\bibfnamefont {A.}~\bibnamefont
  {Miranowicz}}, \bibinfo {author} {\bibfnamefont {S.}~\bibnamefont
  {De~Liberato}}, \bibinfo {author} {\bibfnamefont {S.}~\bibnamefont
  {Savasta}},\ and\ \bibinfo {author} {\bibfnamefont {F.}~\bibnamefont
  {Nori}},\ }\bibfield  {title} {\bibinfo {title} {{Ultrastrong coupling
  between light and matter}},\ }\href
  {https://doi.org/10.1038/s42254-018-0006-2} {\bibfield  {journal} {\bibinfo
  {journal} {Nature Reviews Physics}\ }\textbf {\bibinfo {volume} {1}},\
  \bibinfo {pages} {19} (\bibinfo {year} {2019})}\BibitemShut {NoStop}%
\bibitem [{\citenamefont {Blais}\ \emph {et~al.}(2021)\citenamefont {Blais},
  \citenamefont {Grimsmo}, \citenamefont {Girvin},\ and\ \citenamefont
  {Wallraff}}]{RevModPhys.93.025005}%
  \BibitemOpen
  \bibfield  {author} {\bibinfo {author} {\bibfnamefont {A.}~\bibnamefont
  {Blais}}, \bibinfo {author} {\bibfnamefont {A.~L.}\ \bibnamefont {Grimsmo}},
  \bibinfo {author} {\bibfnamefont {S.~M.}\ \bibnamefont {Girvin}},\ and\
  \bibinfo {author} {\bibfnamefont {A.}~\bibnamefont {Wallraff}},\ }\bibfield
  {title} {\bibinfo {title} {{Circuit quantum electrodynamics}},\ }\href
  {https://doi.org/10.1103/RevModPhys.93.025005} {\bibfield  {journal}
  {\bibinfo  {journal} {Rev. Mod. Phys.}\ }\textbf {\bibinfo {volume} {93}},\
  \bibinfo {pages} {025005} (\bibinfo {year} {2021})}\BibitemShut {NoStop}%
\bibitem [{\citenamefont {Qin}\ \emph {et~al.}(2024)\citenamefont {Qin},
  \citenamefont {Kockum}, \citenamefont {Muñoz}, \citenamefont {Miranowicz},\
  and\ \citenamefont {Nori}}]{QIN20241}%
  \BibitemOpen
  \bibfield  {author} {\bibinfo {author} {\bibfnamefont {W.}~\bibnamefont
  {Qin}}, \bibinfo {author} {\bibfnamefont {A.~F.}\ \bibnamefont {Kockum}},
  \bibinfo {author} {\bibfnamefont {C.~S.}\ \bibnamefont {Muñoz}}, \bibinfo
  {author} {\bibfnamefont {A.}~\bibnamefont {Miranowicz}},\ and\ \bibinfo
  {author} {\bibfnamefont {F.}~\bibnamefont {Nori}},\ }\bibfield  {title}
  {\bibinfo {title} {Quantum amplification and simulation of strong and
  ultrastrong coupling of light and matter},\ }\href
  {https://doi.org/https://doi.org/10.1016/j.physrep.2024.05.003} {\bibfield
  {journal} {\bibinfo  {journal} {Physics Reports}\ }\textbf {\bibinfo {volume}
  {1078}},\ \bibinfo {pages} {1} (\bibinfo {year} {2024})}\BibitemShut
  {NoStop}%
\bibitem [{\citenamefont {Feynman}(1982)}]{Feynman1982}%
  \BibitemOpen
  \bibfield  {author} {\bibinfo {author} {\bibfnamefont {R.~P.}\ \bibnamefont
  {Feynman}},\ }\bibfield  {title} {\bibinfo {title} {Simulating physics with
  computers},\ }\href {https://doi.org/10.1007/BF02650179} {\bibfield
  {journal} {\bibinfo  {journal} {International Journal of Theoretical
  Physics}\ }\textbf {\bibinfo {volume} {21}},\ \bibinfo {pages} {467}
  (\bibinfo {year} {1982})}\BibitemShut {NoStop}%
\bibitem [{\citenamefont {Lloyd}(1996)}]{Lloyd:1996}%
  \BibitemOpen
  \bibfield  {author} {\bibinfo {author} {\bibfnamefont {S.}~\bibnamefont
  {Lloyd}},\ }\bibfield  {title} {\bibinfo {title} {Universal quantum
  simulators},\ }\href {https://doi.org/10.1126/science.273.5278.1073}
  {\bibfield  {journal} {\bibinfo  {journal} {Science}\ }\textbf {\bibinfo
  {volume} {273}},\ \bibinfo {pages} {1073} (\bibinfo {year}
  {1996})}\BibitemShut {NoStop}%
\bibitem [{\citenamefont {Weimer}\ \emph {et~al.}(2010)\citenamefont {Weimer},
  \citenamefont {M{\"u}ller}, \citenamefont {Lesanovsky}, \citenamefont
  {Zoller},\ and\ \citenamefont {B{\"u}chler}}]{Weimer2010}%
  \BibitemOpen
  \bibfield  {author} {\bibinfo {author} {\bibfnamefont {H.}~\bibnamefont
  {Weimer}}, \bibinfo {author} {\bibfnamefont {M.}~\bibnamefont {M{\"u}ller}},
  \bibinfo {author} {\bibfnamefont {I.}~\bibnamefont {Lesanovsky}}, \bibinfo
  {author} {\bibfnamefont {P.}~\bibnamefont {Zoller}},\ and\ \bibinfo {author}
  {\bibfnamefont {H.~P.}\ \bibnamefont {B{\"u}chler}},\ }\bibfield  {title}
  {\bibinfo {title} {A rydberg quantum simulator},\ }\href
  {https://doi.org/10.1038/nphys1614} {\bibfield  {journal} {\bibinfo
  {journal} {Nature Physics}\ }\textbf {\bibinfo {volume} {6}},\ \bibinfo
  {pages} {382} (\bibinfo {year} {2010})}\BibitemShut {NoStop}%
\bibitem [{\citenamefont {Bassman~Oftelie}\ \emph {et~al.}(2021)\citenamefont
  {Bassman~Oftelie}, \citenamefont {Urbanek}, \citenamefont {Metcalf},
  \citenamefont {Carter}, \citenamefont {Kemper},\ and\ \citenamefont
  {de~Jong}}]{BassmanOftelie_2021}%
  \BibitemOpen
  \bibfield  {author} {\bibinfo {author} {\bibfnamefont {L.}~\bibnamefont
  {Bassman~Oftelie}}, \bibinfo {author} {\bibfnamefont {M.}~\bibnamefont
  {Urbanek}}, \bibinfo {author} {\bibfnamefont {M.}~\bibnamefont {Metcalf}},
  \bibinfo {author} {\bibfnamefont {J.}~\bibnamefont {Carter}}, \bibinfo
  {author} {\bibfnamefont {A.~F.}\ \bibnamefont {Kemper}},\ and\ \bibinfo
  {author} {\bibfnamefont {W.~A.}\ \bibnamefont {de~Jong}},\ }\bibfield
  {title} {\bibinfo {title} {Simulating quantum materials with digital quantum
  computers},\ }\href {https://doi.org/10.1088/2058-9565/ac1ca6} {\bibfield
  {journal} {\bibinfo  {journal} {Quantum Science and Technology}\ }\textbf
  {\bibinfo {volume} {6}},\ \bibinfo {pages} {043002} (\bibinfo {year}
  {2021})}\BibitemShut {NoStop}%
\bibitem [{\citenamefont {Bravyi}\ \emph {et~al.}(2024)\citenamefont {Bravyi},
  \citenamefont {Cross}, \citenamefont {Gambetta}, \citenamefont {Maslov},
  \citenamefont {Rall},\ and\ \citenamefont {Yoder}}]{Bravyi2024}%
  \BibitemOpen
  \bibfield  {author} {\bibinfo {author} {\bibfnamefont {S.}~\bibnamefont
  {Bravyi}}, \bibinfo {author} {\bibfnamefont {A.~W.}\ \bibnamefont {Cross}},
  \bibinfo {author} {\bibfnamefont {J.~M.}\ \bibnamefont {Gambetta}}, \bibinfo
  {author} {\bibfnamefont {D.}~\bibnamefont {Maslov}}, \bibinfo {author}
  {\bibfnamefont {P.}~\bibnamefont {Rall}},\ and\ \bibinfo {author}
  {\bibfnamefont {T.~J.}\ \bibnamefont {Yoder}},\ }\bibfield  {title} {\bibinfo
  {title} {High-threshold and low-overhead fault-tolerant quantum memory},\
  }\href {https://doi.org/10.1038/s41586-024-07107-7} {\bibfield  {journal}
  {\bibinfo  {journal} {Nature}\ }\textbf {\bibinfo {volume} {627}},\ \bibinfo
  {pages} {778} (\bibinfo {year} {2024})}\BibitemShut {NoStop}%
\bibitem [{\citenamefont {Miessen}\ \emph {et~al.}(2024)\citenamefont
  {Miessen}, \citenamefont {Egger}, \citenamefont {Tavernelli},\ and\
  \citenamefont {Mazzola}}]{PRXQuantum.5.040320}%
  \BibitemOpen
  \bibfield  {author} {\bibinfo {author} {\bibfnamefont {A.}~\bibnamefont
  {Miessen}}, \bibinfo {author} {\bibfnamefont {D.~J.}\ \bibnamefont {Egger}},
  \bibinfo {author} {\bibfnamefont {I.}~\bibnamefont {Tavernelli}},\ and\
  \bibinfo {author} {\bibfnamefont {G.}~\bibnamefont {Mazzola}},\ }\bibfield
  {title} {\bibinfo {title} {Benchmarking digital quantum simulations above
  hundreds of qubits using quantum critical dynamics},\ }\href
  {https://doi.org/10.1103/PRXQuantum.5.040320} {\bibfield  {journal} {\bibinfo
   {journal} {PRX Quantum}\ }\textbf {\bibinfo {volume} {5}},\ \bibinfo {pages}
  {040320} (\bibinfo {year} {2024})}\BibitemShut {NoStop}%
\bibitem [{\citenamefont {Fauseweh}(2024)}]{Fauseweh2024}%
  \BibitemOpen
  \bibfield  {author} {\bibinfo {author} {\bibfnamefont {B.}~\bibnamefont
  {Fauseweh}},\ }\bibfield  {title} {\bibinfo {title} {Quantum many-body
  simulations on digital quantum computers: State-of-the-art and future
  challenges},\ }\href {https://doi.org/10.1038/s41467-024-46402-9} {\bibfield
  {journal} {\bibinfo  {journal} {Nature Communications}\ }\textbf {\bibinfo
  {volume} {15}},\ \bibinfo {pages} {2123} (\bibinfo {year}
  {2024})}\BibitemShut {NoStop}%
\bibitem [{\citenamefont {Macridin}\ \emph
  {et~al.}(2018{\natexlab{a}})\citenamefont {Macridin}, \citenamefont
  {Spentzouris}, \citenamefont {Amundson},\ and\ \citenamefont
  {Harnik}}]{Macridin:2018}%
  \BibitemOpen
  \bibfield  {author} {\bibinfo {author} {\bibfnamefont {A.}~\bibnamefont
  {Macridin}}, \bibinfo {author} {\bibfnamefont {P.}~\bibnamefont
  {Spentzouris}}, \bibinfo {author} {\bibfnamefont {J.}~\bibnamefont
  {Amundson}},\ and\ \bibinfo {author} {\bibfnamefont {R.}~\bibnamefont
  {Harnik}},\ }\bibfield  {title} {\bibinfo {title} {Electron-phonon systems on
  a universal quantum computer},\ }\href
  {https://doi.org/10.1103/PhysRevLett.121.110504} {\bibfield  {journal}
  {\bibinfo  {journal} {Phys. Rev. Lett.}\ }\textbf {\bibinfo {volume} {121}},\
  \bibinfo {pages} {110504} (\bibinfo {year} {2018}{\natexlab{a}})}\BibitemShut
  {NoStop}%
\bibitem [{\citenamefont {Macridin}\ \emph
  {et~al.}(2018{\natexlab{b}})\citenamefont {Macridin}, \citenamefont
  {Spentzouris}, \citenamefont {Amundson},\ and\ \citenamefont
  {Harnik}}]{Macridin:2018a}%
  \BibitemOpen
  \bibfield  {author} {\bibinfo {author} {\bibfnamefont {A.}~\bibnamefont
  {Macridin}}, \bibinfo {author} {\bibfnamefont {P.}~\bibnamefont
  {Spentzouris}}, \bibinfo {author} {\bibfnamefont {J.}~\bibnamefont
  {Amundson}},\ and\ \bibinfo {author} {\bibfnamefont {R.}~\bibnamefont
  {Harnik}},\ }\bibfield  {title} {\bibinfo {title} {Digital quantum
  computation of fermion-boson interacting systems},\ }\href
  {https://doi.org/10.1103/PhysRevA.98.042312} {\bibfield  {journal} {\bibinfo
  {journal} {Phys. Rev. A}\ }\textbf {\bibinfo {volume} {98}},\ \bibinfo
  {pages} {042312} (\bibinfo {year} {2018}{\natexlab{b}})}\BibitemShut
  {NoStop}%
\bibitem [{\citenamefont {Mezzacapo}\ \emph {et~al.}(2014)\citenamefont
  {Mezzacapo}, \citenamefont {Las~Heras}, \citenamefont {Pedernales},
  \citenamefont {DiCarlo}, \citenamefont {Solano},\ and\ \citenamefont
  {Lamata}}]{Mezzacapo:2014aa}%
  \BibitemOpen
  \bibfield  {author} {\bibinfo {author} {\bibfnamefont {A.}~\bibnamefont
  {Mezzacapo}}, \bibinfo {author} {\bibfnamefont {U.}~\bibnamefont
  {Las~Heras}}, \bibinfo {author} {\bibfnamefont {J.~S.}\ \bibnamefont
  {Pedernales}}, \bibinfo {author} {\bibfnamefont {L.}~\bibnamefont {DiCarlo}},
  \bibinfo {author} {\bibfnamefont {E.}~\bibnamefont {Solano}},\ and\ \bibinfo
  {author} {\bibfnamefont {L.}~\bibnamefont {Lamata}},\ }\bibfield  {title}
  {\bibinfo {title} {{Digital Quantum Rabi and Dicke Models in Superconducting
  Circuits}},\ }\href {https://doi.org/10.1038/srep07482} {\bibfield  {journal}
  {\bibinfo  {journal} {Scientific Reports}\ }\textbf {\bibinfo {volume} {4}},\
  \bibinfo {pages} {7482} (\bibinfo {year} {2014})}\BibitemShut {NoStop}%
\bibitem [{\citenamefont {Langford}\ \emph {et~al.}(2017)\citenamefont
  {Langford}, \citenamefont {Sagastizabal}, \citenamefont {Kounalakis},
  \citenamefont {Dickel}, \citenamefont {Bruno}, \citenamefont {Luthi},
  \citenamefont {Thoen}, \citenamefont {Endo},\ and\ \citenamefont
  {DiCarlo}}]{Langford:2017aa}%
  \BibitemOpen
  \bibfield  {author} {\bibinfo {author} {\bibfnamefont {N.~K.}\ \bibnamefont
  {Langford}}, \bibinfo {author} {\bibfnamefont {R.}~\bibnamefont
  {Sagastizabal}}, \bibinfo {author} {\bibfnamefont {M.}~\bibnamefont
  {Kounalakis}}, \bibinfo {author} {\bibfnamefont {C.}~\bibnamefont {Dickel}},
  \bibinfo {author} {\bibfnamefont {A.}~\bibnamefont {Bruno}}, \bibinfo
  {author} {\bibfnamefont {F.}~\bibnamefont {Luthi}}, \bibinfo {author}
  {\bibfnamefont {D.~J.}\ \bibnamefont {Thoen}}, \bibinfo {author}
  {\bibfnamefont {A.}~\bibnamefont {Endo}},\ and\ \bibinfo {author}
  {\bibfnamefont {L.}~\bibnamefont {DiCarlo}},\ }\bibfield  {title} {\bibinfo
  {title} {{Experimentally simulating the dynamics of quantum light and matter
  at deep-strong coupling}},\ }\href
  {https://doi.org/10.1038/s41467-017-01061-x} {\bibfield  {journal} {\bibinfo
  {journal} {Nature Communications}\ }\textbf {\bibinfo {volume} {8}},\
  \bibinfo {pages} {1715} (\bibinfo {year} {2017})}\BibitemShut {NoStop}%
\bibitem [{\citenamefont {Crane}\ \emph {et~al.}(2024)\citenamefont {Crane},
  \citenamefont {Smith}, \citenamefont {Tomesh}, \citenamefont {Eickbusch},
  \citenamefont {Martyn}, \citenamefont {K\"uhn}, \citenamefont {Funcke},
  \citenamefont {DeMarco}, \citenamefont {Chuang}, \citenamefont {Wiebe},
  \citenamefont {Schuckert},\ and\ \citenamefont {Girvin}}]{crane2024}%
  \BibitemOpen
  \bibfield  {author} {\bibinfo {author} {\bibfnamefont {E.}~\bibnamefont
  {Crane}}, \bibinfo {author} {\bibfnamefont {K.~C.}\ \bibnamefont {Smith}},
  \bibinfo {author} {\bibfnamefont {T.}~\bibnamefont {Tomesh}}, \bibinfo
  {author} {\bibfnamefont {A.}~\bibnamefont {Eickbusch}}, \bibinfo {author}
  {\bibfnamefont {J.~M.}\ \bibnamefont {Martyn}}, \bibinfo {author}
  {\bibfnamefont {S.}~\bibnamefont {K\"uhn}}, \bibinfo {author} {\bibfnamefont
  {L.}~\bibnamefont {Funcke}}, \bibinfo {author} {\bibfnamefont {M.~A.}\
  \bibnamefont {DeMarco}}, \bibinfo {author} {\bibfnamefont {I.~L.}\
  \bibnamefont {Chuang}}, \bibinfo {author} {\bibfnamefont {N.}~\bibnamefont
  {Wiebe}}, \bibinfo {author} {\bibfnamefont {A.}~\bibnamefont {Schuckert}},\
  and\ \bibinfo {author} {\bibfnamefont {S.~M.}\ \bibnamefont {Girvin}},\
  }\href {https://arxiv.org/abs/2409.03747} {\bibinfo {title} {Hybrid
  oscillator-qubit quantum processors: Simulating fermions, bosons, and gauge
  fields}} (\bibinfo {year} {2024}),\ \Eprint
  {https://arxiv.org/abs/2409.03747} {arXiv:2409.03747 [quant-ph]} \BibitemShut
  {NoStop}%
\bibitem [{\citenamefont {Kumar}\ \emph {et~al.}(2025)\citenamefont {Kumar},
  \citenamefont {Hegade}, \citenamefont {Visuri}, \citenamefont {Bhargava},
  \citenamefont {Hernandez}, \citenamefont {Solano}, \citenamefont
  {Albarr{\'a}n-Arriagada},\ and\ \citenamefont {Barrios}}]{Kumar2025}%
  \BibitemOpen
  \bibfield  {author} {\bibinfo {author} {\bibfnamefont {S.}~\bibnamefont
  {Kumar}}, \bibinfo {author} {\bibfnamefont {N.~N.}\ \bibnamefont {Hegade}},
  \bibinfo {author} {\bibfnamefont {A.-M.}\ \bibnamefont {Visuri}}, \bibinfo
  {author} {\bibfnamefont {B.~A.}\ \bibnamefont {Bhargava}}, \bibinfo {author}
  {\bibfnamefont {J.~F.~R.}\ \bibnamefont {Hernandez}}, \bibinfo {author}
  {\bibfnamefont {E.}~\bibnamefont {Solano}}, \bibinfo {author} {\bibfnamefont
  {F.}~\bibnamefont {Albarr{\'a}n-Arriagada}},\ and\ \bibinfo {author}
  {\bibfnamefont {G.~A.}\ \bibnamefont {Barrios}},\ }\bibfield  {title}
  {\bibinfo {title} {Digital-analog quantum computing of fermion-boson models
  in superconducting circuits},\ }\href
  {https://doi.org/10.1038/s41534-025-01001-4} {\bibfield  {journal} {\bibinfo
  {journal} {npj Quantum Information}\ }\textbf {\bibinfo {volume} {11}},\
  \bibinfo {pages} {43} (\bibinfo {year} {2025})}\BibitemShut {NoStop}%
\bibitem [{\citenamefont {Huber}\ \emph {et~al.}(2025)\citenamefont {Huber},
  \citenamefont {Roy}, \citenamefont {Koch}, \citenamefont {Tsitsilin},
  \citenamefont {Schirk}, \citenamefont {Glaser}, \citenamefont {Bruckmoser},
  \citenamefont {Schweizer}, \citenamefont {Romeiro}, \citenamefont {Krylov},
  \citenamefont {Singh}, \citenamefont {Haslbeck}, \citenamefont {Knudsen},
  \citenamefont {Marx}, \citenamefont {Pfeiffer}, \citenamefont {Schneider},
  \citenamefont {Wallner}, \citenamefont {Bunch}, \citenamefont {Richard},
  \citenamefont {S\"odergren}, \citenamefont {Liegener}, \citenamefont
  {Werninghaus},\ and\ \citenamefont
  {Filipp}}]{huber2024parametricmultielementcouplingarchitecture}%
  \BibitemOpen
  \bibfield  {author} {\bibinfo {author} {\bibfnamefont {G.}~\bibnamefont
  {Huber}}, \bibinfo {author} {\bibfnamefont {F.}~\bibnamefont {Roy}}, \bibinfo
  {author} {\bibfnamefont {L.}~\bibnamefont {Koch}}, \bibinfo {author}
  {\bibfnamefont {I.}~\bibnamefont {Tsitsilin}}, \bibinfo {author}
  {\bibfnamefont {J.}~\bibnamefont {Schirk}}, \bibinfo {author} {\bibfnamefont
  {N.}~\bibnamefont {Glaser}}, \bibinfo {author} {\bibfnamefont
  {N.}~\bibnamefont {Bruckmoser}}, \bibinfo {author} {\bibfnamefont
  {C.}~\bibnamefont {Schweizer}}, \bibinfo {author} {\bibfnamefont
  {J.}~\bibnamefont {Romeiro}}, \bibinfo {author} {\bibfnamefont
  {G.}~\bibnamefont {Krylov}}, \bibinfo {author} {\bibfnamefont
  {M.}~\bibnamefont {Singh}}, \bibinfo {author} {\bibfnamefont
  {F.}~\bibnamefont {Haslbeck}}, \bibinfo {author} {\bibfnamefont
  {M.}~\bibnamefont {Knudsen}}, \bibinfo {author} {\bibfnamefont
  {A.}~\bibnamefont {Marx}}, \bibinfo {author} {\bibfnamefont {F.}~\bibnamefont
  {Pfeiffer}}, \bibinfo {author} {\bibfnamefont {C.}~\bibnamefont {Schneider}},
  \bibinfo {author} {\bibfnamefont {F.}~\bibnamefont {Wallner}}, \bibinfo
  {author} {\bibfnamefont {D.}~\bibnamefont {Bunch}}, \bibinfo {author}
  {\bibfnamefont {L.}~\bibnamefont {Richard}}, \bibinfo {author} {\bibfnamefont
  {L.}~\bibnamefont {S\"odergren}}, \bibinfo {author} {\bibfnamefont
  {K.}~\bibnamefont {Liegener}}, \bibinfo {author} {\bibfnamefont
  {M.}~\bibnamefont {Werninghaus}},\ and\ \bibinfo {author} {\bibfnamefont
  {S.}~\bibnamefont {Filipp}},\ }\bibfield  {title} {\bibinfo {title}
  {Parametric multielement coupling architecture for coherent and dissipative
  control of superconducting qubits},\ }\href
  {https://doi.org/10.1103/9shv-l4cx} {\bibfield  {journal} {\bibinfo
  {journal} {PRX Quantum}\ }\textbf {\bibinfo {volume} {6}},\ \bibinfo {pages}
  {030313} (\bibinfo {year} {2025})}\BibitemShut {NoStop}%
\bibitem [{\citenamefont {Reagor}\ \emph {et~al.}(2016)\citenamefont {Reagor},
  \citenamefont {Pfaff}, \citenamefont {Axline}, \citenamefont {Heeres},
  \citenamefont {Ofek}, \citenamefont {Sliwa}, \citenamefont {Holland},
  \citenamefont {Wang}, \citenamefont {Blumoff}, \citenamefont {Chou},
  \citenamefont {Hatridge}, \citenamefont {Frunzio}, \citenamefont {Devoret},
  \citenamefont {Jiang},\ and\ \citenamefont
  {Schoelkopf}}]{PhysRevB.94.014506}%
  \BibitemOpen
  \bibfield  {author} {\bibinfo {author} {\bibfnamefont {M.}~\bibnamefont
  {Reagor}}, \bibinfo {author} {\bibfnamefont {W.}~\bibnamefont {Pfaff}},
  \bibinfo {author} {\bibfnamefont {C.}~\bibnamefont {Axline}}, \bibinfo
  {author} {\bibfnamefont {R.~W.}\ \bibnamefont {Heeres}}, \bibinfo {author}
  {\bibfnamefont {N.}~\bibnamefont {Ofek}}, \bibinfo {author} {\bibfnamefont
  {K.}~\bibnamefont {Sliwa}}, \bibinfo {author} {\bibfnamefont
  {E.}~\bibnamefont {Holland}}, \bibinfo {author} {\bibfnamefont
  {C.}~\bibnamefont {Wang}}, \bibinfo {author} {\bibfnamefont {J.}~\bibnamefont
  {Blumoff}}, \bibinfo {author} {\bibfnamefont {K.}~\bibnamefont {Chou}},
  \bibinfo {author} {\bibfnamefont {M.~J.}\ \bibnamefont {Hatridge}}, \bibinfo
  {author} {\bibfnamefont {L.}~\bibnamefont {Frunzio}}, \bibinfo {author}
  {\bibfnamefont {M.~H.}\ \bibnamefont {Devoret}}, \bibinfo {author}
  {\bibfnamefont {L.}~\bibnamefont {Jiang}},\ and\ \bibinfo {author}
  {\bibfnamefont {R.~J.}\ \bibnamefont {Schoelkopf}},\ }\bibfield  {title}
  {\bibinfo {title} {{Quantum memory with millisecond coherence in circuit
  QED}},\ }\href {https://doi.org/10.1103/PhysRevB.94.014506} {\bibfield
  {journal} {\bibinfo  {journal} {Phys. Rev. B}\ }\textbf {\bibinfo {volume}
  {94}},\ \bibinfo {pages} {014506} (\bibinfo {year} {2016})}\BibitemShut
  {NoStop}%
\bibitem [{\citenamefont {Ganjam}\ \emph {et~al.}(2024)\citenamefont {Ganjam},
  \citenamefont {Wang}, \citenamefont {Lu}, \citenamefont {Banerjee},
  \citenamefont {Lei}, \citenamefont {Krayzman}, \citenamefont {Kisslinger},
  \citenamefont {Zhou}, \citenamefont {Li}, \citenamefont {Jia}, \citenamefont
  {Liu}, \citenamefont {Frunzio},\ and\ \citenamefont
  {Schoelkopf}}]{Ganjam2024}%
  \BibitemOpen
  \bibfield  {author} {\bibinfo {author} {\bibfnamefont {S.}~\bibnamefont
  {Ganjam}}, \bibinfo {author} {\bibfnamefont {Y.}~\bibnamefont {Wang}},
  \bibinfo {author} {\bibfnamefont {Y.}~\bibnamefont {Lu}}, \bibinfo {author}
  {\bibfnamefont {A.}~\bibnamefont {Banerjee}}, \bibinfo {author}
  {\bibfnamefont {C.~U.}\ \bibnamefont {Lei}}, \bibinfo {author} {\bibfnamefont
  {L.}~\bibnamefont {Krayzman}}, \bibinfo {author} {\bibfnamefont
  {K.}~\bibnamefont {Kisslinger}}, \bibinfo {author} {\bibfnamefont
  {C.}~\bibnamefont {Zhou}}, \bibinfo {author} {\bibfnamefont {R.}~\bibnamefont
  {Li}}, \bibinfo {author} {\bibfnamefont {Y.}~\bibnamefont {Jia}}, \bibinfo
  {author} {\bibfnamefont {M.}~\bibnamefont {Liu}}, \bibinfo {author}
  {\bibfnamefont {L.}~\bibnamefont {Frunzio}},\ and\ \bibinfo {author}
  {\bibfnamefont {R.~J.}\ \bibnamefont {Schoelkopf}},\ }\bibfield  {title}
  {\bibinfo {title} {Surpassing millisecond coherence in on chip
  superconducting quantum memories by optimizing materials and circuit
  design},\ }\href {https://doi.org/10.1038/s41467-024-47857-6} {\bibfield
  {journal} {\bibinfo  {journal} {Nature Communications}\ }\textbf {\bibinfo
  {volume} {15}},\ \bibinfo {pages} {3687} (\bibinfo {year}
  {2024})}\BibitemShut {NoStop}%
\bibitem [{\citenamefont {Um}\ \emph {et~al.}(2016)\citenamefont {Um},
  \citenamefont {Zhang}, \citenamefont {Lv}, \citenamefont {Lu}, \citenamefont
  {An}, \citenamefont {Zhang}, \citenamefont {Nha}, \citenamefont {Kim},\ and\
  \citenamefont {Kim}}]{Um2016}%
  \BibitemOpen
  \bibfield  {author} {\bibinfo {author} {\bibfnamefont {M.}~\bibnamefont
  {Um}}, \bibinfo {author} {\bibfnamefont {J.}~\bibnamefont {Zhang}}, \bibinfo
  {author} {\bibfnamefont {D.}~\bibnamefont {Lv}}, \bibinfo {author}
  {\bibfnamefont {Y.}~\bibnamefont {Lu}}, \bibinfo {author} {\bibfnamefont
  {S.}~\bibnamefont {An}}, \bibinfo {author} {\bibfnamefont {J.-N.}\
  \bibnamefont {Zhang}}, \bibinfo {author} {\bibfnamefont {H.}~\bibnamefont
  {Nha}}, \bibinfo {author} {\bibfnamefont {M.~S.}\ \bibnamefont {Kim}},\ and\
  \bibinfo {author} {\bibfnamefont {K.}~\bibnamefont {Kim}},\ }\bibfield
  {title} {\bibinfo {title} {Phonon arithmetic in a trapped ion system},\
  }\href {https://doi.org/10.1038/ncomms11410} {\bibfield  {journal} {\bibinfo
  {journal} {Nature Communications}\ }\textbf {\bibinfo {volume} {7}},\
  \bibinfo {pages} {11410} (\bibinfo {year} {2016})}\BibitemShut {NoStop}%
\bibitem [{\citenamefont {Wang}\ \emph {et~al.}(2020)\citenamefont {Wang},
  \citenamefont {Crain}, \citenamefont {Fang}, \citenamefont {Zhang},
  \citenamefont {Huang}, \citenamefont {Liang}, \citenamefont {Leung},
  \citenamefont {Brown},\ and\ \citenamefont {Kim}}]{PhysRevLett.125.150505}%
  \BibitemOpen
  \bibfield  {author} {\bibinfo {author} {\bibfnamefont {Y.}~\bibnamefont
  {Wang}}, \bibinfo {author} {\bibfnamefont {S.}~\bibnamefont {Crain}},
  \bibinfo {author} {\bibfnamefont {C.}~\bibnamefont {Fang}}, \bibinfo {author}
  {\bibfnamefont {B.}~\bibnamefont {Zhang}}, \bibinfo {author} {\bibfnamefont
  {S.}~\bibnamefont {Huang}}, \bibinfo {author} {\bibfnamefont
  {Q.}~\bibnamefont {Liang}}, \bibinfo {author} {\bibfnamefont {P.~H.}\
  \bibnamefont {Leung}}, \bibinfo {author} {\bibfnamefont {K.~R.}\ \bibnamefont
  {Brown}},\ and\ \bibinfo {author} {\bibfnamefont {J.}~\bibnamefont {Kim}},\
  }\bibfield  {title} {\bibinfo {title} {High-fidelity two-qubit gates using a
  microelectromechanical-system-based beam steering system for individual qubit
  addressing},\ }\href {https://doi.org/10.1103/PhysRevLett.125.150505}
  {\bibfield  {journal} {\bibinfo  {journal} {Phys. Rev. Lett.}\ }\textbf
  {\bibinfo {volume} {125}},\ \bibinfo {pages} {150505} (\bibinfo {year}
  {2020})}\BibitemShut {NoStop}%
\bibitem [{\citenamefont {Shapiro}\ \emph {et~al.}(2025)\citenamefont
  {Shapiro}, \citenamefont {Weber}, \citenamefont {Bode}, \citenamefont
  {Wilhelm},\ and\ \citenamefont {Bagrets}}]{shapiro_2025_16581022}%
  \BibitemOpen
  \bibfield  {author} {\bibinfo {author} {\bibfnamefont {D.~S.}\ \bibnamefont
  {Shapiro}}, \bibinfo {author} {\bibfnamefont {Y.}~\bibnamefont {Weber}},
  \bibinfo {author} {\bibfnamefont {T.}~\bibnamefont {Bode}}, \bibinfo {author}
  {\bibfnamefont {F.~K.}\ \bibnamefont {Wilhelm}},\ and\ \bibinfo {author}
  {\bibfnamefont {D.}~\bibnamefont {Bagrets}},\ }\href@noop {} {\bibinfo
  {title} {{Data for Digital-Analog Simulations of Schr\"odinger Cat States in
  the Dicke-Ising Model}}},\ \bibinfo {howpublished}
  {\url{https://zenodo.org/records/16581022}} (\bibinfo {year}
  {2025})\BibitemShut {NoStop}%
\bibitem [{\citenamefont {Leghtas}\ \emph {et~al.}(2013)\citenamefont
  {Leghtas}, \citenamefont {Kirchmair}, \citenamefont {Vlastakis},
  \citenamefont {Devoret}, \citenamefont {Schoelkopf},\ and\ \citenamefont
  {Mirrahimi}}]{PhysRevA.87.042315}%
  \BibitemOpen
  \bibfield  {author} {\bibinfo {author} {\bibfnamefont {Z.}~\bibnamefont
  {Leghtas}}, \bibinfo {author} {\bibfnamefont {G.}~\bibnamefont {Kirchmair}},
  \bibinfo {author} {\bibfnamefont {B.}~\bibnamefont {Vlastakis}}, \bibinfo
  {author} {\bibfnamefont {M.~H.}\ \bibnamefont {Devoret}}, \bibinfo {author}
  {\bibfnamefont {R.~J.}\ \bibnamefont {Schoelkopf}},\ and\ \bibinfo {author}
  {\bibfnamefont {M.}~\bibnamefont {Mirrahimi}},\ }\bibfield  {title} {\bibinfo
  {title} {{Deterministic protocol for mapping a qubit to coherent state
  superpositions in a cavity}},\ }\href
  {https://doi.org/10.1103/PhysRevA.87.042315} {\bibfield  {journal} {\bibinfo
  {journal} {Phys. Rev. A}\ }\textbf {\bibinfo {volume} {87}},\ \bibinfo
  {pages} {042315} (\bibinfo {year} {2013})}\BibitemShut {NoStop}%
\bibitem [{\citenamefont {Vlastakis}\ \emph {et~al.}(2013)\citenamefont
  {Vlastakis}, \citenamefont {Kirchmair}, \citenamefont {Leghtas},
  \citenamefont {Nigg}, \citenamefont {Frunzio}, \citenamefont {Girvin},
  \citenamefont {Mirrahimi}, \citenamefont {Devoret},\ and\ \citenamefont
  {Schoelkopf}}]{doi:10.1126/science.1243289}%
  \BibitemOpen
  \bibfield  {author} {\bibinfo {author} {\bibfnamefont {B.}~\bibnamefont
  {Vlastakis}}, \bibinfo {author} {\bibfnamefont {G.}~\bibnamefont
  {Kirchmair}}, \bibinfo {author} {\bibfnamefont {Z.}~\bibnamefont {Leghtas}},
  \bibinfo {author} {\bibfnamefont {S.~E.}\ \bibnamefont {Nigg}}, \bibinfo
  {author} {\bibfnamefont {L.}~\bibnamefont {Frunzio}}, \bibinfo {author}
  {\bibfnamefont {S.~M.}\ \bibnamefont {Girvin}}, \bibinfo {author}
  {\bibfnamefont {M.}~\bibnamefont {Mirrahimi}}, \bibinfo {author}
  {\bibfnamefont {M.~H.}\ \bibnamefont {Devoret}},\ and\ \bibinfo {author}
  {\bibfnamefont {R.~J.}\ \bibnamefont {Schoelkopf}},\ }\bibfield  {title}
  {\bibinfo {title} {{Deterministically Encoding Quantum Information Using
  100-Photon Schr{\"o}dinger Cat States}},\ }\href
  {https://doi.org/10.1126/science.1243289} {\bibfield  {journal} {\bibinfo
  {journal} {Science}\ }\textbf {\bibinfo {volume} {342}},\ \bibinfo {pages}
  {607} (\bibinfo {year} {2013})}\BibitemShut {NoStop}%
\bibitem [{\citenamefont {Grimm}\ \emph {et~al.}(2020)\citenamefont {Grimm},
  \citenamefont {Frattini}, \citenamefont {Puri}, \citenamefont {Mundhada},
  \citenamefont {Touzard}, \citenamefont {Mirrahimi}, \citenamefont {Girvin},
  \citenamefont {Shankar},\ and\ \citenamefont {Devoret}}]{Grimm2020}%
  \BibitemOpen
  \bibfield  {author} {\bibinfo {author} {\bibfnamefont {A.}~\bibnamefont
  {Grimm}}, \bibinfo {author} {\bibfnamefont {N.~E.}\ \bibnamefont {Frattini}},
  \bibinfo {author} {\bibfnamefont {S.}~\bibnamefont {Puri}}, \bibinfo {author}
  {\bibfnamefont {S.~O.}\ \bibnamefont {Mundhada}}, \bibinfo {author}
  {\bibfnamefont {S.}~\bibnamefont {Touzard}}, \bibinfo {author} {\bibfnamefont
  {M.}~\bibnamefont {Mirrahimi}}, \bibinfo {author} {\bibfnamefont {S.~M.}\
  \bibnamefont {Girvin}}, \bibinfo {author} {\bibfnamefont {S.}~\bibnamefont
  {Shankar}},\ and\ \bibinfo {author} {\bibfnamefont {M.~H.}\ \bibnamefont
  {Devoret}},\ }\bibfield  {title} {\bibinfo {title} {{Stabilization and
  operation of a Kerr-cat qubit}},\ }\href
  {https://doi.org/10.1038/s41586-020-2587-z} {\bibfield  {journal} {\bibinfo
  {journal} {Nature}\ }\textbf {\bibinfo {volume} {584}},\ \bibinfo {pages}
  {205} (\bibinfo {year} {2020})}\BibitemShut {NoStop}%
\bibitem [{\citenamefont {Lescanne}\ \emph {et~al.}(2020)\citenamefont
  {Lescanne}, \citenamefont {Villiers}, \citenamefont {Peronnin}, \citenamefont
  {Sarlette}, \citenamefont {Delbecq}, \citenamefont {Huard}, \citenamefont
  {Kontos}, \citenamefont {Mirrahimi},\ and\ \citenamefont
  {Leghtas}}]{Lescanne2020}%
  \BibitemOpen
  \bibfield  {author} {\bibinfo {author} {\bibfnamefont {R.}~\bibnamefont
  {Lescanne}}, \bibinfo {author} {\bibfnamefont {M.}~\bibnamefont {Villiers}},
  \bibinfo {author} {\bibfnamefont {T.}~\bibnamefont {Peronnin}}, \bibinfo
  {author} {\bibfnamefont {A.}~\bibnamefont {Sarlette}}, \bibinfo {author}
  {\bibfnamefont {M.}~\bibnamefont {Delbecq}}, \bibinfo {author} {\bibfnamefont
  {B.}~\bibnamefont {Huard}}, \bibinfo {author} {\bibfnamefont
  {T.}~\bibnamefont {Kontos}}, \bibinfo {author} {\bibfnamefont
  {M.}~\bibnamefont {Mirrahimi}},\ and\ \bibinfo {author} {\bibfnamefont
  {Z.}~\bibnamefont {Leghtas}},\ }\bibfield  {title} {\bibinfo {title}
  {{Exponential suppression of bit-flips in a qubit encoded in an
  oscillator}},\ }\href {https://doi.org/10.1038/s41567-020-0824-x} {\bibfield
  {journal} {\bibinfo  {journal} {Nature Physics}\ }\textbf {\bibinfo {volume}
  {16}},\ \bibinfo {pages} {509} (\bibinfo {year} {2020})}\BibitemShut
  {NoStop}%
\bibitem [{\citenamefont {Xu}\ \emph {et~al.}(2021)\citenamefont {Xu},
  \citenamefont {Fallas~Padilla},\ and\ \citenamefont
  {Pu}}]{PhysRevA.104.043708}%
  \BibitemOpen
  \bibfield  {author} {\bibinfo {author} {\bibfnamefont {Y.}~\bibnamefont
  {Xu}}, \bibinfo {author} {\bibfnamefont {D.}~\bibnamefont {Fallas~Padilla}},\
  and\ \bibinfo {author} {\bibfnamefont {H.}~\bibnamefont {Pu}},\ }\bibfield
  {title} {\bibinfo {title} {{Multicriticality and quantum fluctuation in a
  generalized Dicke model}},\ }\href
  {https://doi.org/10.1103/PhysRevA.104.043708} {\bibfield  {journal} {\bibinfo
   {journal} {Phys. Rev. A}\ }\textbf {\bibinfo {volume} {104}},\ \bibinfo
  {pages} {043708} (\bibinfo {year} {2021})}\BibitemShut {NoStop}%
\bibitem [{\citenamefont {Salath\'e}\ \emph {et~al.}(2015)\citenamefont
  {Salath\'e}, \citenamefont {Mondal}, \citenamefont {Oppliger}, \citenamefont
  {Heinsoo}, \citenamefont {Kurpiers}, \citenamefont
  {Poto\ifmmode~\check{c}\else \v{c}\fi{}nik}, \citenamefont {Mezzacapo},
  \citenamefont {Las~Heras}, \citenamefont {Lamata}, \citenamefont {Solano},
  \citenamefont {Filipp},\ and\ \citenamefont {Wallraff}}]{PhysRevX.5.021027}%
  \BibitemOpen
  \bibfield  {author} {\bibinfo {author} {\bibfnamefont {Y.}~\bibnamefont
  {Salath\'e}}, \bibinfo {author} {\bibfnamefont {M.}~\bibnamefont {Mondal}},
  \bibinfo {author} {\bibfnamefont {M.}~\bibnamefont {Oppliger}}, \bibinfo
  {author} {\bibfnamefont {J.}~\bibnamefont {Heinsoo}}, \bibinfo {author}
  {\bibfnamefont {P.}~\bibnamefont {Kurpiers}}, \bibinfo {author}
  {\bibfnamefont {A.}~\bibnamefont {Poto\ifmmode~\check{c}\else
  \v{c}\fi{}nik}}, \bibinfo {author} {\bibfnamefont {A.}~\bibnamefont
  {Mezzacapo}}, \bibinfo {author} {\bibfnamefont {U.}~\bibnamefont
  {Las~Heras}}, \bibinfo {author} {\bibfnamefont {L.}~\bibnamefont {Lamata}},
  \bibinfo {author} {\bibfnamefont {E.}~\bibnamefont {Solano}}, \bibinfo
  {author} {\bibfnamefont {S.}~\bibnamefont {Filipp}},\ and\ \bibinfo {author}
  {\bibfnamefont {A.}~\bibnamefont {Wallraff}},\ }\bibfield  {title} {\bibinfo
  {title} {{Digital Quantum Simulation of Spin Models with Circuit Quantum
  Electrodynamics}},\ }\href {https://doi.org/10.1103/PhysRevX.5.021027}
  {\bibfield  {journal} {\bibinfo  {journal} {Phys. Rev. X}\ }\textbf {\bibinfo
  {volume} {5}},\ \bibinfo {pages} {021027} (\bibinfo {year}
  {2015})}\BibitemShut {NoStop}%
\bibitem [{\citenamefont {Wecker}\ \emph {et~al.}(2015)\citenamefont {Wecker},
  \citenamefont {Hastings}, \citenamefont {Wiebe}, \citenamefont {Clark},
  \citenamefont {Nayak},\ and\ \citenamefont {Troyer}}]{PhysRevA.92.062318}%
  \BibitemOpen
  \bibfield  {author} {\bibinfo {author} {\bibfnamefont {D.}~\bibnamefont
  {Wecker}}, \bibinfo {author} {\bibfnamefont {M.~B.}\ \bibnamefont
  {Hastings}}, \bibinfo {author} {\bibfnamefont {N.}~\bibnamefont {Wiebe}},
  \bibinfo {author} {\bibfnamefont {B.~K.}\ \bibnamefont {Clark}}, \bibinfo
  {author} {\bibfnamefont {C.}~\bibnamefont {Nayak}},\ and\ \bibinfo {author}
  {\bibfnamefont {M.}~\bibnamefont {Troyer}},\ }\bibfield  {title} {\bibinfo
  {title} {{Solving strongly correlated electron models on a quantum
  computer}},\ }\href {https://doi.org/10.1103/PhysRevA.92.062318} {\bibfield
  {journal} {\bibinfo  {journal} {Phys. Rev. A}\ }\textbf {\bibinfo {volume}
  {92}},\ \bibinfo {pages} {062318} (\bibinfo {year} {2015})}\BibitemShut
  {NoStop}%
\bibitem [{\citenamefont {Lutterbach}\ and\ \citenamefont
  {Davidovich}(1997)}]{PhysRevLett.78.2547}%
  \BibitemOpen
  \bibfield  {author} {\bibinfo {author} {\bibfnamefont {L.~G.}\ \bibnamefont
  {Lutterbach}}\ and\ \bibinfo {author} {\bibfnamefont {L.}~\bibnamefont
  {Davidovich}},\ }\bibfield  {title} {\bibinfo {title} {{Method for Direct
  Measurement of the Wigner Function in Cavity QED and Ion Traps}},\ }\href
  {https://doi.org/10.1103/PhysRevLett.78.2547} {\bibfield  {journal} {\bibinfo
   {journal} {Phys. Rev. Lett.}\ }\textbf {\bibinfo {volume} {78}},\ \bibinfo
  {pages} {2547} (\bibinfo {year} {1997})}\BibitemShut {NoStop}%
\end{thebibliography}

%

     \onecolumngrid
   \newcounter{defcounter}
\setcounter{defcounter}{0}

\appendix

  \section{Mean-field free energy for Dicke-Ising Hamiltonian at $\omega_z=0$}
  \label{App_JW}
In this part of the Appendix, we derive the free energy (\ref{F_D}) from the Hamiltonian (\ref{H_DI}) assuming $\omega_z=0$. In the limit of zero $\omega_z$, only two types of  spin operators remain in the Hamiltonian, $\sigma^x_q$ and $\sigma^z_q \sigma^z_{q+1}$. Applying the Jordan-Wigner  representation to them gives
 \begin{equation}
\hat \sigma^x_q  = \hat c^\dagger_q \hat c_q -\hat c_q\hat c^\dagger_q  , \quad\hat  \sigma^z_q \hat \sigma^z_{q+1}=  \hat c^\dagger_q\hat c_{q+1} + \hat c^\dagger_{q+1}\hat c_{q} + \hat c^\dagger_{q}\hat c_{q+1}^\dagger + \hat c_{q+1}\hat c_{q} .  \label{Maj}
\end{equation}
The Hamiltonian (\ref{H_DI}) after this transformation  has
  bilinear fermion combinations,
\begin{equation}
\hat H_{\rm DI}= \omega_0 \hat a^\dagger \hat a-J\sum_{q=1}^N  \big(     \hat c_q^\dagger \hat c_{q+1}+ \hat  c_{q+1}^\dagger \hat c_{q}+   \hat  c_q^\dagger \hat   c_{q+1}^\dagger+ \hat c_{q+1}  \hat   c_{q}\big) +\frac{ g}{\sqrt{N}} (\hat  a^\dagger +\hat a)\sum_{q=1}^N(\hat  c_q^\dagger\hat  c_q- \hat  c_q \hat  c_q^\dagger) .
\end{equation}
We note that for   $\omega_z\neq 0$ additional terms $\sim \hat \sigma^z_q$ arise,  yielding nonlocal  fermion strings and resulting in a more complicated derivation of the free energy.
Consider the partition function $Z={\rm tr} \big(e^{-\hat H_{\rm DI}/T}\big)$ at finite temperature $T$. It is  reduced to the Matsubara path integral
\begin{equation}
Z=\int  d [  a, \bar a, c, \bar c] e^{-S[  a, \bar a, c, \bar c] }\label{Z_0}
\end{equation}
 over complex boson fields $ a(\tau), \bar a(\tau)$ and Grassmann fields $ c_q(\tau), \bar c_q(\tau)$, where $\tau$ is the   imaginary  time $\tau\in[0, 1/T]$.  These fields describe, respectively, photons  and Jordan-Wigner fermions.   The Matsubara action in (\ref{Z_0}) is 
\begin{equation}
S=\int\limits_0^{1/T}d\tau \Big( \bar a \partial_\tau a+ \sum_{q=1}^N \bar c\partial_\tau c +H_{\rm DI}[a, \bar a, c, \bar c]\Big ) .  \label{action_0}
\end{equation}
Assuming periodic boundary conditions for the Ising chain, we introduce the wave numbers ${\bf k}=\frac{2\pi n}{N}-\pi$ with $0\!\leq \!n< \! N$   spanning a Brillouin zone. The Fourier transformation into   $\bf{k}$-space for Grassmann fields reads $c_{\bf k}= \frac{1}{\sqrt N} \sum_{q=1}^N e^{-i{\bf k} q }c_q$.
The action (\ref{action_0}) can be parametrized via  Nambu vectors  $\Psi_{\bf k} = \begin{bmatrix}
c_{\bf k} &
 \bar c_{-\bf k}
 \end{bmatrix}^T$ in $\gamma$-space resulting in the following form:
\begin{equation}
 S = S_{\rm ph}-\frac{1}{2}\sum_{\bf k}  \int\limits_0^{1/T} d\tau   \Psi_{-\bf k}^T \gamma_x  G_{\bf k}^{-1}[a,\bar a] \Psi_{\bf k} , \label{action}
\end{equation}
where the inverted Green function (Lagrangian) reads 
\begin{equation}
-  G_{\bf k}^{-1}[a(\tau),\bar a(\tau)] = 
  \begin{bmatrix}
\partial_\tau - 2J\cos {\bf k}  + 2g (\bar a (\tau)+a(\tau))/\sqrt{N} &  -2 i J \sin {\bf k}   \\ \\
 2 i J \sin {\bf k}  & \partial_\tau  +2J\cos {\bf k}  -2g (\bar a (\tau)+a(\tau))/\sqrt{N}
  \end{bmatrix} \label{G_inv}
\end{equation}
and 
\begin{equation}S_{\rm ph}= \int\limits_0^{1/T} d\tau \left(\bar a \partial_\tau a +\omega_0\bar a a\right) 
\end{equation} is the free photon action.
Note, that $G_{\bf k}^{-1}[a,\bar a] $  is a nonstationary matrix because of time-depending $a$-fields and, therefore, an inversion  is a nontrivial task.  
Below we tackle this problem in a mean-field approximation.
    
We parametrize the complex  $\bar a$ and $a$ through real fields $u$ and $v$,
\begin{equation}
 a(\tau) =\sqrt{N}\frac{u(\tau) +i v(\tau) }{2} , \quad \bar a(\tau)=\sqrt{N}\frac{u(\tau) -i v(\tau) }{2}.
\end{equation}
After a Fourier transformation defined as $a(\tau)=\sum_n a_n e^{-i 2\pi n T \tau}$ and  $\bar a(\tau)=\sum_n a_n e^{i 2\pi n T \tau}$, we  take the Gaussian integrals over real boson $v_n$ and real  Grassmann $\Psi_{\bf k}$ fields. As a result, we receive
an effective action for the real boson $u(\tau)$, which  is a sum of the free boson action $S_u$ and the spin contribution given by the logarithm of the fermion  determinant,
\begin{equation}
S_{\rm eff}=\frac{1}{4}N\int\limits_0^{1/T} \left(\frac{1}{\omega_0}\left(\partial_\tau u\right)^2 +\omega_0u^2\right) d\tau \ - \ \frac{1}{2} {\rm  ln} \ \! {\rm   det}(-G^{-1}_{\bf k}[u]) . \label{trln_inst}
\end{equation}
It features the inverted Green function (\ref{G_inv}) written as $G^{-1}_{\bf k}[u]=-\gamma_0\partial_\tau -H_{\bf k}(\tau)$, where the $\tau$-dependent Hamiltonian matrix in Nambu $\gamma$-space reads 
\begin{equation}
H_{\bf k}(\tau)= -2J(\gamma_z\cos {\bf k}-\gamma_y\sin {\bf k}) +2g\gamma_z u(\tau). \label{H_tau}
\end{equation}
In the mean-field approximation, we assume $u(\tau)={\rm const}$ neglecting temporal fluctuations.
The fermion determinant in (\ref{trln_inst}) can be found analytically through an infinite product over Matsubara frequencies with 2$\times$2 $\gamma$-matrix determinants,
\begin{equation}
{\rm  ln} \ \! {\rm   det}(G^{-1}_{\bf k}[u]G_{\bf k}[0])= N\int\limits_{-\pi}^\pi\frac{d{\bf k}}{2\pi}\ln\left ( \prod_n \frac{{\rm det}_\gamma(-i2\pi n T\gamma_0 -2J(\gamma_z\cos {\bf k}-\gamma_y\sin {\bf k}) +2g\gamma_z u)}{{\rm det}_\gamma(-i2\pi n T\gamma_0 -2J(\gamma_z\cos {\bf k}-\gamma_y\sin {\bf k}))}\right) . \label{H_tau}
\end{equation}
We added $G_{\bf k}[0]$ to regularize the action. This factor emerges from  normalizing the partition sum by its value at  $g=0$. To compute the infinite product we use the identity 
\be \prod_{n\geq 1} (1+x^2/n^2)=\frac{1}{\pi \sqrt x}\sinh(\pi  |x|). \label{product}
\ee
 Taking  the leading term $\sim \frac{1}{T}$ in the limit of low temperatures, and then integrating over ${\bf k}$, we arrive at the mean-field action
  \begin{equation}
S_{\rm mf}= N\int\limits_0^{1/T} \left(\frac{\left(\partial_\tau u\right)^2}{4\omega_0} +\mathcal{F}(u) \right)d\tau \label{S_mf_app}
\end{equation} 
with the normalized free energy $\mathcal{F}(u)=\frac{1}{N}F_{\rm DI}(u)$ provided in Eq.~(\ref{F_DI}),
 \begin{equation}
\mathcal{F}(u) = \frac{1}{4}\omega_0u^2 -  \frac{2}{\pi} (J+g |u|) {\rm E}\left[\frac{4 g J |u|}{(J+g |u|)^2}\right]  . \label{F_DI_app}
\end{equation}
 
 \section{Instanton trajectory}
 \label{App_instanton}
Consider the mean-field part of the action  (\ref{S_mf_app}). When the system is in the superradiant  phase, a variation of $S_{\rm mf}$ by $u$ yields the equation for an instanton saddle-point trajectory, 
 \begin{equation}
\partial^2_\tau u - 2\omega_0\partial_u \mathcal{F}(u)=0 , \label{instanton_eq} 
\end{equation}
with boundary conditions  $u(\tau\!=\!0)=-u(\tau\!=\!1/T)\!=\!- u_0$, where $u_0\!>\!0$ is a nonzero solution of the equation $\partial_u \mathcal{F} (u)=0$ for the superradiant order parameter.
There is also an  integral motion, which is  analogous to the full energy in classical mechanics. It reads
 \begin{equation}
 - \frac{\left(\partial_\tau u\right)^2}{4\omega_0} +\mathcal{F}(u) =\mathcal{F}(-u_0)  \label{instanton_eq_1}
 \end{equation}
 where the constant $\mathcal{F}(-u_0)$ is given by the free energy minimum at     $u=-u_0$.
 The instanton solution is given by an \textit{implicit} function $u_{\rm inst}(\tau)$, which follows from Eq.~(\ref{instanton_eq_1}) as
\begin{equation}
    \tau = \int\limits_{-u_0}^{u_{\rm inst}(\tau)} \frac{d u}{\sqrt{2\omega_0(\mathcal{F}(u) -\mathcal{F}(-u_0))}}  . \label{instanton_appendix}
\end{equation}
The schematic shape of the solution that follows from this integral is presented in Fig.~\ref{angular}(c).

{\color{black} \section{Derivation of the Circuit Structure}\label{s_di_gate_app}

In this Appendix, we comment on the derivation of the Rabi gate~\eqref{JC} and the sequence of gates~\eqref{S_D_gate0} that simulate the evolution of the Dicke–Ising model.
We begin with the Rabi model. For its Trotterized evolution operator,
\be
\hat U_R(t) = \prod_{k=1}^{N} \hat U_R(t_{k+1}, t_k), \qquad t_k = (k-1)\Delta t,
\ee
each step is approximated by a sequence of three evolutions governed by the JC Hamiltonian~\eqref{HR}, as shown in Eq.~\eqref{rabi_gate_evol} of the main text.
It is advantageous to rewrite this equation in the following form:
\be
\label{eq:U_R_Trotter}
\hat U_R(t_{k+1}, t_k) = \hat U_{\rm JC}(t_k + \Delta t, t_k + \tfrac{3}{4} \Delta t)\, \hat\sigma^x \, \hat U_{\rm JC}(t_k + \tfrac{3}{4} \Delta t, t_k + \tfrac{1}{4} \Delta t) \, \hat\sigma^x\,
\hat U_{\rm JC}(t_k + \tfrac{1}{4} \Delta t, t_k),
\ee
where $\hat U_{\rm JC}(t)$ denotes the evolution operator of the JC model.

Our goal is to relate this Trotterized evolution to a sequence of quantum gates that can be implemented on quantum hardware, including the JC gate~\eqref{JC} introduced in the main text. To this end, we switch to the interaction picture with respect to the free Hamiltonian $\hat H_0$, see Eqs.\eqref{eq:H_0_def} and \eqref{eq:Int_Rep}. Using the identity
\be
e^{i \hat H_0 t} \hat U_{\rm JC} (t, t') e^{- i \hat H_0 t'} = \hat S_{\rm JC} (\theta), \qquad \theta = g (t - t'),
\ee
we find that the above unitary transformation of $\hat U_{\rm JC}$ yields the JC gate~\eqref{JC} with the phase determined by a product of the time interval and the strength of interaction.
Applying this to the full Trotterized step~\eqref{eq:U_R_Trotter}, we obtain
\be
\hat S_{R} (t_{k+1}, t_k) := e^{i \hat H_0 t_{k+1}} \hat U_{R}(t_{k+1}, t_k) e^{- i \hat H_0 t_k} = \hat S_{\rm JC}(\theta/2) \hat X_\pi(\varphi_+) \hat S_{\rm JC}(\theta) \hat X_\pi(\varphi_-) \hat S_{\rm JC}(\theta/2),
\ee
where the rotated $X$-gates $\hat X_\pi(\varphi_\pm)$ are defined in Eq.~\eqref{x_gate} with phases  $\varphi_- = \omega_0 (t_k + \Delta t /4)$ and $\varphi_+ = \omega_0 (t_k  + 3\Delta t/4)$. These arise from expressing the Pauli $\hat\sigma^x$ operators in the rotating frame:
\be
\hat X_\pi (\varphi_+) = e^{- \tfrac i2 \omega_0 ( t_k + \tfrac 3 4 \Delta t) \hat\sigma^z } \hat \sigma^x e^{ \tfrac i2 \omega_0 ( t_k + \tfrac 3 4 \Delta t) \hat\sigma^z}  \equiv \exp{(-i \varphi_+ \hat\sigma^z)} \hat\sigma^x,
\ee
with a similar expression for  $\hat X_\pi (\varphi_-)$.
In this way, we have derived the unitary gate~\eqref{SRabi}, which emulates a single Trotter step of the Rabi model in the rotating frame.

Let us now discuss how the gate sequence~\eqref{S_DI_gate0}, which approximates the evolution of the Dicke–Ising model, can be justified. To this end, we begin by describing the construction of the Dicke gate $\hat S_{\rm D}$, defined in Eq.~\eqref{S_D_gate0}, in the case of two spins. 
It will become clear that a generalization to the case with an arbitrary number of spins $N$ is straightforward. 
For $N = 2$, we rewrite the Dicke Hamiltonian as a sum of three terms:
\be
\hat H_D = \hat H^{(1)} +  \hat H^{(2)} - \hat h_0, \qquad  \hat H^{(j)} = \hat h_0 + g(\hat a +\hat a ^\dagger)\hat \sigma^x_j, \qquad  \hat h_0 = \omega_0 \hat a^\dagger \hat a.
\ee
This decomposition naturally suggests a Trotterization scheme of the form:
\be
\label{eq:U_D_N2}
\hat U_D(t_{k+1}, t_k) \approx \hat U_R^{(2)}(t_k + \Delta t, t_k)  \, e^{i \hat h_0 \Delta t}\,   \hat U_R^{(1)}(t_k + \Delta t, t_k),
\ee
where each $\hat U_R^{(j)}$ denotes the evolution operator generated by the Rabi Hamiltonian $\hat H^{(j)}$. For these, we apply the same approximation as previously used,
see Eq.~\eqref{eq:U_R_Trotter}.
It is important to emphasize that there are multiple ways to Trotterize the evolution operator corresponding to $\hat H_D$. Our specific choice of Trotterization is motivated by its favorable error properties: the discretization error scales as ${\cal O}(\Delta t^3)$, which matches that of the Rabi gate~\eqref{SRabi}.

It is now instructive to move to the rotating frame defined by the free bosonic Hamiltonian $\hat h_0$. Let us denote evolution operators in this frame by $\widetilde S(t, t')$, in contrast to $\hat S(t, t')$, which refers to the interaction representation with respect to the Hamiltonian $\hat H_0$ used in the main text.
Noting that $e^{i \hat h_0 \Delta t} \equiv \hat U_0 (t_k, t_k + \Delta t)$ represents backward time evolution under the free Hamiltonian, 
we can deduce from Eq.~\eqref{eq:U_D_N2} the following relation:
\be
\label{eq:S_D_tilde}
\widetilde S_D(t_{k+1}, t_k) = \widetilde S_R^{(2)}(t_k + \Delta t, t_k) \widetilde S_R^{(1)}(t_k + \Delta t, t_k).
\ee
Here, the Rabi gates in the new rotating frame are related to the basic gates~\eqref{SRabi} by simple $Z$-rotations:
\be
\widetilde S_R^{(j)}(t_k + \Delta t, t_k) = e^{ \tfrac i 2 \phi_{k+1} \hat \sigma_j^z} \hat S_R^{(j)}(t_k + \Delta t, t_k) 
e^{ - \tfrac i 2 \phi_k \hat \sigma_j^z}, \qquad \phi_k = \omega_0 t_k.
\ee
The same applies to the full gate $\widetilde S_D$. This gate differs from its counterpart $\hat S_D$ by a similarity transformation, 
consisting of $Z$-rotations acting on both spins. Consequently, the structure of Eq.~\eqref{eq:S_D_tilde} also holds in the rotating frame 
defined with respect to $\hat H_0$:
\be
\hat S_D(t_{k+1}, t_k) = \hat S_R^{(2)}(t_k + \Delta t, t_k) \hat S_R^{(1)}(t_k + \Delta t, t_k).
\ee
Thus, we have arrived at the result~\eqref{S_D_gate0}, provided the qubit–boson architecture permits interactions between any qubit and the resonator. 
In the case of restricted connectivity, SWAP gates must be used, leading to the equivalent expression:
\be
\hat S_D(t_{k+1}, t_k) =   {\rm SWAP}^{(12)}\hat S_R^{(1)}(t_k + \Delta t, t_k) {\rm SWAP}^{(12)} \hat S_R^{(1)}(t_k + \Delta t, t_k).
\ee

The above considerations can be readily extended to the case of an arbitrary number of qubits $N$. Indeed, the Dicke Hamiltonian can be rewritten as follows:
\be
\hat H_D = \sum_{j=1}^N \hat H^{(j)} - (N-1) \hat h_0,
\ee
and the corresponding Trotterization scheme takes the form:
\be
\label{eq:U_D_N}
\hat U_D(t_{k+1}, t_k) \approx \hat U^{(N)}(t_k + \Delta t, t_k)  \prod_{j=1}^{N-1}\, e^{i \hat h_0 \Delta t}\,   \hat U^{(j)}(t_k + \Delta t, t_k),
\ee
which is a straightforward generalization of Eq.\eqref{eq:U_D_N2}. 
The result~\eqref{S_D_gate0} from the main text then follows by applying the same reasoning as in the case of $N=2$.

It remains to discuss how the Ising-type terms in the Hamiltonian~\eqref{H_DI} can be incorporated into the above Trotterized scheme. 
To justify the appearance of $Z$- and $ZZ$-gates in Eq.\eqref{S_DI_gate0}, we assume that the Ising Hamiltonian is applied during the time interval 
$[t_{k} + \Delta t, t_k]$ in the form of $\delta$-pulses:
\be
\hat H(t) = \hat H_D  - \eta\, \delta(t-t_{k+1})\sum_{j=1}^{N-1}\hat\sigma_j^z\hat\sigma_{j+1}^z - \beta\, \delta(t-t_{k+1}) \sum_{j=1}^N\hat \sigma^z_j, 
\qquad \eta = J\Delta t, \quad \beta = \omega_z \Delta t,
\ee
so that the dynamics depends only on the effective phases $\eta$ and $\beta$. 
This ansatz corresponds to a separate Trotterization of the Dicke and Ising Hamiltonians and directly leads to the gate sequence given in Eq.\eqref{S_DI_gate0}.

}

 {\color{black}
\section{Model of a Dissipation in the Circuit}\label{noise_model}
 In the emulation of noise effects upon Trotterization, we apply Lindbladian dissipative dynamics after each of the $L$   unitariy $\hat S_{\rm DI}$-gates. This model of noise  takes into account dephasing ($\Gamma_\phi$) and relaxation ($\Gamma_1$) rates of the qubits, and assumes  Rabi gate duration  {\it physical} time $\tau_{\rm Rabi}$. When a qubit is detuned from the resonator, it is assumed to remain fully coherent. Photon relaxation during a single $\hat S_{\rm DI}$-gate occurs at a rate $\kappa$ at a time interval  $N\tau_{\rm R}$, due to $N$ Rabi gates  being applied sequentially. We do not account for errors introduced by  CNOT or SWAP gates, their operation times are also omitted from the noise model. 

Consider the first Trotter step where the initial state $\hat \rho_0=|\rm FM\rangle\langle\rm FM|$ evolves into a mixed state $\hat \rho_1$. This step proceeds as follows: (i)  a unitary transformation is applied to obtain  $\hat \rho_0'=\hat S_{\rm DI}\hat \rho_0\hat S_{\rm DI}^\dagger$, (ii)  the  state $\hat \rho_0'$ then undergoes dissipative evolution according to the Lindblad master equation
\begin{equation}
\frac{d\hat \rho}{dt} = N\kappa \left( \hat a\hat \rho \hat a^\dagger - \frac{1}{2} \{ \hat a^\dagger \hat a, \hat \rho \} \right)+ \frac{\Gamma_\phi}{2} \sum_j\left(\hat  \sigma_j^z \hat \rho \hat \sigma_j^z - \hat \rho \right)+
 \Gamma_1 \sum_j\left( \hat \sigma_j^- \hat \rho \hat \sigma_j^+ - \frac{1}{2} \{ \hat \sigma_j^+ \hat \sigma_j^-, \hat \rho \} \right). \label{lindblad}
\end{equation}
The evolution governed by this equation is applied for a time interval $\tau_{\rm R}$,  transforming $\hat \rho_0'$ into $\hat \rho_1$.
The prefactor $N$ in front of $\kappa$ reflects that fact that the Rabi gate is applied $N$ times to the resonator.}

  \end{document}